\documentclass[aps,prd,twocolumn,showpacs,nofootinbib,floatfix,superscriptaddress]{revtex4-1}

\newcommand{\Rmnum}[1]{\expandafter\@slowromancap\romannumeral #1@}
\makeatother
\usepackage[colorlinks,linkcolor=blue,anchorcolor=blue,citecolor=blue,urlcolor=blue,breaklinks=true]{hyperref}

\usepackage[font=normal]{subfig}
\usepackage{epsfig}
\usepackage{natbib}
\usepackage{epstopdf}
\usepackage{mathrsfs}
\usepackage{slashed}
\usepackage{amsmath}
\usepackage{verbatim}
\usepackage{graphicx}
\usepackage{amssymb}
\usepackage{psfrag}
\usepackage{array}
\usepackage{subfig}
\usepackage{url}

\usepackage{caption}
\usepackage{soul}
\usepackage{ulem}

\newcommand{\bef}{\begin{figure}[htb]}
\newcommand{\eef}{\end{figure}}

\def\bea#1\eea{\begin{align}#1\end{align}}

\begin{document}
\title{Parton rescattering and gluon saturation in dijet production at EIC}

\author{Yuan-Yuan Zhang}
\affiliation{Key Laboratory of Quark and Lepton Physics (MOE) and Institute of Particle Physics, Central China Normal University, Wuhan 430079, China}
\affiliation{School of Science and Engineering, The Chinese University of Hong Kong, Shenzhen, Shenzhen, Guangdong, 518172, P.R. China}
\affiliation{University of Science and Technology of China, Hefei, Anhui, 230026, P.R.China}

\author{Xin-Nian Wang}
\email{xnwang@lbl.gov}
\affiliation{Key Laboratory of Quark and Lepton Physics (MOE) and Institute of Particle Physics, Central China Normal University, Wuhan 430079, China}
\affiliation{Nuclear Science Division, Lawrence Berkeley National Laboratory, Berkeley, CA 94720, USA}
\email{current address}

\begin{abstract}
Large angle gluon radiations induced by multiple parton scatterings contribute to dijet production in deeply inelastic scattering off a large nucleus at the Electron-Ion Collider. Within the generalized high-twist approach to multiple parton scattering, contributions at the leading order and twist-four in perturbative QCD and large Bjorken momentum fraction $x_B$ can be expressed as a convolution of the multiple parton scattering amplitudes and the transverse momentum dependent (TMD) two-parton correlation matrix elements. We study this medium-induced dijet spectrum and its azimuthal angle correlation under the approximation of small longitudinal momentum transfer in the secondary scattering and the factorization of two-parton correlation matrix elements as a product of quark and gluon TMD parton distribution function (PDF). Contributions to dijet cross section from double scattering are power-suppressed and only become sizable for mini-jets at small transverse momentum. We find that the total dijet correlation for these mini-jets, which also includes the contribution from single scattering, is sensitive to the transverse momentum broadening in the quark TMD PDF at large $x$ and saturation in the gluon TMD PDF at small $x$ inside the nucleus. The correlation from double scattering is also found to increase with the dijet rapidity gap and have a quadratic nuclear-size dependence because of the Landau-Pomeranchuk-Migdal (LPM) interference in gluon emission induced by multiple scattering. Experimental measurements of such unique features in the dijet correlation can shed light on the LPM interference in strong interaction and gluon saturation in large nuclei.
\end{abstract}


\maketitle

\section{Introduction}

Jet quenching in high-energy heavy-ion collisions caused by energy loss of energetic partons due to multiple parton scattering and induced gluon radiation in a hot quark-gluon plasma (QGP) has been the subject of intense theoretical and experimental studies \cite{Gyulassy:2003mc,Wang:2004dn,Majumder:2010qh,Muller:2012zq,Mehtar-Tani:2013pia,Qin:2015srf,Cao:2020wlm} over the last several decades. Parton energy loss is dictated by the jet transport coefficient $\hat{q}$, which is defined as the averaged transverse momentum transfer squared per unit length and related to the gluon density distribution in the medium \cite{Baier:1996kr,Baier:1996sk,CasalderreySolana:2007sw,Liang:2008vz}. Recent extraction of the temperature \cite{Chen:2010te,Burke:2013yra,Xie:2019oxg,Feal:2019xfl} and jet energy dependence\cite{Kumar:2017des,Kumar:2019uvu,Soltz:2019aea,JETSCAPE:2021ehl} of $\hat{q}$ through phenomenological studies of jet quenching can provide further insights into the properties of the QGP.

Multiple parton scattering and jet quenching can also occur in deeply inelastic scattering (DIS) off a large nucleus at the electron-ion collider (EIC) when the struck quark propagates inside the cold nuclear medium. Quark energy loss due to induced gluon radiation  has been shown to cause the suppression of leading hadrons in semi-inclusive DIS (SIDIS) off nuclei  \cite{Guo:2000nz,Wang:2001ifa,Wang:2002ri,Zhang:2003wk,Zhang:2003yn,Arleo:2003jz,Majumder:2009ge}. The extracted jet transport coefficent $\hat q$ in cold nuclear matter from experimental data on the suppression of leading hadrons in SIDIS \cite{Chang:2014fba,Li:2020zbk} and transverse momentum broadening \cite{Ru:2019qvz} is about two orders of magnitude smaller than that in the hot QGP in high-energy heavy-ion collisions. This is expected since the jet transport coefficent is directly related to the gluon number density of the medium which is much smaller in a confined cold nuclear matter than a deconfined hot QGP. It is, nevertheless, of great interest to explore the physics mechanism behind such a small value of jet transport coefficient in the cold nuclear matter.

Several approaches have been developed to study parton energy loss in a medium with strong interaction over the last several decades. These includes Baier-Dokshitzer-Mueller-Peigne-Schiff-Zakharov (BDMPS-Z)~\cite{Baier:1996sk,Baier:1996kr,Zakharov:1996fv}, Arnold-Moore-Yaffe (AMY)~\cite{Arnold:2000dr, Arnold:2003zc},  Gyulassy-Levai-Vitev-Wiedemann (GLV-W)~\cite{Gyulassy:1999zd,Gyulassy:2000er, Gyulassy:2000fs,Wiedemann:2000za}, high-twist (HT) \cite{Guo:2000nz,Wang:2001ifa,Zhang:2003yn,Zhang:2003wk,Qin:2012fua} and  more recent Soft Collinear Effective Theory gluon (SCET$_{\rm G}$)~\cite{Ovanesyan:2011xy,Ovanesyan:2011kn,Kang:2014xsa} approaches with different approximations about the interaction between the propagating parton and the medium. Within the high-twist approach, one assumes the transverse momentum of the medium partons is small relative to that of the radiated gluon. Under such a collinear approximation an expansion in the medium parton transverse momentum is carried out and the collinear twist-four matrix elements can be factorized out \cite{Guo:2000nz,Wang:2001ifa,Kang:2013raa,Kang:2014ela,Kang:2016ron} in both DIS and Drell-Yan lepton pair production in p+A collisions. The collinear twist-four matrix elements can be related to the jet transport coefficient. Such an approach has been recently extended to a generalized high-twist approach \cite{Zhang:2018kkn,Zhang:2018nie,Zhang:2019toi} without collinear expansion in the transverse momentum of medium partons, by relaxing the approximation that medium parton transverse momentum is smaller than that of the radiated gluon. The final radiated gluon spectra can be expressed in terms of the transverse momentum dependent (TMD) jet transport coefficient which is just TMD gluon distribution density of the medium. In the soft radiation and small longitudinal momentum exchange limit, the final result becomes the same as that of the GLV approach under the first opacity approximation.

Within the generalized high-twist approach, one can also study multiple parton scattering in the opposite limit of the collinear exchanged gluon approximation. In this case, the initial transverse momentum from the medium gluon can be large and comparable to the transverse momentum of the final radiated gluon. Therefore, the final dijet spectra from multiple scattering will carry the information of the medium gluon TMD distribution. This will be the focus of this study.
What’s more, the hard splitting and inclusion of initial quark transverse momentum in the generalized High-Twist approach made the dijet a natural observable to entangle the medium gluon TMD distribution. The dijet with low transverse momentum and large angle capture the medium information better, since under our approximation final quark and gluon transverse momentum should be same order as initial quark and gluon transverse momentum. 
mong several approaches to induced gluon radiation due to multiple scattering in high-energy heavy-ion collisions, BDMPS-Z and GLV can be adapted for the application to DIS eA collisions. BDMPS-Z assumes multiple soft scattering while GLV considers the leading order in opacity expansion. Our approach is very similar to GLV and we can recover GLV results in soft and static scattering limit [46]. These two approaches are a better approximation in DIS since the number of multiple scatterings in DIS eA is very small.  Our result should be valid at both small and large transverse momentum of the dijiet $l_T$ relative to the momentum imbalance $k_T$. This is especially important because as we will show the nuclear enhanced nuclear modification of dijet cross section is measurable at $l_T $~ a few GeV. At much larger  $l_T$, such nuclear modification is negligible for experimental observation.

Dijet production in proton-nucleus ($p$+A) collisions and electron-nucleus ($e$+A) DIS have been proposed to study the collinear nuclear parton distribution functions (PDF) \cite{Klasen:2018gtb,Guzey:2020zza}, nuclear TMD PDF, multi-gluon correlations and gluon saturation in large nuclei \cite{Marquet:2007vb,Dominguez:2011wm,Altinoluk:2015dpi,Hatta:2016dxp,Mantysaari:2019hkq,Salazar:2019ncp,Hatta:2020bgy}. In this paper, we will consider contributions to dijet production from multiple parton scattering in $e$+A DIS at the future EIC within the generalized high-twist approach. At large Bjorken $x_B$, the momentum fraction of the  struck quark from the virtual photon scattering in DIS,  dijet production at the leading order (LO) is dominated by large angle gluon splitting from the struck quark after the photon-quark (single) scattering.  Such large angle splitting can also be induced by a secondary scattering between the struck quark and a medium gluon from the nucleus target. Within the generalized high-twist approach, the spectrum of such medium-induced dijet production is related to the TMD PDF of soft gluons from the nuclear medium. The medium-induced dijet spectrum has a nuclear enhancement that is quadratic in the nuclear size $R_A$ and increases with the rapidity gap of the dijet, two unique features caused by the Landau-Pomeranchuk-Migdal (LPM) interference. We will study the sensitivity of this nuclear enhanced LO contribution to the medium gluon TMD PDF at small $x$, the scale of the gluon saturation and the transverse momentum broadening of the quark TMD PDF at large $x_B$.

 

The remainder of this paper is organized as follows. In Section \ref{sec-dijetcross}, we calculate the dijet cross sections for both single and double scattering (medium-induced splitting) in $e$+A DIS at large $x_B$. We will discuss the structure of the medium-induced splitting induced by different processes of multiple scattering with the LPM interference and how the final result depends on the two-parton correlation function which can be approximated as a  product of quark and gluon TMD PDF.  In Section \ref{sec-qhatTMDsaturation}, we will introduce a simple model for implementing gluon saturation in the parameterized gluon TMD PDF inside the nucleus and its relation to the jet transport coefficient $\hat q$. We will study the nuclear modifications to dijet spectra, from both single and double scattering through numerical calculations in Section \ref{sec-numerical}. We will also explore the azimuthal angle, rapidity gap and nuclear size dependence of these nuclear modifications. A summary and an outlook are given in Section \ref{sec-conclusion}.

\section{Dijet production in DIS}
\label{sec-dijetcross}

\subsection{Single scattering}
\label{subsec-single}

\bef
 \hspace*{-0.8cm}
 \includegraphics[scale=0.4]{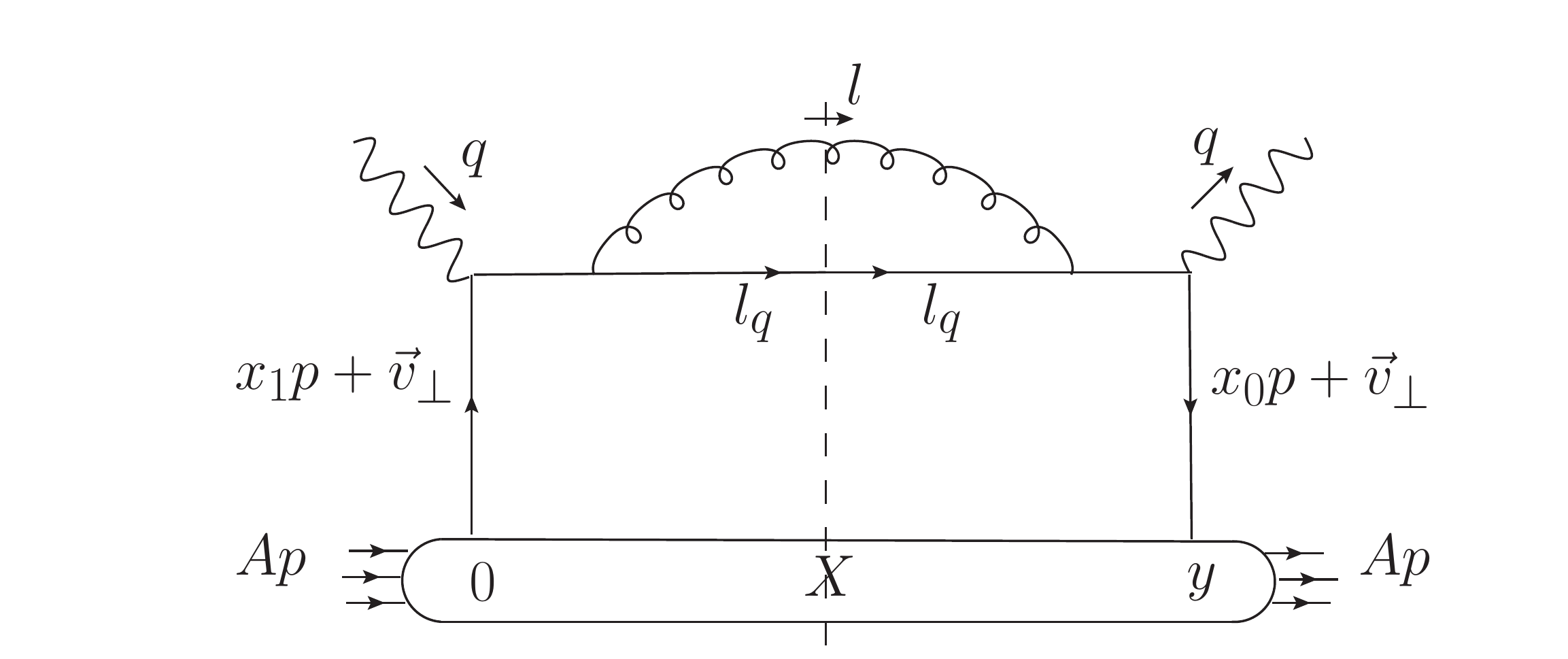} 
 \caption{Dijet production in single scattering at LO.}
 \label{fig:singlescattering}
\eef

We consider dijet production in DIS at large $x_B$ in this study. At LO, the dominant contribution in the single photon-quark  scattering is from large angle gluon splitting from the struck quark as illustrated in Fig.\ref{fig:singlescattering}. The contribution from photon-gluon fusion $\gamma*+g\rightarrow q+\bar q$ should be small because of the small gluon distribution at large $x_B$. 
With the kinematics of dijet production we consider here, the initial state radiation will mainly contribute to the QCD evolution of the quark distribution in the nuclear target. Interference between initial and final state radiation is power-suppressed ($k_T^2/Q^2$) which we neglect in this study. Similar approximations are considered when we calculate contributions from multiple parton scattering.

The kinematics in the Breit frame (see Fig.\ref{fig:singlescattering}) by our convention are,
\begin{equation}
\begin{split}
 &p = [p^+, 0, \vec{0}_{\perp}], \\
 &q = [-\dfrac{Q^2}{2q^-},q^-,\vec{0}_{\perp}]\equiv [-x_Bp^+,q^-,\vec 0_\perp],\\
 &l = [\dfrac{l_{\perp}^2}{2(1-z)q^-}, (1-z)q^-, \vec{l}_{\perp}], \\
 & l_q=[\dfrac{l_{q\perp}^2}{2zq^-}, zq^-, \vec{l}_{q\perp}],\\
 &\epsilon (l) = [\dfrac{\vec{\epsilon}_{\perp}\cdot \vec{l}_{\perp}}{(1-z)q^-}, 0, \vec{\epsilon}_{\perp}],
\end{split}
\label{eq:kinematics}
\end{equation}
where $p$ is the momentum per nucleon of the nuclear target, $q$ is the momentum of the virtual photon, $l_q$ is the momentum of the final quark, $l$ is the momentum of the radiative gluon with the polarization vector $\epsilon (l)$ in an axial gauge $A^- = 0$ and $\vec{v}_{\perp} = \vec{l}_{\perp}+ \vec{l}_{q\perp}$ is the initial transverse momentum of the struck quark that determines the momentum imbalance of the final dijet in the single scattering. Under this convention and the collinear approximation $Q\gg l_{q\perp}, l_\perp, v_\perp$, the hadronic tensor for gluon radiation from a single scattering in Fig.\ref{fig:singlescattering} is,
\begin{eqnarray}
\frac{d W_S^{\mu\nu}}{dz d^2l_{\perp}d^2l_{q\perp}} &=&\int d x_0    \frac{\alpha_{s}}{2 \pi} \frac{1+z^{2}}{1-z} H_{(0)}^{\mu \nu}(x_0) \nonumber \\
&\times& \frac{C_F}{\pi}  \frac{q_{A}(x_0, \vec{l}_{\perp}+ \vec{l}_{q\perp}) }{[\vec{l}_{\perp}-(1-z)(\vec{l}_{\perp}+ \vec{l}_{q\perp})]^{2}},
\end{eqnarray}
where,
\begin{eqnarray}
H^{\mu\nu}_{(0)} (x_0)&=& \frac{\pi \delta(x_0-x_B-x_L-x_E) }{2p^+q^-} \\
&&\times{\rm Tr}[\gamma\cdot p \gamma^{\mu} \gamma \cdot (q+x_0p)\gamma^{\nu}  ] ,\nonumber \label{eq:H0munu} \\
x_L &=& \frac{l_{\perp}^2}{2p^+q^-z(1-z)}, \quad
x_E = \frac{v_{\perp}^2 - 2\vec{v}_{\perp}\cdot \vec{l}_{\perp}}{2p^+q^- z} .
\end{eqnarray}
In the above equation, 
\begin{eqnarray}
q_A(x,\vec v_\perp) &=& \int \frac{dy^-}{2\pi} \frac{d^2y_\perp}{(2\pi)^2}
e^{-ixp^+y^- +i\vec v_\perp\cdot \vec y_\perp} \nonumber \\
&& \times \langle A \mid \bar\psi(y^-,\vec y_\perp)\frac{\gamma^+}{2} \psi(0,\vec 0_\perp )
\mid A \rangle ,
\label{eq:nqpdf}
\end{eqnarray}
is the quark TMD PDF inside the nucleus target. This factorized form of the dijet cross section in DIS has been proven  both within the traditional pQCD approach \cite{Boer:2011fh} and SCET approach \cite{Kang:2020xgk,delCastillo:2020omr} when the dijets are back-to-back.  Since we focus on nuclear modification of dijet cross section in the leading order in this study, we also assume that such factorization can be applied to the double hard scattering processes with large transverse momentum imbalance in the following section. The rigorous proof of the factorization and study of the associated soft functions arising from soft interactions is beyond the scope of this paper.

In the large $x_B$ region, we assume  $x_B \gg x_L , x_E $. The quark TMD PDF can be approximated as $q_A(x_B+x_L+x_E, \vec{l}_{\perp}+ \vec{l}_{q\perp})\approx q_A(x_B, \vec{l}_{\perp}+ \vec{l}_{q\perp})$. Contracting the above hadronic tensor with the leptonic tensor, one can get the dijet cross section from gluon radiation associated with a single scattering in $e$+A DIS,
\begin{widetext}
\begin{equation}
 \frac{d\sigma_{e{\rm A}}^{S}}{dx_B dQ^2 dz d^2l_{\perp}d^2l_{q\perp}}  \approx \frac{2\pi \alpha_{\text{em}}^2}{Q^4}\sum_{q} e_{q}^{2}   [1+(1-\frac{Q^2}{x_Bs})^2]
  \frac{\alpha_{s}}{2 \pi} \frac{ 1+z^{2}}{1-z} \frac{C_F}{\pi}  \frac{q_A(x_B, \vec{l}_{\perp}+ \vec{l}_{q\perp})}{[\vec{l}_{\perp}-(1-z)(\vec{l}_{\perp}+ \vec{l}_{q\perp})]^{2}},
  \label{eq:single-crsec}
\end{equation}
\end{widetext}
where  $s=(p+q)^2=2p^+q^-$ is the center-of-mass energy squared of the photon-nucleon scattering,  $\alpha_{\text{em}}$ is the fine-structure constant in quantum electrodynamics (QED) and $e_q$ is the electric charge of the quark. 

In the definition of the quark TMD PDF in Eq.~(\ref{eq:nqpdf}), we have omitted the Wilson gauge link between two quark field operators due to multiple soft interaction between the quark and the nucleus target which guarantees the gauge invariance of the quark TMD PDF \cite{Ji:2004wu}. Such soft interactions in a nuclear target embedded in the quark TMD PDF leads to an effective transverse momentum broadening of the initial quark \cite{Liang:2008vz},
\begin{equation}
q_A(x,\vec v_\perp) =\int dy_0^- d^2b_\perp \rho_A(y_0^-, b_\perp)
q_N(x,\vec v_\perp, b_\perp),
\label{eq:qpdfa}
\end{equation}
where $\rho_A$ is the nuclear density distribution with the normalization, 
\begin{equation}
\int dy_0^- d^2b_\perp \rho_A(y_0^-,b_\perp)=A,
\end{equation}
$y_0^-$ is the light-cone coordinate in the Breit frame for the primary photon-quark scattering and $b_\perp$ is the impact-parameter of the photon-nucleus interaction.  The effective quark TMD PDF per nucleon inside a nucleus, with transverse momentum broadening from multiple soft collinear or eikonal interaction between the quark and the nucleus, is given at LO \cite{Liang:2008vz} by the convolution of the nucleon's quark TMD PDF in vacuum $q_N^0(x,\vec u_\perp)$ and a Gaussian broadening:
\begin{eqnarray}
 q_N(x,\vec v_\perp, b_\perp) &\equiv&
\frac{R_A^q(x,b_\perp)}{\pi \Delta_F(b_\perp)} 
\int d^2u_\perp e^{-\frac{(\vec v_\perp-\vec u_\perp)^2}{\Delta_F(b_\perp)}}q_N^0(x,\vec u_\perp), \nonumber \\ 
\Delta_{F}(b_\perp)&=&\int dy_0^-\hat q_F(y_0^-,b_\perp).
\label{eq:qpdf}
\end{eqnarray}
 The broadening is characterized by the quark transport coefficient \cite{Liang:2008vz},
\begin{eqnarray}
\hat{q}_F(y_0^-,b_\perp) &=&\rho_A(y_0^-,b_\perp) \frac{2\pi^2}{N_c} \int \frac{d^2 \vec{k}_{\perp}}{(2\pi)^2}  \alpha_{\rm s}
\phi_N(x_G, \vec{k}_{\perp})|_{x_G\approx 0} \nonumber \\
&\approx& \frac{2\pi^2}{N_c} \alpha_{\rm s} \rho_A(y_0^-,b_\perp) x_Gg(x_G)|_{x_G\approx 0},
\label{eq:qhat1}
\end{eqnarray}
which is approximately proportional to the soft gluon distribution density $xg(x)\rho_A=\rho_A\int d^2k_\perp\phi_N(x, \vec{k}_{\perp})/(2\pi)^2$.
Here, $\alpha_{\rm s}$ is the strong coupling constant and $\phi_N(x,\vec{k}_{\perp})$ is the gluon TMD PDF per nucleon inside the nucleus, 
\begin{eqnarray}
\phi_N(x,\vec{k}_{\perp}) &=& \int \frac{d y_{12}^-}{2\pi p^+} 
\int d^2  \vec{y}_{12\perp} e^{-ixp^+ y_{12}^- + i\vec{k}_{\perp}\cdot \vec{y}_{12\perp}} \nonumber \\
&\times& \langle N|  F_{\alpha}^{\;\;+}( y_{12}^-, \vec{y}_{12\perp}) F^{+\alpha}(0, \vec{0}_{\perp}) |N\rangle.
\label{equ:gluonTMD}
\end{eqnarray}
We also take into account of the nuclear modification of the collinear or transverse-momentum-integrated PDF through a modification factor $R_A^q(x,b_\perp)$ in Eq.~(\ref{eq:qpdf}). Such a modification factor has been parameterized through global fits to experimental data \cite{Hirai:2007cx,Eskola:2009uj}.  In the numerical studies in this paper we always consider $x_B=0.2$ and $Q^2=200$ GeV$^2$.  The modification factor $R_A^q(x_B,b_\perp)\approx 1$ in this kinematics.

\subsection{Double scattering}

In $e$+A DIS, as the struck quark from the photon-quark interaction goes through multiple interactions inside the nucleus, a secondary hard scattering with another nucleon inside the nucleus can induce a gluon radiation leading to the medium-induced dijet production.
 We refer this secondary scattering as hard since the medium gluon carries small but finite momentum fraction as compared to the soft interactions that lead to the Wilson gauge link in the quark TMD PDF and the quark transverse momentum broadening.
This process will interfere with the dijet production from the initial photon-quark scattering in which the final quark or radiated gluon scatters with another nucleon inside the nucleus, leading to the LPM interference. In Ref.~\cite{Zhang:2019toi}, we have calculated the radiative gluon spectrum induced by double scattering in DIS including LPM interference. For this study, we extend the calculation to include the initial transverse momentum of the quark $\vec v_\perp$ in the induced gluon spectra. In the large $x_B$ region, one can treat the longitudinal momentum transfer in the second scattering as small. When this momentum fraction comes from the medium gluon, the steep falling gluon TMD PDF effectively cuts off contributions when the momentum fraction is large. The hadronic tensor for gluon radiation induced by double scattering can be factorized in terms of the hard part of the photon-quark scattering $H^{\mu\nu}_{(0)}$, the transverse part of the induced gluon spectra per mean-free-path ${\cal N}_g$ and the TMD quark-gluon correlation function $T_{qg}^A$: 
\begin{widetext}
\begin{eqnarray}
 \dfrac{dW^{\mu \nu}_{D}}{dz} &=& \int dx_0 dy_{0}^- dy_{1}^- d^2b_\perp \int d^2l_\perp \int d^2 v_\perp \int \frac{d^2k_\perp}{(2\pi)^2}  H^{\mu\nu}_{(0)}(x_0)\frac{\alpha_{\rm s}}{2\pi} \frac{1+z^2}{1-z} \frac{2\pi \alpha_{\rm s}}{N_c}{\cal N}_g \nonumber \\
 &\times& T_{qg}^A(y_{0}^-, y_{1}^-, \vec b_\perp, x_0,x_1,x_2,\vec v_\perp,\vec k_\perp) ,
\label{eq:hadronic-novp}
\end{eqnarray}

\begin{eqnarray}
\hspace{-0.2in} T_{qg}^A(y_{0}^-, y_{1}^-, \vec b_\perp, x_0,x_1,x_2,\vec v_\perp,\vec k_\perp) & \equiv & \int \frac{dy^-}{2\pi} \frac{ d^2y_\perp}{(2\pi)^2}  dy_{12}^-  d^2\vec{y}_{12\perp}  e^{-ix_0 p^+y^- +i\vec v_\perp\cdot\vec y_\perp} e^{-ix_2p^+y_{12}^- + i\vec{k}_{\perp}\cdot\vec{y}_{12\perp}} e^{i(x_0-x_1)p^+y_1^-}  \nonumber  \\
 & & \hspace{-0.7in} \times  \langle A |\bar{\psi}(y_0^-+y^{-},\vec{b}_{\perp}+\vec y_\perp) \frac{\gamma^{+}}{2} A^{+}(y_{1}^{-}+y_{12}^-, \vec{b}_{\perp}+\vec{y}_{12\perp}) A^{+}(y_{1}^{-}, \vec{b}_\perp) \psi(y_0^-,\vec{b}_\perp)| A \rangle\theta(f_1)\theta(f_2),
 \label{eq:Tqgcorr}
\end{eqnarray}

\begin{equation}
\theta(f_1) \theta(f_2)=\left\{\begin{array}{lll}{\theta(y_2^-)\theta(y_1^- -y^-):} & {\text {central,}} \\ {\theta(y_2^- -y_1^-)\theta(y_1^- -y^-):} & {\text {left,}}\\{\theta(y_1^- -y_2^-)\theta(y_2^-):}& {\text {right.}}
\end{array} \right.
\label{eq:theta_func}
\end{equation}

\end{widetext}
where the dijet momentum imbalance is $\vec{l}_{\perp}+\vec{l}_{q\perp}=\vec{v}_{\perp}+\vec{k}_{\perp}$ because of the momentum conservation,  $b_\perp$ is the impact-parameter of the photon-nucleus collisions, $y_0^-$ and $y_1^-$ are the light-cone coordinates in the Breit frame of the primary photon-quark and the secondary quark-gluon scattering, respectively. The above result is the same as that in our last study \cite{Zhang:2019toi} without the initial transverse momentum of the struck quark after the substitution $\vec{l}_{\perp} \rightarrow \vec{l}_{\perp} - (1-z)\vec{v}_{\perp}$. 

The hadronic tensors for 23 different diagrams are listed in Appendix \ref{append-HadronicTensor}. Take the central cut double scattering diagram in Fig. \ref{fig:example-diagram} as an example, its hadronic tensor has two terms, with two different TMD quark gluon correlation functions:

 \begin{figure}[h]
\centering
{
  \includegraphics[scale=0.4]{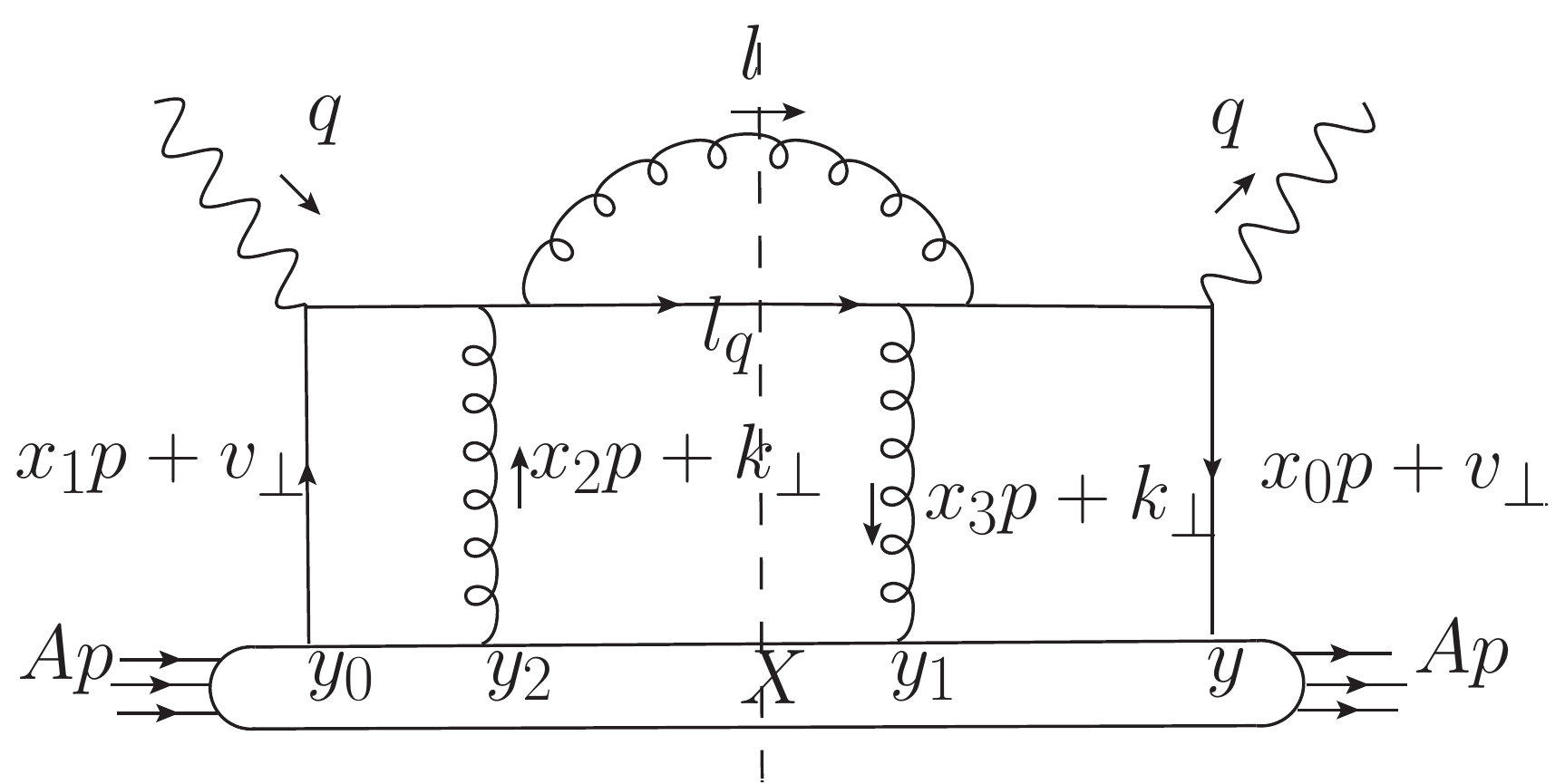}
 }
  \caption{Central Cut 12}
  \label{fig:example-diagram}
\end{figure}

\begin{widetext}

\begin{eqnarray}
 \dfrac{dW^{\mu \nu}_{D,\text{Fig.}\ref{fig:example-diagram}}}{dz} &=& \int dx_0 dy_{0}^- dy_{1}^- d^2b_\perp \int d^2l_\perp \int d^2 v_\perp \int \frac{d^2k_\perp}{(2\pi)^2} H^{\mu\nu}_{(0)}(x_0) \frac{\alpha_{\rm s}}{2\pi} \frac{1+z^2}{1-z} \frac{2\pi \alpha_{\rm s}}{N_c} {\cal N}_g^{\text{Fig.} \ref{fig:example-diagram}} \nonumber \\
 & & \times  \left[T_{qg}^A(y_{0}^-, y_{1}^-, \vec b_\perp, x_B+x_F,x+x_F,x_L+x_S-x_F,\vec v_\perp,\vec k_\perp)\right.\\
 & &\left. +T_{qg}^A(y_{0}^-, y_{1}^-, \vec b_\perp, x_B+x_L+x_E,x+x_F,x_L+x_S-x_F,\vec v_\perp,\vec k_\perp) \right], \nonumber
\label{eq:hadronic-fig2}
\end{eqnarray}

\begin{equation}
    {\cal N}_g^{\text{Fig.} \ref{fig:example-diagram}} = \frac{1}{2N_c} \frac{[\vec{l}_{\perp} -(1- z)\vec{v}_{\perp}]\cdot [\vec{l}_{\perp} -(1- z)(\vec{v}_{\perp}+\vec{k}_{\perp})]}{[\vec{l}_{\perp} -(1- z)\vec{v}_{\perp}]^2[\vec{l}_{\perp} -(1- z)(\vec{v}_{\perp}+\vec{k}_{\perp})]^2} 
\end{equation}

\end{widetext}

 The above result is obtained after integration over the light-cone momentum fractions of initial parton lines in the Feynman diagram $x_0, x_1,x_2$ and $x_3$. One of these momentum fractions, $x_3$, is fixed through momentum conservation $x_1+x_2=x_0+x_3$.  The momentum fraction $x_0$ is fixed by the on-shell condition of the final quark which appears as a $\delta$ function in the hard part $H0^{\mu\nu}$.  Two poles in the quark propagators effectively put two of the intermediate quark lines on shell through a contour integration, fixing both momentum fractions $x_1$ and $x_2$, in terms of the momentum fractions defined as
\begin{equation}
\begin{aligned}
& x_F = \frac{{v}_{\perp}^2}{2p^+q^-},\\
 &x_D = \frac{k_{\perp}^2 -2\vec{k}_{\perp}\cdot \vec{l}_{\perp}}{2p^+q^-z} ,  \quad x_L = \frac{l_{\perp}^2}{2p^+q^-z(1-z)},\\
 & x_E = \frac{v_{\perp}^{2}-2 \vec{v}_{\perp} \cdot \vec{l}_{\perp}}{2 p^{+} q^{-} z}, \quad x_H = \frac{(\vec{v}_{\perp}+\vec{k}_{\perp})^2}{2p^+q^-}, \\
& x_S =  \frac{(\vec{v}_{\perp}+\vec{k}_{\perp})^2-2(\vec{v}_{\perp}+\vec{k}_{\perp})\cdot \vec{l}_{\perp}}{2p^+q^-z}. 
\end{aligned}
\label{equ:x-variables}
\end{equation} 
These typically are related to the momentum fractions that the initial partons need to carry in order to produce the final state gluon and quark with the given transverse momentum $l_T$, $k_T$ and $v_T$. The two contributions in Eq.~(\ref{eq:hadronic-fig2}) from Fig.~\ref{fig:example-diagram} correspond to two different sets of poles in the contour integration: $x_0=x_B+x_F, x_1 = x_B+x_F, x_2 = x_L+x_S-x_F$; $x_0=x_B+x_L+x_E, x_1 = x_B+x_F, x_2 = x_L+x_S-x_F$. We refer readers to Refs.~\cite{Wang:2001ifa} and \cite{Zhang:2019toi} for details about these contour integration and the different choices of the momentum fractions.

In the above result, we have assumed that the momentum fraction $x_B$ carried by the initial quark, usually referred to as the Bjorken variable, is much larger than other longitudinal momentum fractions carried by the medium gluon in different scattering amplitudes as shown in the Feynman diagrams in the Appendix \ref{append-HadronicTensor}.

We consider the dominant case for a large nucleus target in which the primary photon-quark scattering happens at the light-cone coordinate $(y_0^-, \vec{b}_{\perp})$ of one nucleon while the secondary quark-gluon scattering happens at $(y_1^-,\vec{b}_{\perp})$ of another nucleon inside the nucleus in the Breit frame. We neglect the case that two scatterings happen inside the same nucleon which is not enhanced by the nucleus size. We further assume that the TMD quark-gluon correlation function in Eq.~(\ref{eq:Tqgcorr}) can be factorized as a product of the initial quark TMD PDF of a nucleon and the soft gluon TMD PDF of another nucleon inside the nucleus \cite{Zhang:2019toi,Osborne:2002st},

\begin{eqnarray}
 T_{qg}^A(y_{0}^-, y_{1}^-, b_\perp, x_0,x_1,x_2,\vec v_\perp,\vec k_\perp)
 &=&\rho_A(y_0^-, b_\perp) \rho_A(y_1, b_\perp)  \nonumber \\
 & & \hspace{-1.8in} \times  q_N(x_0,\vec v_\perp, b_\perp) e^{i(x_0-x_1)p^+y_1^-} \frac{\phi_N(x_2,\vec k_\perp)}{k_\perp^2},
 \label{eq:tqgcorr2}
\end{eqnarray}
where $q_N(x,\vec v_\perp, b_\perp)$ is the effective quark TMD PDF per nucleon inside the nucleus as defined in Eq.~(\ref{eq:qpdf}) and $\phi_N(x_2,\vec k_\perp)$ is the gluon TMD PDF per nucleon inside the nucleus as defined in Eq.~(\ref{equ:gluonTMD}). If one also considers nuclear effects on the medium gluon TMD distribution $\phi_N(x_2,\vec k_\perp)$ such as gluon saturation, the gluon TMD distribution also depends on the impact-parameter $b_\perp$ indirectly through the saturation scale as we will discuss later. This is implicitly assumed in the following calculations.

With the above factorized TMD quark and gluon correlation function, the medium-induced dijet cross section from double scattering in $e$+A DIS can be expressed as,
\begin{widetext}
\begin{eqnarray}
\frac{d\sigma_{e{\rm A}}^D}{dx_B dQ^2 dz d^2l_{\perp}d^2l_{q\perp}}  &=& \frac{2\pi \alpha_{\text{em}}^2}{Q^4}\sum_{q} e_{q}^{2}   [1+(1-\frac{Q^2}{x_Bs})^2]\frac{\alpha_{\rm s}}{2 \pi} \frac{1+z^{2}}{1-z} \frac{2\pi \alpha_{\rm s}}{ N_c} \int \frac{d^2k_{\perp}}{(2\pi)^2} \int d^2b_{\perp} dy_0^- dy_1^- \nonumber \\
& & \hspace{-1.0in} \times \rho_A(y_0^-, b_{\perp}) \rho_A(y_1^-, b_{\perp})
q_N(x_B,\vec v_\perp,b_\perp)
\frac{\phi_N(x_G,\vec{k}_{\perp})}{{k}_{\perp}^2} \left[{\cal N}_g^{\text{qLPM}} + {\cal N}_g^{\text{gLPM}} +{\cal N}_g^{\text{nonLPM}}  \right].
\label{equ:nXsection}
\end{eqnarray}

\begin{eqnarray}
 && {\cal N}_g^{\text{qLPM}} = \frac{1}{N_c}\left( \frac{[\vec{l}_{\perp} -(1- z)\vec{v}_{\perp}]\cdot [\vec{l}_{\perp} -(1- z)(\vec{l}_{\perp}+\vec{l}_{q\perp})]}{[\vec{l}_{\perp} -(1- z)\vec{v}_{\perp}]^2[\vec{l}_{\perp} -(1- z)(\vec{l}_{\perp}+\vec{l}_{q\perp})]^2} - \frac{1}{[\vec{l}_{\perp} - (1-z)\vec{v}_{\perp}]^2}  \right)  \nonumber \\
  & & \hspace{1.0in}\times\left(1-\cos[(x_L+x_E-x_{F})p^+(y_1^- -y_0^-)]\right),
  \label{eq:qLPM} \\
 && {\cal N}_g^{\text{gLPM}} = C_A\left( \frac{2}{[\vec{l}_{\perp} -(1- z)\vec{v}_{\perp}-\vec{k}_{\perp}]^2}-\frac{[\vec{l}_{\perp} -(1- z)\vec{v}_{\perp}-\vec{k}_{\perp}]\cdot [\vec{l}_{\perp} -(1- z)(\vec{l}_{\perp}+\vec{l}_{q\perp})]}{[\vec{l}_{\perp} -(1- z)\vec{v}_{\perp}-\vec{k}_{\perp}]^2[\vec{l}_{\perp} -(1- z)(\vec{l}_{\perp}+\vec{l}_{q\perp})]^2}     \right.  \nonumber \\
   & & \hspace{0.6in} - \left.\frac{[\vec{l}_{\perp} - (1-z)\vec{v}_{\perp}]\cdot [\vec{l}_{\perp} - (1-z)\vec{v}_{\perp}-\vec{k}_{\perp}]}{[\vec{l}_{\perp} - (1-z)\vec{v}_{\perp}]^2[\vec{l}_{\perp} - (1-z)\vec{v}_{\perp}-\vec{k}_{\perp}]^2 } \right)  \times (1-\cos[(x_L+\frac{z}{1-z}x_D+x_S-x_F)p^+(y_1^- -y_0^-)]),
    \label{eq:gLPM} \\
    && {\cal N}_g^{\text{nonLPM}}=C_F\left(\frac{1}{[\vec{l}_{\perp}-(1-z)(\vec{l}_{\perp}+\vec{l}_{q\perp})]^{2}}  -    \frac{1}{[\vec{l}_{\perp} - (1-z)\vec{v}_{\perp}]^2} \right),
\label{eq:nonLPM}
\end{eqnarray}

\end{widetext}

The medium gluon in different processes of double scattering and induced radiation (see the Appendix \ref{append-HadronicTensor}) carries different longitudinal momentum fractions as listed in Eq.~(\ref{equ:x-variables}). They are assumed to be small as compared to that of the primary struck quark $x_B$. 
 For $l_{\perp}^2 \sim k_{\perp}^2 \sim v_{\perp}^2 \ll Q^2$, we use $x_G$ to represent the small values of the longitudinal momentum fractions of the medium gluon in the medium-induced dijet cross section,
 \begin{equation}
     x_G = \frac{k_{\perp}^2}{2p^+q^-}.
 \end{equation}

 Because of the interference among different radiation amplitudes, the final results for ${\cal N}_g^{\rm qLPM}$, ${\cal N}_g^{\rm gLPM}$, and ${\cal N}_g^{\rm nonLPM}$ all vanish  when $k_T\rightarrow 0$. This cancels the diverging behavior of the $1/k_T^2$ factor associated with the gluon TMD in the final dijet spectra in Eq.~(\ref{equ:nXsection}).

\begin{figure}[h]
\centering
{
  \includegraphics[scale=0.4]{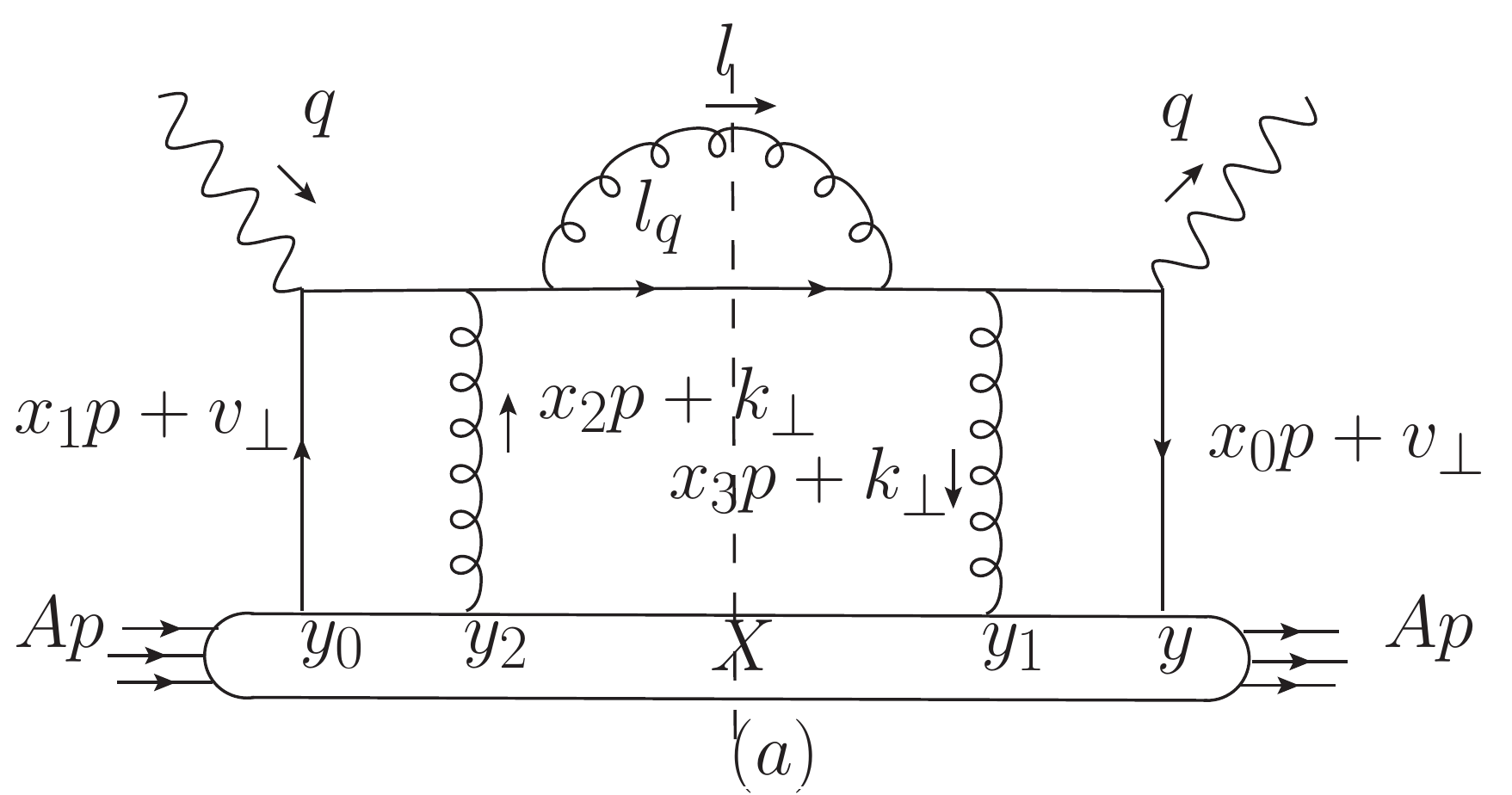} 
  \includegraphics[scale=0.4]{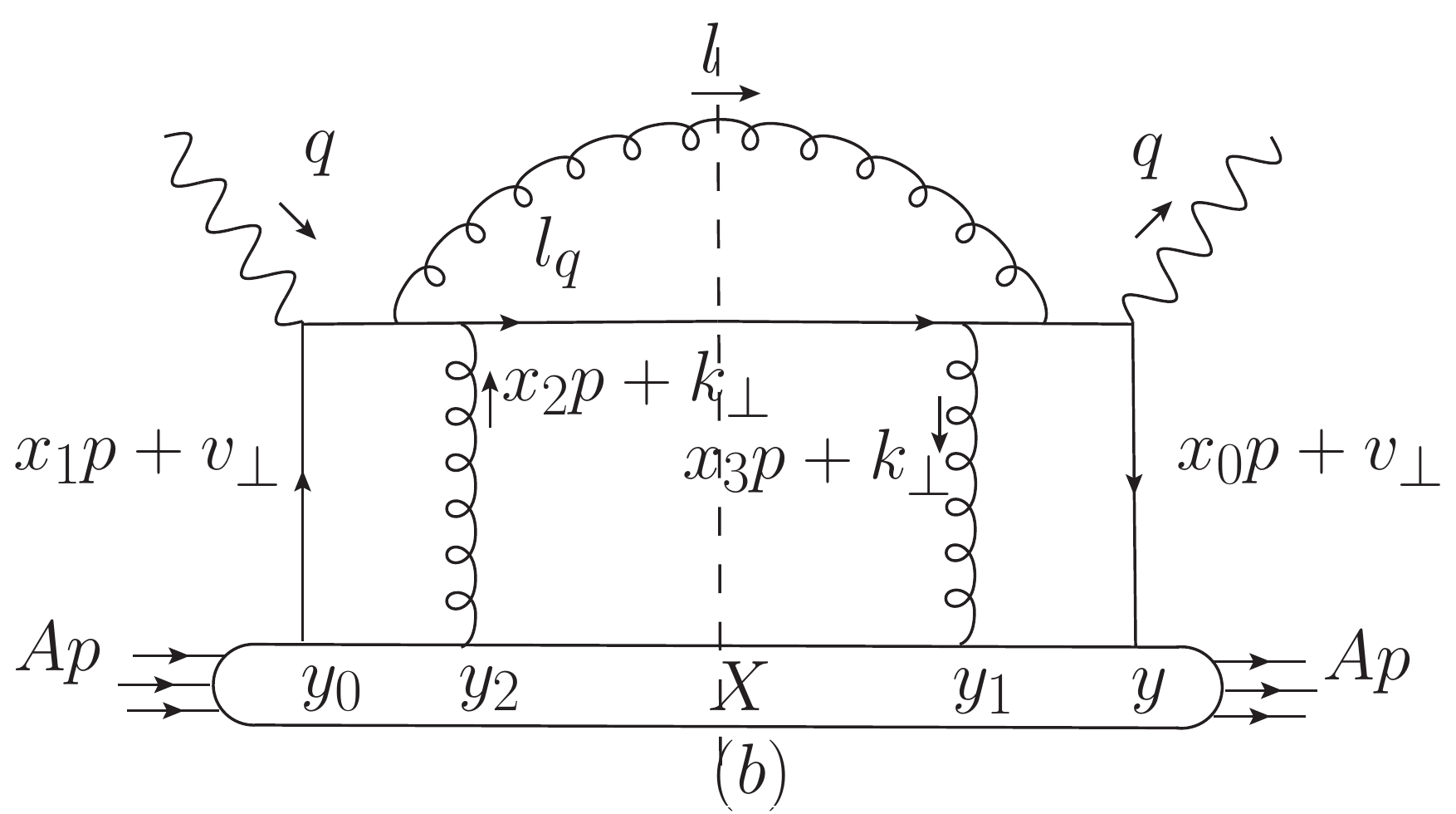} 
  \includegraphics[scale=0.4]{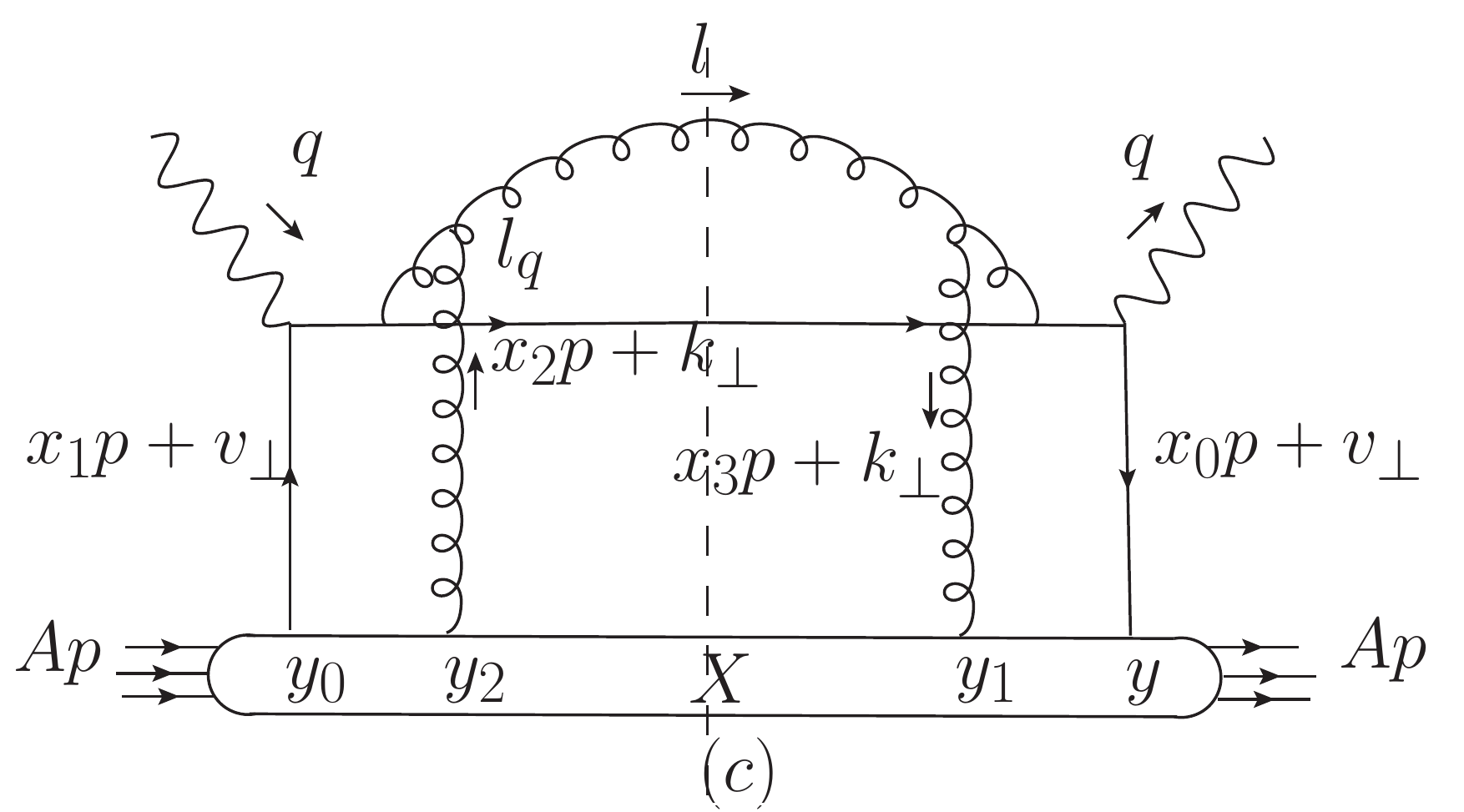}
  }
  \caption{Symmetric central-cut diagrams that contribute to (a) non-LPM term (${\cal N}_g^{\rm nonLPM}$),  (b) q-LPM term (${\cal N}_g^{\rm qLPM}$) and  (c) g-LPM term (${\cal N}_g^{\rm gLPM}$) in the medium-induced gluon spectrum rate.}
  \label{fig:diffterm}
\end{figure}

We have separated the transverse part of the induced gluon spectrum per mean-free-path or induced gluon spectrum rate  ${\cal N}_g$ into three different terms according to how the gluon is radiated in the symmetric central-cut diagrams. The symmetric central-cut diagrams for these three terms are illustrated in Figs.~\ref{fig:diffterm}(a), (b), and (c). The corresponding amplitudes of the radiative processes in these central-cut diagrams are illustrated in Figs.~\ref{fig:amplitude}(a), (b), and (c).  We note that the central-cut diagram in Fig.~\ref{fig:diffterm}(a) contains the final-state gluon radiation of the quark-gluon scattering, while Fig.~\ref{fig:diffterm}(b) contains both the final-state gluon radiation of the primary quark-photon scattering and the initial-state gluon radiation of the secondary quark-gluon scattering. Similarly, the central-cut diagram in Fig.~\ref{fig:diffterm}(c) contains the final-state radiation from  photon-quark scattering followed by gluon-gluon scattering and gluon radiation from the gluon propagator in the secondary quark-gluon scattering. In addition, one has to include all the interference among the three diagrams in Figs.~\ref{fig:amplitude}(a), (b), and (c) and the interference between radiation amplitudes induced by single (vacuum radiation) and triple scattering (quark-photon scattering followed by two quark-gluon scatterings) in left/right-cut diagrams. Our separation of the three different contributions to the induced gluon spectrum rate and their physics interpretations are based on the central-cut diagrams.

 \bef
 \centering
  \includegraphics[scale=0.5]{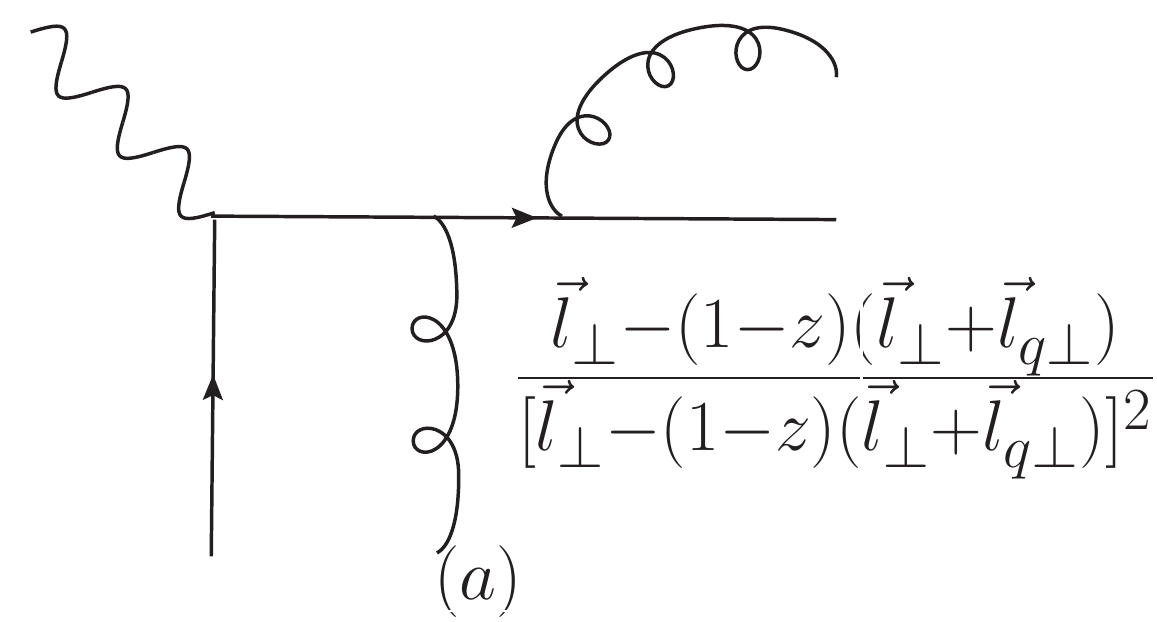} \\
 \includegraphics[scale=0.5]{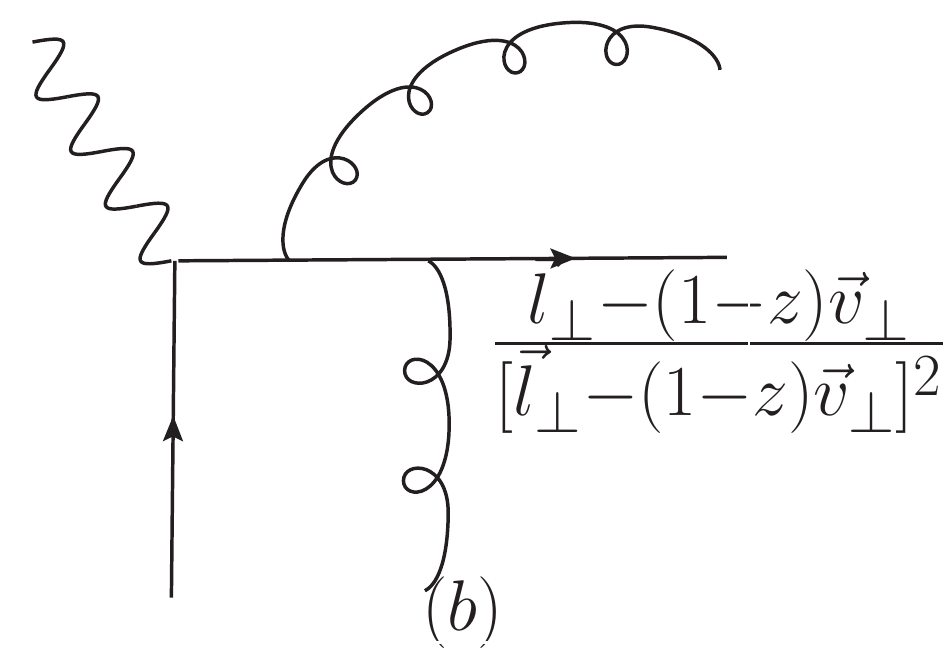} \\
 \includegraphics[scale=0.5]{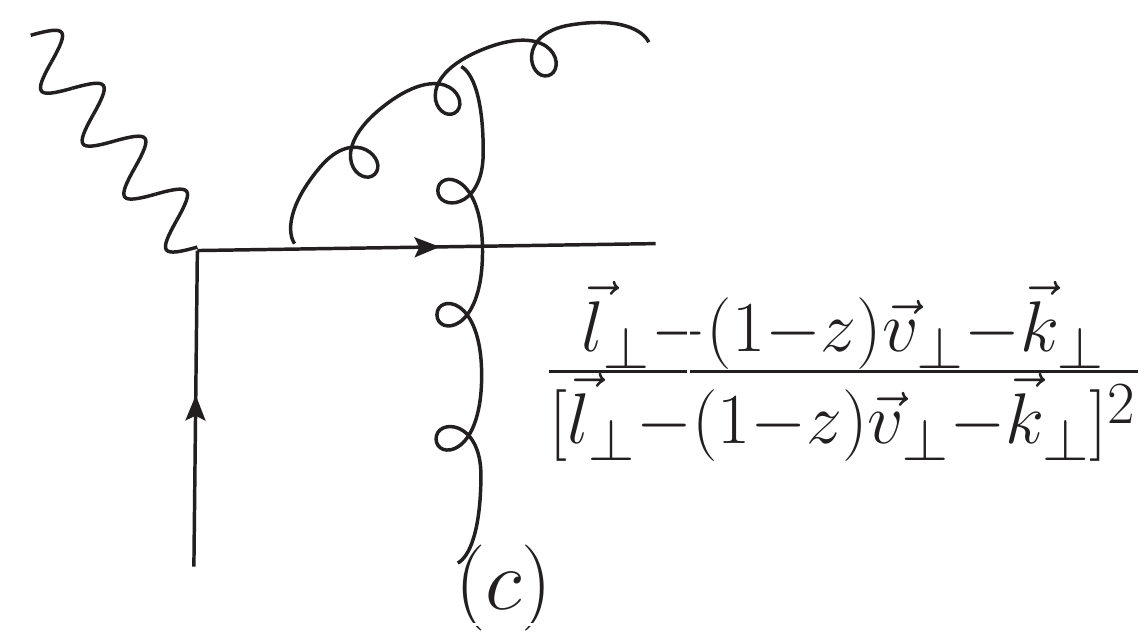}
 \caption{Diagrams for gluon radiation induced by double scattering that are responsible for (a) non-LPM amplitude, (b) q-LPM amplitude and (c) g-LPM amplitude.}
  \label{fig:amplitude}
 \eef

The first term of the gluon spectrum rate ${\cal N}_g^{\rm qLPM}$ in Eq.~(\ref{eq:qLPM}) comes from the final-state radiation of the photon-quark scattering in  Fig.~\ref{fig:amplitude}(b) and its interference with the amplitude of the final-state gluon radiation in  Fig.~\ref{fig:amplitude}(a).
The LPM interference between gluon radiation induced by photon-quark scattering at $y_0^-$ and quark-gluon scattering at $y_1^-$ leads to the suppression factor $1-\cos[(y_1^- -y_0^-)/\tau_{qf}]$ when the distance $y_1^- -y_0^-$ is smaller than the formation time, 
\begin{equation}
\tau_{qf}=\frac{1}{(x_L+x_E-x_{F})p^+} =  \frac{2q^{-} z(1-z)}{[\vec{l}_{\perp} - (1-z)\vec{v}_{\perp}]^2},
\end{equation}
which can be understood as the formation time of the radiated gluon from the struck quark in Fig.~\ref{fig:amplitude}(b).

The second term of the gluon spectrum rate ${\cal N}_g^{\rm gLPM}$ in Eq.~(\ref{eq:gLPM}) comes from the gluon radiation off the gluon propagator in the quark-gluon scattering in Fig.~\ref{fig:amplitude}(c) and its interference with the amplitudes of gluon radiation from the quark in Figs.~\ref{fig:amplitude}(a) and (b). The formation time $\tau_{gf}$ in the LPM suppression factor $1-\cos[(y_1^- -y_0^-)/\tau_{gf}]$ depends on the transverse momentum $\vec k_\perp$ of the medium gluon,
\begin{eqnarray}
 \tau_{gf}&=&\frac{1}{(x_L+\frac{z}{1-z}x_D+x_S-x_F)p^+}\nonumber \\
 &=& \frac{2q^{-} z(1-z)} {[\vec{l}_{\perp}-(1-z)\vec{v}_{\perp} - \vec{k}_{\perp}]^2},
\end{eqnarray}
which can be interpreted as the formation time of the intermediate gluon in Fig.~\ref{fig:amplitude}(c).

The third term of the induced gluon spectrum rate ${\cal N}_g^{\rm nonLPM}$ in Eq.~(\ref{eq:nonLPM}) comes from the final-state radiation in  Fig.~\ref{fig:amplitude}(a)  minus the initial-state radiation of the quark-gluon scattering  in Fig.~\ref{fig:amplitude}(b), where the  minus sign arises because of the space-like nature of the initial-state radiation.  This term contains radiation amplitudes from the beginning and the end of multiple scatterings
that do not participate in the LPM interference and is negligible when the number of scatterings is large. We keep this finite term since only two scatterings are considered in our study here.

In the soft radiation limit $z\rightarrow 1$, both ${\cal N}_g^{\rm nonLPM}$ and ${\cal N}_g^{\rm qLPM}$ vanish, only ${\cal N}_g^{\rm gLPM}$ remains and one recovers  the GLV result \cite{Gyulassy:1999zd,Gyulassy:2000fs}  for induced gluon spectra in the leading opacity approximation. In our study of medium-induced dijet production, we will keep all three contributions though ${\cal N}_g^{\rm gLPM}$ is the most dominant contribution as we will show in the final numerical results.

\section{Jet transport coefficient, TMD gluon distribution and saturation}
\label{sec-qhatTMDsaturation}

Because of the momentum conservation, the transverse momentum imbalance of the final dijet is related to the transverse momentum of the initial quark and medium gluon,
\begin{eqnarray}
    \vec l_\perp + \vec l_{q\perp}&=&\vec v_\perp \;\;\; ({\rm single \;\; scattering}),
     \\
    \vec l_\perp + \vec l_{q\perp}&=&\vec v_\perp + \vec k_\perp  \;\;\; ({\rm double \;\; scattering}).
\end{eqnarray}
Therefore, the transverse momentum broadening in the nuclear quark TMD PDF will affect the LO pQCD result for the dijet spectrum due to single scattering in $e$+A DIS through the jet transport coefficient $\hat q_F$ which is related to the gluon TMD PDF according to Eq.~(\ref{eq:qhat1}). The contribution from medium-induced dijet production due to double scattering, on the other hand,  will depend on the jet transport coefficient through quark transverse momentum broadening as well as directly on the medium gluon TMD PDF $\phi_N(x_G,\vec k_\perp)$ inside the nucleus. Therefore, the nuclear modification of the dijet spectrum in $e$+A DIS will be sensitive, both directly and indirectly,  to the medium gluon TMD PDF inside the nucleus.

\subsection{TMD PDF}
In this study, we will use the TMDlib package\cite{Hautmann:2014kza,Martinez:2018jxt} for the quark and gluon TMD PDF in nucleons and their scale evolution which are parameterized from global fits to experimental data. The gluon TMD distribution $\phi_N(x, k_{\perp},\mu^2)$ defined in our study is related to the TMD PDF $x\mathcal{A}(x, k_{\perp}, \mu^2)$ in TMDlib as,
\begin{equation}
\phi_N^0(x, k_{\perp}, \mu^2) = 4\pi x\mathcal{A}(x, k_\perp, \mu^2),
\end{equation}
which are related to the collinear gluon PDF as
\begin{eqnarray}
xg\left(x, \mu^{2}\right)&=&\int \frac{d^2k_\perp}{(2\pi)^2}\phi_N^0(x,k_\perp,\mu^2)
\nonumber \\
&=&\int \frac{d^2k_\perp}{\pi} x\mathcal{A}\left(x, k_\perp, \mu^{2}\right) .
\end{eqnarray}
This is an empirical formula for the relation between gluon TMD distribution and collinear distribution, since the connection between them is not very rigorous and fraught with theoretical subtleties \cite{Collins:2003fm,Collins:2016hqq,Gamberg:2017jha}.
Shown in Fig.~\ref{fig:tmdpdfpdf} are the collinear gluon PDF from HERA2.0PDF parameterization ``HERAPDF20-NLO-ALPHAS-118" in the LHAPDF package  \cite{Buckley:2014ana} (solid lines) and 
``PB-NLO-HERAI+II-2018-set1" \cite{Hautmann:2014kza,Martinez:2018jxt} parameterization for gluon TMD PDF in the TMDlib package after integrating over the transverse momentum (dashed line) at different scales $\mu^2 = 100, 1000, 10000$ GeV$^2$. Numerically they agree with each other reasonably well. 

\bef
\centering
\includegraphics[scale=0.55]{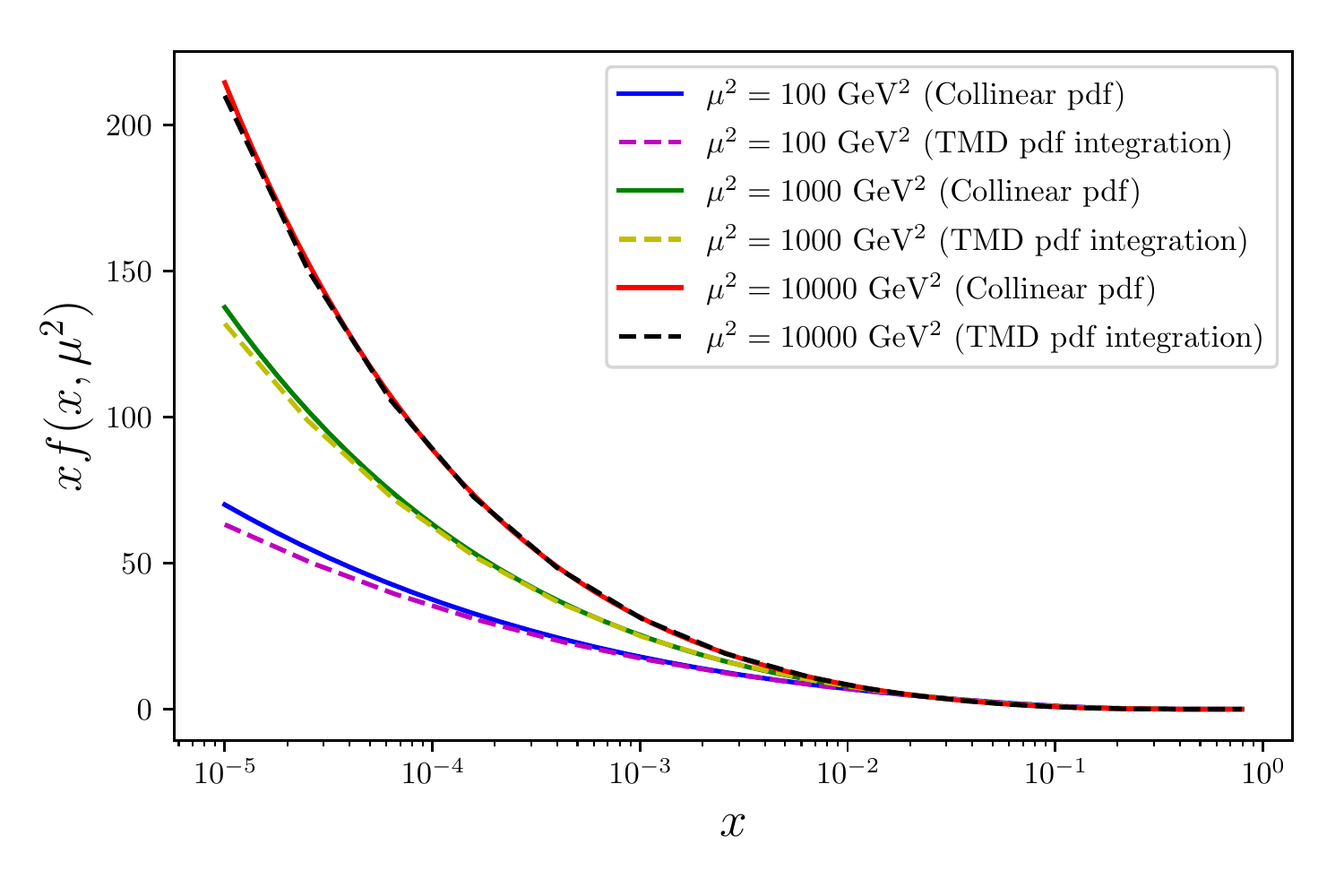}
\caption{The collinear gluon PDF from HERA2.0PDF ``HERAPDF20-NLO-ALPHAS-118" in the LHAPDF package (solid) as compared to that from the gluon TMD PDF  ``PB-NLO-HERAI+II-2018-set1" in the TMDlib package after integration over the transverse momentum (dashed) for evolution scales at $\mu^2$=100, 1000 and 10000 GeV$^2$.}
\label{fig:tmdpdfpdf}
\eef

Note that the TMDlib package has a minimum starting evolution scale $\mu_0^2$ for TMD PDF's. For the ``PB-NLO-HERAI+II-2018-set1" that we use, $\mu_0^2=1.9$ GeV$^2$.

\subsection{Gluon saturation}

Parton distributions inside a nucleus will have nuclear modifications because of the multiple interaction in the initial state. These initial state interactions can lead to transverse momentum broadening \cite{Liang:2008vz}, gluon saturation \cite{Gribov:1984tu,Mueller:1985wy,Mueller:1989st,McLerran:1993ni,McLerran:1993ka,Mueller:1999wm} and parton shadowing \cite{Qiu:1986wh,Brodsky:1989qz,Eskola:1993mb}  as well as nuclear modification of the PDF at large $x$ known as the EMC effect \cite{Ashman:1988bf,Arneodo:1988aa}.  The nuclear modification of the quark TMD distribution in this study will be given by Eq.~(\ref{eq:qpdf}) with a transverse momentum broadening. The nuclear modification factor $R^q_A(x_B,b_{\perp})$  [see Eq.~(\ref{eq:qpdf})] for collinear quark PDF will be given by the EPS09 parameterization \cite{Eskola:2009uj}. For gluon TMD PDF, we will consider the effect of gluon saturation.

To study the dijet spectrum in $e$+A DIS including the gluon splitting induced by double scattering, we propose a simple model for gluon saturation in the medium gluon TMD distribution,

\begin{equation}
\phi_N(x_G, k_\perp,\mu^2)=  \left \{
    \begin{array}{lr}
    \phi_N^0(\frac{Q_s^2}{Q^2}x_B, Q_s,\mu^2)|_{\mu^2=Q_s^2}, k_\perp<Q_s;   \\ \\
    \phi_N^0(x_G, k_\perp,\mu^2)|_{\mu^2=k_\perp^2},\quad\; k_\perp>Q_s,
     \end{array}
    \right .
    \label{equ:saturated_phi}
\end{equation}
where $\phi_N^0(x_G, k_\perp,\mu^2)$ is the nucleon gluon TMD distribution in vacuum as given by the parameterizations in TMDlib and $Q_s$ is the saturation scale. This model is similar to the ansatz for unintegrated gluon distribution function with saturation in Refs.~\cite{Kharzeev:2001gp,Dumitru:2011wq}. We will use the scale $\mu^2=k_\perp^2$ in the above equation to include the scale dependence of the medium gluon TMD distribution that is involved in the secondary quark-gluon scattering. The typical small longitudinal momentum fraction carried by the medium gluon in the double scattering in $e$+A DIS can be written as,
\begin{equation}
x_G = \frac{k_{\perp}^2}{2p^+q^-} = \frac{k_{\perp}^2}{Q^2}x_B.
\end{equation}
which is bounded by a lower limit $x_B{Q_s^2}/{Q^2}$ when the transverse momentum becomes smaller than the saturation scale $k_\perp\le Q_s$.

Gluon saturation happens in scattering processes in a nucleus or nucleon when the gluon density at small $x$ becomes high enough so that gluon fusion starts to overcome gluon splitting and the gluon density reaches a saturation limit. The saturation scale in this scenario can be related to the gluon density \cite{Mueller:1999wm},
\begin{eqnarray}
  & &Q_s^2(x_B,Q^2, b_\perp) \equiv \int dy^-  \hat{q}_A(y^-) \nonumber  \\ 
  &&\hspace{0.15in} =\frac{4\pi^2 C_A}{N_c^2-1} t_A(b_\perp)
  \int \frac{d^2k_\perp}{(2\pi)^2} \alpha_{\rm s}(\mu) \phi_N(x_G, k_{\perp},\mu^2),
  \label{equ:Q2} 
  \end{eqnarray}
where
\begin{equation}
t_A(b_\perp)=\int dy_0^-\rho_A(y_0^-,b_\perp)
\end{equation}
is the nuclear thickness function and $\hat{q}_A$ is the gluon jet transport coefficient. The integration range for $k_{\perp}$ is bounded by the kinematic constraint $(x_Gp^+)^2 +2k_{\perp}^2 \leq {p^+}^2$ or $k_\perp \le \sqrt{2Q^2(\sqrt{1+{1}/{(4x_B^2)}}-1)}$. A running coupling constant,
\begin{equation}
\alpha_{\rm s}(\mu) =  \frac{2\pi}{11-2N_f/3}\frac{1}{\ln \mu/\Lambda_{\rm QCD}},
\end{equation}
is used in the above equation for the saturation scale and in the dijet cross section, where $N_f = 3$ and $\Lambda_{\rm QCD} =0.246$ GeV. With the model for saturated gluon TMD distribution in Eq.~(\ref{equ:saturated_phi}) as probed by the propagating quark or gluon through multiple soft interaction, Eq.~(\ref{equ:Q2}) also measures the transverse momentum broadening of a gluon traveling through a large nucleus. Eq.~(\ref{equ:Q2}) is a self-consistent equation which can be solved for given values of $x_B$, $Q^2$ and the impact parameter $b_\perp$ in a nucleus $A$. 
 
\bef
\centering
\includegraphics[scale=0.6]{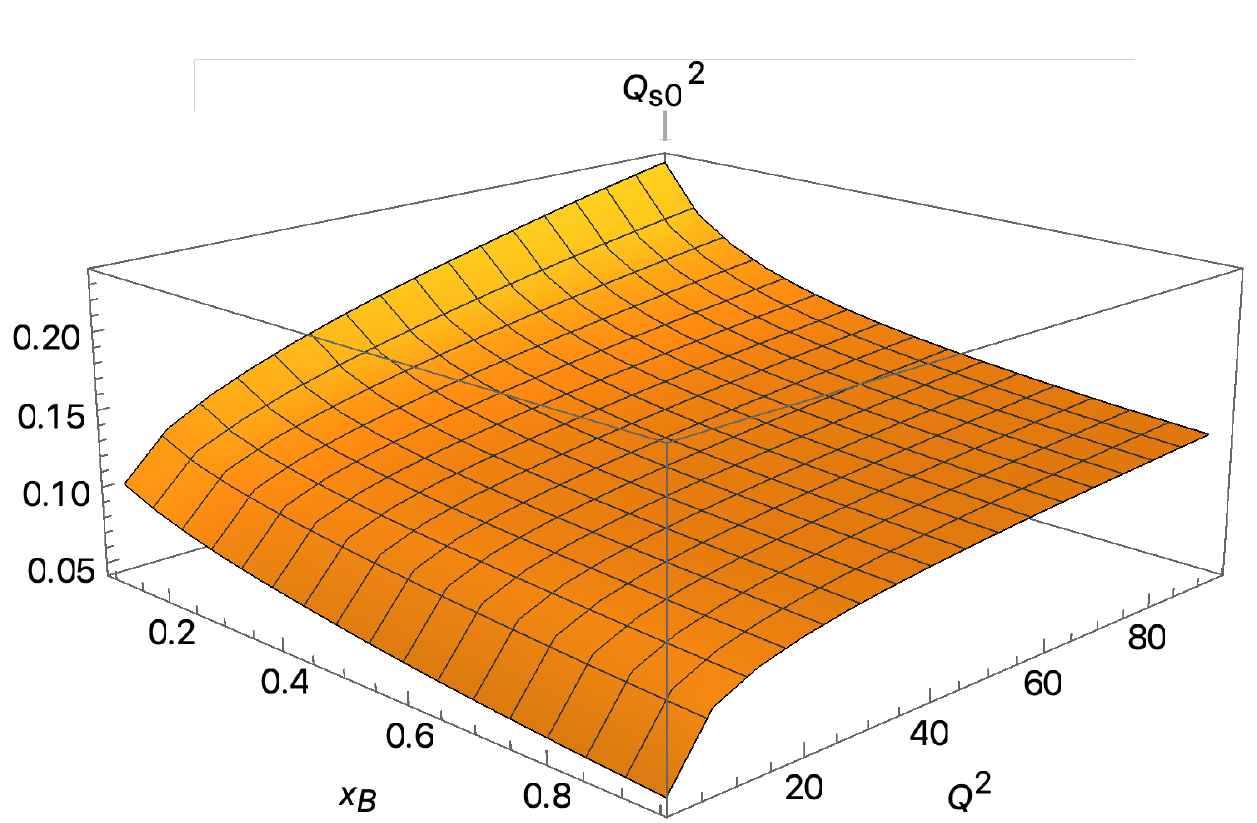}
\caption{The $x_B$ and $Q^2$ dependence of the scaled saturation scale $Q_{s0}^2(x_B,Q^2)$ inside Pb from solving Eq.~(\ref{equ:Q2}).}
\label{fig:Qs_x_Q2}
\eef

 We consider a hard-sphere model for the nuclear distribution in the rest frame,
 \begin{equation}
     \rho_A(r)=\frac{3}{4\pi r_0^3} \theta(R_A-r).
 \end{equation}
The nuclear thickness function is then,
 \begin{equation}
     t_A(b_\perp)=\frac{3R_A}{2\pi r_0^3} \sqrt{1-\frac{b_\perp^2}{R_A^2}},
 \end{equation}
 where $R_A=r_0 A^{1/3}$ is the radius of the nucleus and $r_0=1.12$ fm. The solution for $Q_s^2$ in Eq.~(\ref{equ:Q2}) should scale approximately with the nuclear thickness function $t_A(b_\perp)$, which is proportional to the length of  the quark propagation inside the nucleus. We can therefore approximately factor out the impact-parameter dependence of the saturation scale,
 \begin{equation}
     Q_s^2(x_B,Q^2, b_\perp)\approx Q_{s0}^2(x_B,Q^2) A^{1/3} \sqrt{1-\frac{b_\perp^2}{R_A^2}}.
     \label{eq:scaledQs}
 \end{equation}
 Shown in Fig.~\ref{fig:Qs_x_Q2} is the scaled saturation scale $Q_{s0}^2(x_B,Q^2)$ as a function of $Q^2$ and $x_B$ from solving Eq.~(\ref{equ:Q2})  with the nucleon gluon TMD PDF in vacuum given by the “PB-NLO-HERAI+II-2018-set1” in TMDlib. We can see in Fig.~\ref{fig:Qs_x_Q2} there is a weak dependence on $x_B$ and scale $Q^2$ in the region of large $x_B \sim 0.1 - 1$ and moderate to large scale $Q^2\sim 2-100$ GeV$^2$. For a large nucleus such as lead (Pb) ($A=208$), the saturation scale in this region of kinematics is $Q_s^2=Q_{s0}^2A^{1/3}=0.25 - 1.29$ GeV$^2$ in $e$+A DIS at zero impact-parameter $b_\perp=0$.  We note that these values are much smaller than the saturation scale $Q_{s0}^2\sim 1$ GeV$^2$ \cite{GolecBiernat:1999qd} one expects at very small $x_B$ where the saturation scale in a large nucleus becomes large so that pQCD can be applied to calculate parton distributions with gluon saturation \cite{McLerran:1993ni,McLerran:1993ka,Mueller:1999wm}.  Within our simple model for gluon saturation in this study, the saturation scale $Q_{s0}^2$ can reach 1 GeV$^2$ at about $x_B \sim 0.001$ and $Q^2\sim 100$ GeV$^2$.

\subsection{Jet transport coefficient and $p_\perp$ broadening}

Since we consider $e$+A DIS in the large $x_B$ region, the quark TMD PDF will not be affected by gluon saturation at small $x$. However, one should consider the effect of transverse momentum ($p_\perp$) broadening due to multiple soft interaction in addition to the EMC nuclear modification  of quark PDF as we model in Eq.~(\ref{eq:qpdf}). This $p_\perp$ broadening is controlled by the quark transport coefficient $\hat q_F$.

The quark transport coefficient in cold nuclei has been extracted from phenomenological studies of $p_\perp$ broadening and suppression of the final-state hadrons in semi-inclusive DIS (SIDIS).  Analyses of the suppression of single inclusive hadrons in $e$+A SIDIS by Chang et al.~\cite{Chang:2014fba} within the high-twist model of parton energy loss and medium modification of the fragmentation functions give the quark transport coefficient at the center of a cold nucleus in its rest frame $\hat q_F^0\approx 0.02$ GeV$^2$/fm which was assumed to be independent of $x_B$ and $Q^2$. Similar work by Li, Liu and Vitev \cite{Li:2020zbk} on the suppression of single inclusive hadrons in $e$+A SIDIS within the SCET$_{\rm G}$ approach gives $\hat q_F^0\approx 0.03$ GeV$^2$/fm.

One can also extract the quark transport coefficient $\hat q_F$ from the transverse momentum broadening of hadrons in $e$+A SIDIS. At LO and neglecting the radiative corrections \cite{Liou:2013qya,Wu:2014nca,Kang:2013raa,Kang:2014ela,Kang:2016ron,Blaizot:2014bha,Ghiglieri:2015zma,Blaizot:2019muz}, the average transverse momentum broadening of a quark can be related to $\hat q_F$ according to Eq.~(\ref{eq:qpdf})
,
\begin{eqnarray}
    \langle\Delta p_{\perp q}^2\rangle_{eA}&=&\langle p_{\perp q}^2\rangle_{eA}- \langle p_{\perp q}^2\rangle_{ep} \nonumber \\
    &=&\frac{\int d^2b_\perp t_A(b_\perp) \int dy_0^- \hat q_F(x_B,Q^2,y_0^-,b_\perp)}{\int d^2b_\perp t_A(b_\perp)} \nonumber \\
    &=& \frac{3}{2}R_A \hat q_F^0(x_B,Q^2),
    \label{eq:pT_BR}
\end{eqnarray}
in the hard-sphere model of the nuclear distribution, where $\hat q_F^0$ is the quark transport coefficient at the center of a cold nucleus. The transverse momentum broadening of leading hadrons is $\langle\Delta p_{\perp h}^2\rangle_{eA}=\langle z_h^2\rangle \langle\Delta p_{\perp q}^2\rangle_{eA}$ and $z_h$ is the momentum fraction of hadrons in the quark fragmentation. A recent comprehensive analysis of the experimental data on the transverse momentum broadening of a variety of hadrons in  $e$+A and Drell-Yan dilepton in p+A collisions  \cite{Ru:2019qvz} gives $\hat q_F^0\approx 0.015$ GeV$^2$/fm  in $e$+A DIS with a weak dependence on $x_B$ and the scale $Q^2$ in the range $0.05<x_B<0.4$  and $1<Q^2<10$ GeV$^2$.

In general, one can define the TMD jet transport coefficient \cite{CasalderreySolana:2007sw,Zhang:2019toi} for a parton with color representation $R$ as,
\begin{eqnarray}
\hat{q}_R(y^-) &=& \int \frac{d^2 \vec{k}_{\perp}}{(2\pi)^2} \hat{q}_R({k}_{\perp},y^-), \nonumber \\
\hat{q}_R({k}_{\perp},y^-)& = &  \int dx_G \delta(x_G - \frac{k_{\perp}^2}{2p^+q^-})  \frac{4\pi^2 C_R}{N_c^2-1} \nonumber \\
& & \times\rho_A(y^-) \alpha_s \phi_N(x_G, \vec{k}_{\perp}, \mu^2),
\label{equ:qhat}
\end{eqnarray}
where $C_R$ is the Casimir color factor, $C_F=(N_c^2-1)/2N_c$ for a quark and $C_A=N_c$ for a gluon.  This definition of the gluon transport coefficient $\hat q_A$ is the same as that for the gluon saturation scale $Q_s^2$ in Eq.~(\ref{equ:Q2}). After integrating $\hat q_A$ over the parton propagation path, we get essentially the total transverse momentum broadening squared of a propagating gluon which is the same as the gluon saturation scale $Q_s^2$. Therefore, the gluon transport coefficient $\hat q_A^0$ at the center of a nucleus can be related to the scaled gluon saturation $Q^2_{s0}$ in Eq.~(\ref{eq:scaledQs}),
\begin{equation}
    \hat q_A^0(x_B,Q^2)=\frac{A^{1/3}}{2R_A} Q^2_{s0}(x_B,Q^2).
    \label{eq:qa0}
\end{equation}
The quark transport coefficient $\hat q_F$ is a factor $C_F/C_A=4/9$ smaller than that of a gluon $\hat q_A$. Using this relation,  one can also obtain the quark transport coefficient from the numerical solution to Eq.~(\ref{equ:Q2}) for the gluon saturation scale as shown in Fig.~\ref{fig:Qs_x_Q2}. For $x_B=0.1-0.4$ and $Q^2=2-6$ GeV$^2$, one gets $\hat q_F^0\approx 0.013 - 0.023$ GeV$^2$/fm which is consistent with the value $\hat q_F^0\approx 0.015$ GeV$^2$/fm extracted from the transverse momentum broadening of hadrons in $e$+A SIDIS ~\cite{Ru:2019qvz}  and $\hat q_F^0\approx 0.02 - 0.03$ GeV$^2$/fm from the suppression of single inclusive hadrons in $e$+A SIDIS \cite{Chang:2014fba,Li:2020zbk}. Such momentum broadening can also be measured through azimuthal angle correlation between a single jet and the lepton in $e$+A DIS as proposed recently by Liu {\it et al.} \cite{Liu:2018trl}. For self-consistency, we will use the simple model for both the gluon TMD PDF in Eq.~(\ref{equ:saturated_phi}) with saturation and $\hat q_F^0$ as obtained from the gluon saturation scale in Eq.~(\ref{eq:qa0}) for the quark transverse momentum broadening in Eq.~(\ref{eq:qpdf}) in our following calculation of dijet spectrum in $e$+A DIS.

\section{Nuclear modification of dijet spectra in $e$+A DIS}
\label{sec-numerical}
\subsection{Kinematics}

Using the transverse momentum broadening and gluon saturation for TMD PDF inside nuclei as modeled in the above section, we will evaluate the nuclear modification of the dijet spectrum at LO in $e$+A DIS. We will focus on the region of relative large $x_B \geq 0.2$, which is within the kinematic coverage in experiments at the proposed EIC \cite{AbdulKhalek:2021gbh,DIS} at the Brookhaven National Laboratory as shown in Fig.~\ref{fig:xQ2coverage}.
\bef
\centering
\includegraphics[width=\linewidth]{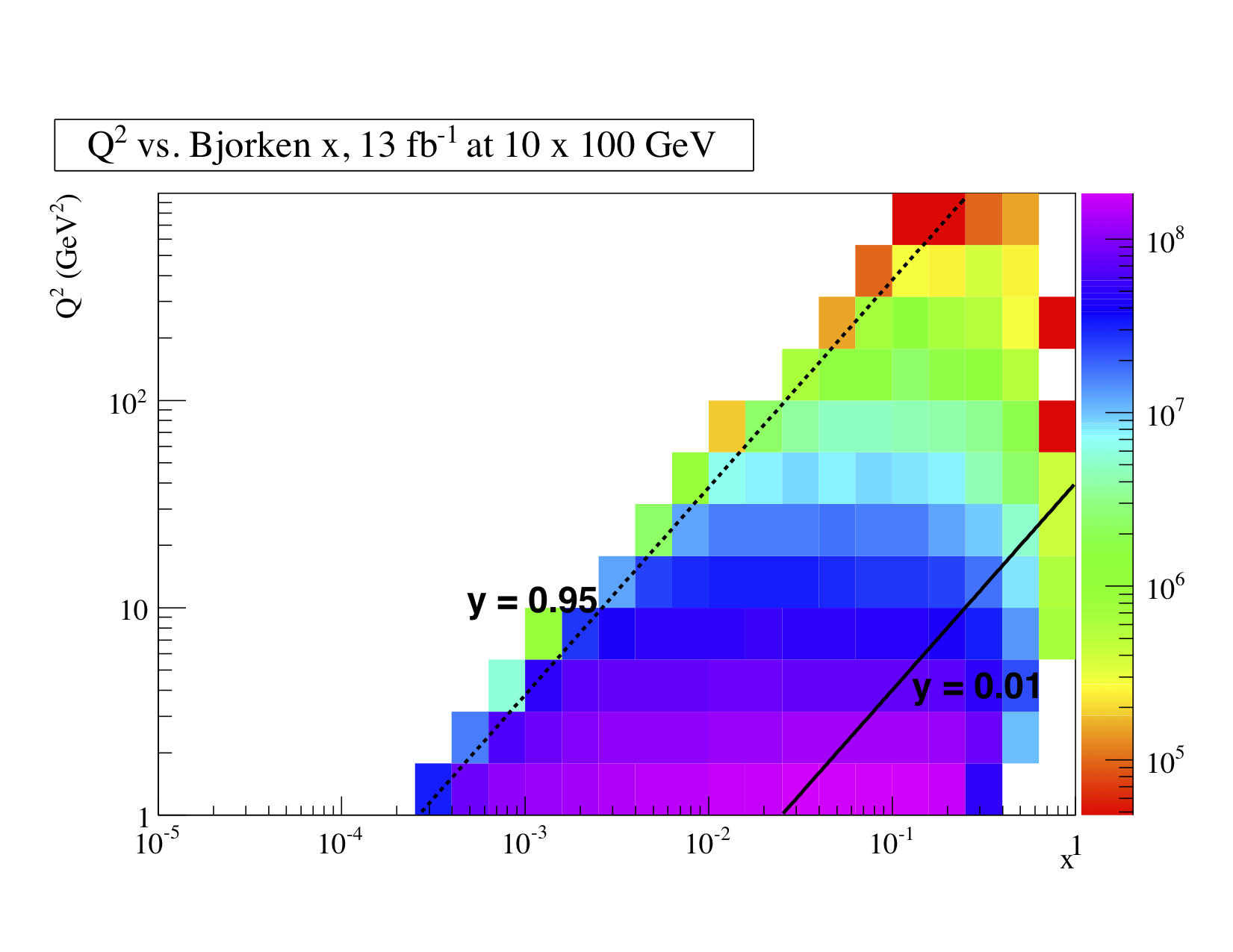}
\caption{The $Q^2$ and $x$ coverage of EIC with the electron beam energy $E_e =10$ GeV and the ion beam energy per nucleon $E_N = 100$ GeV~\cite{AbdulKhalek:2021gbh,DIS}.}
\label{fig:xQ2coverage} 
\eef
We assume the proposed EIC will have the electron beam energy $E_e = 10$ GeV, the highest ion beam energy per nucleon  $E_N = 100$ GeV and the center-of-mass energy is $\sqrt{s_{eN}}=63.2$ GeV. We will assume a typical set of kinematics $x_B=0.2$ and $Q^2=200$ GeV$^2$ for all the numerical calculations of the dijet spectrum in the following, unless specifically stated otherwise.

We will work in the Breit frame of the initial quark in which the rapidity of the radiated gluon $y_l$ and the final quark $y_{l_q}$
are,
\begin{eqnarray}
y_{l_q} &=& \frac{1}{2}\ln \frac{l_q^+}{l_q^-} =  
\frac{1}{2}\ln \frac{l_{q\perp}^2}{2(1-z)^2(q^-)^2}, \nonumber \\
y_{l} &=&\frac{1}{2}\ln \frac{l^+}{l^-} = \frac{1}{2} \ln \frac{l_{\perp}^2}{2z^2(q^-)^2},
\label{equ:dijet-rapidity}
\end{eqnarray}
respectively, according to Eq.~(\ref{eq:kinematics}). To have two well separated jets in the dijet production, we require $(y_{l_q} - y_l)^2+\Delta\phi^2 > \Delta R^2$, where $\Delta\phi$ is the azimuthal angle difference between the two jets. Such a requirement will constrain the range of the transverse momenta $l_{\perp}$ and $l_{q\perp}$, momentum fraction $z$ and azimuthal angle difference $\Delta\phi$.

 In the calculation of the hadronic tensor, we have made the collinear approximation which requires $Q^2\gg (\vec{l}_{\perp}+\vec{l}_{q\perp})^2$ ($Q^2 \geq 4(\vec{l}_{\perp}+\vec{l}_{q\perp})^2$). This provides an additional constraint on the dijet azimuthal angle $\Delta\phi$ and transverse momenta $l_{\perp}$ and $l_{q\perp}$ where our calculations are applicable. Among the longitudinal momentum fractions of the medium gluon, the largest fraction $x_L ={l_{\perp}^2}/{[2p^+q^-z(1-z)]} $ should still be $x_L \ll x_B$ ($x_L\leq x_B/2$).  This will provide an upper bound for the jet transverse momentum $l_{\perp}^2\leq (1-z)zQ^2/2$ .

In the transverse part of the induced gluon spectrum per mean-free-path ${\cal N}_g$ in Eqs.~(\ref{eq:gLPM})-(\ref{eq:nonLPM}),  there are three collinear divergences in the denominators of the radiation amplitudes:
\begin{eqnarray}
& (1): & \;\;\;\;\vec{l}_{\perp} - (1-z)\vec{v}_{\perp}=0, \nonumber \\
& (2): & \;\;\;\; \vec{l}_{\perp}-(1-z)(\vec l_{\perp}+\vec l_{q\perp})=0, \nonumber \\
& (3):& \;\;\;\; \vec{l}_{\perp}-(1-z)\vec v_{\perp}-\vec k {\perp}=0 .
\end{eqnarray}
 The first and the third divergence are canceled by the LPM interference factors in ${\cal N}_g^{\rm qLPM}$ and ${\cal N}_g^{\rm gLPM}$, respectively. The second divergence is regulated by the angular separation in the kinematics of the dijet. However, the first divergence still remains in ${\cal N}_g^{\rm nonLPM}$ and ${\cal N}_g^{\rm gLPM}$. This divergence arises when the radiated [Fig.\ref{fig:amplitude} (b)] or intermediate gluon [Fig.\ref{fig:amplitude} (c)] with the momentum $l = (0, (1-z)q^-, (1-z)\vec{v}_{\perp})$ becomes collinear to the  quark as the emitter with the momentum $(0,q^-,\vec{v}_{\perp})$.  This divergence is normally absorbed into the renormalized TMD quark-gluon correlation function for $[\vec{l}_{\perp} - (1-z)\vec{v}_{\perp}]^2 < \mu_f^2$.  The factorization scale $\mu_f$ will serve to regularize the collinear divergence at $\vec{l}_{\perp}- (1-z)\vec{v}_{\perp}=0$ in the dijet spectrum.
 
We also include the running strong coupling constant $\alpha_{\rm s}(\mu)$ for both gluon radiation and the secondary scattering in the dijet cross section and in the calculation of the saturation scale $Q_s^2$ or the quark jet transport coefficient $\hat{q}_F$. We will set the scale $\mu=l_\perp$ in the running coupling constant $\alpha_s(\mu)$ associated with the gluon radiation, $\mu=\max(k_\perp,Q_s)$ in the secondary scattering and the factorization scale $\mu_f=1$ GeV. The final result is found not sensitive to $\mu_f$ for $\mu_f\le 1$ GeV.

\subsection{Dijet spectrum}

The dijet spectrum in $e$+A DIS from single and double scattering in Sec.\ref{sec-dijetcross} [Eqs.~(\ref{eq:single-crsec}) and (\ref{equ:nXsection})] can be expressed as 
\begin{equation}
    \frac{d\sigma_{e{\rm A}}^{S(D)}}{dx_B dQ^2 dz d^2l_{\perp}d^2l_{q\perp}}
    \equiv A \frac{d\sigma_{ep}^0}{dQ^2} \frac{dN^{S(D)}_{\rm dijet}}{dx_B dz d^2l_{\perp}d^2l_{q\perp}},
\end{equation}
where $d\sigma_{ep}^0/dQ^2 \equiv 2\pi \alpha_{\rm em}^2/Q^4$,
and $dN_{\rm dijet}$ is the dijet spectrum per target nucleon that we will discuss in the remainder of this section.  In $e+p$ DIS, only single scattering without $p_\perp$ broadening contributes to the dijet production at LO.

\begin{widetext}

\bef
\includegraphics[width=8cm]{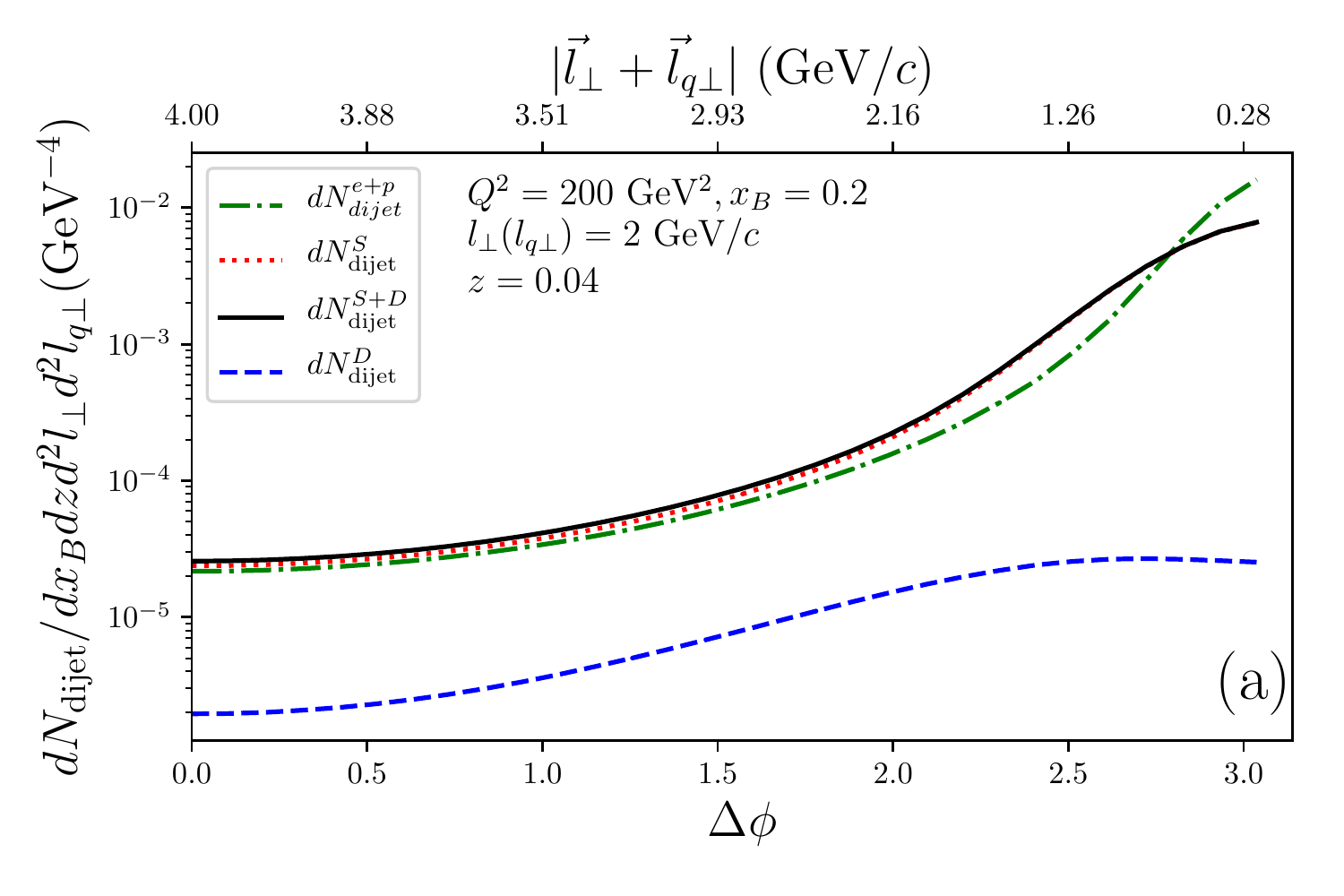}
\includegraphics[width=8cm]{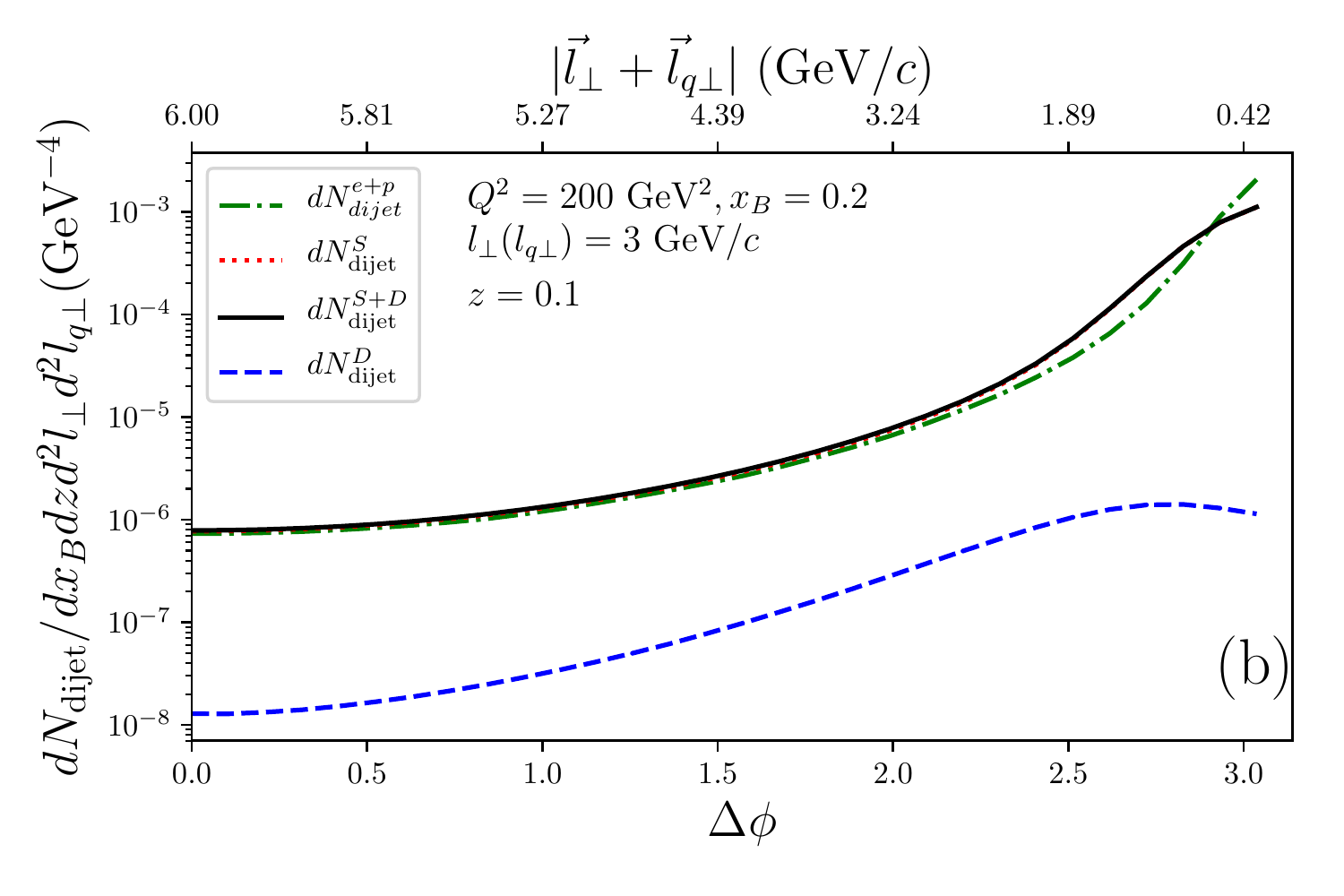}
\includegraphics[width=8cm]{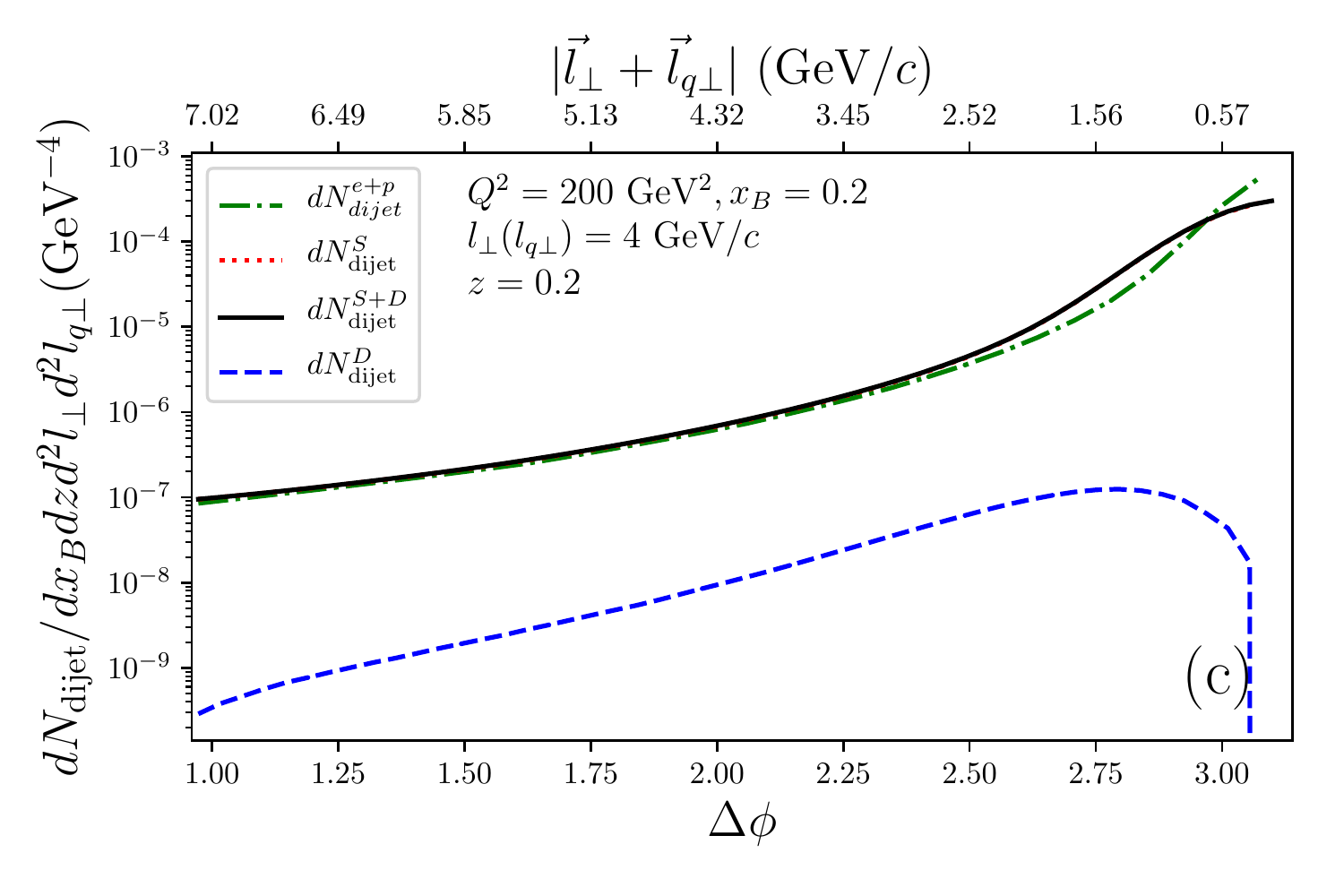}
\includegraphics[width=8cm]{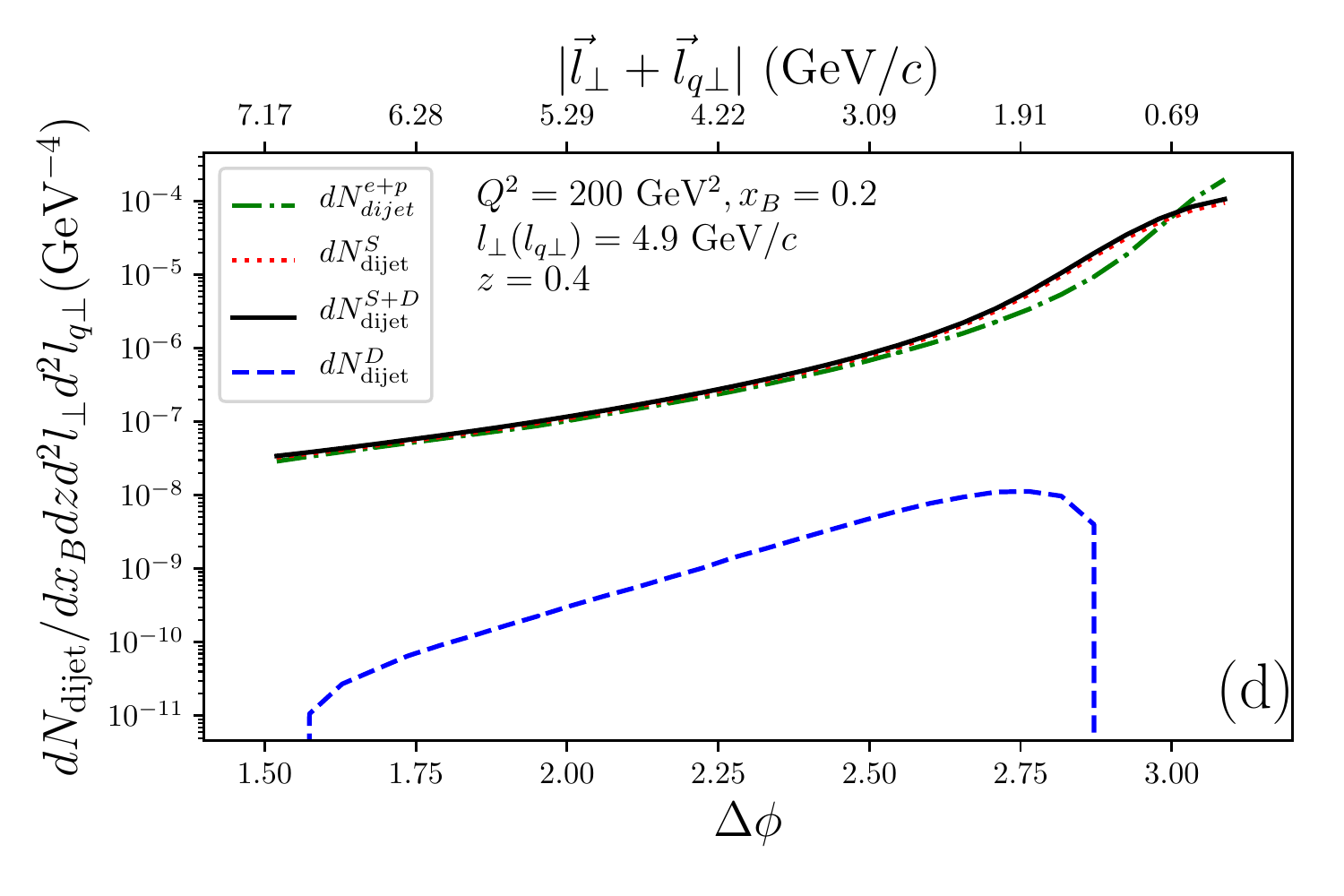}
\caption{Dijet spectra from e+p(dot-dashed) DIS, $e$+Pb DIS single (dotted), double scattering (dashed) and their sum (solid) as a function of the azimuthal angle $\Delta\phi$ or the transverse momentum imbalance $|\vec l_\perp+\vec l_{q\perp}|$  with $x_B=0.2$ and  $Q^2 = 200$ GeV$^2$ for (a) [$z, l_\perp$(GeV/$c$)]=[0.04, 2], (b) [0.1, 3], (c) [0.2, 4] and (d) [0.4, 4.9].}
\label{fig:Xsec_azAngle-distr}
\eef
\end{widetext}

Plotted in Fig.~\ref{fig:Xsec_azAngle-distr} are contributions to the dijet  spectrum from single (dotted), double scattering (dashed) and their sum (solid) in $e$+Pb as well as $e+p$ DIS (dot-dashed) as a function of the azimuthal angle difference $\Delta\phi$ (or the transverse momentum imbalance $|\vec l_\perp+\vec l_{q\perp}|$ on the top legend) between the two jets that have the same transverse momentum $l_\perp=l_{q\perp}$. The dijet spectra from single scattering in both $e$+Pb and $e+p$ DIS peak at the back-to-back direction ($\Delta\phi=\pi$). The change in the single scattering from $e$+Pb to $e+p$ is caused by the quark transverse momentum broadening due to multiple soft scattering inside the nucleus target. 

For large values of the jet transverse momenta $l_\perp$ and $l_{q\perp}$, contributions from the double hard scattering are power-suppressed relative to the single scattering. Therefore
they are much smaller than the contribution from single scattering.  One has to go to smaller values of the jet transverse momentum in order to see the effect of double hard scattering. These two mini-jets don't peak at the back-to-back direction because of the secondary hard scattering. The double scattering contribution can become negative in some region because of the destructive interference between the initial and final-state radiation in ${\cal N}_g^{\rm nonLPM}$. However, the total dijet cross section (single + double scattering) (solid lines) is still positive because it is dominated by the contribution from single scattering.

\subsection{Nuclear modification}
To quantify the nuclear modification to the dijet spectrum in $e$+A DIS, we evaluate numerically the modification factor which is defined as the ratio of the dijet differential cross sections in $e$+A and $e+p$ DIS. The modification factor has contributions from both single and double scattering,
\begin{equation}
    I_{e{\rm A}}^{S+D}=
    I_{e{\rm A}}^{S}(l_\perp,l_{q\perp},\Delta\phi,z)+
    I_{e{\rm A}}^{ D}(l_\perp,l_{q\perp},\Delta\phi,z),
\end{equation}
  the modification factor from single (double) scattering is defined as,
\begin{equation}
I^{S(D)}_{e{\rm A}}(l_{\perp},l_{q\perp}, \Delta\phi, z)= 
\left.\frac{d \hat{\sigma}_{e{\rm A}}^{S(D)}}{d{\cal P}}\middle/
A\frac{d \hat{\sigma}_{ep}}{d{\cal P}} \right. .
\end{equation}
where $d{\cal P}\equiv dx_B dQ^2 dz d^2l_{\perp}d^2l_{q\perp}$.
 We assume that we do not distinguish quark from gluon jet in experiments. The above dijet cross sections $\hat{\sigma}$ are the sum of the cross sections with the kinematics of quark and gluon exchanged,
\begin{equation}
\hat{\sigma}
\equiv \sigma(l_\perp,l_{q\perp},\Delta\phi,z)+ \sigma(l_{q\perp},l_\perp,\Delta\phi,1-z).
\end{equation}
We will examine the azimuthal angle difference $\Delta\phi$, the momentum fraction $z$ or rapidity gap $|y_l-y_{l_q}|$ and the nuclear size $R_A$ dependence of the nuclear modification factor in this section. Since the calculation of $I^S_{\rm eA}$ is straightforward, which contains just the effect of transverse momentum broadening of the initial quark on the process of single scattering, we will focus on the behavior of $I^D_{\rm eA}$ and its dependence on the azimuthal angle, transverse momentum imbalance, rapidity and nuclear size.

\subsubsection{Azimuthal angle $\Delta\phi$ dependence}

We first investigate the azimuthal angle $\Delta\phi$ or the dijet transverse momentum imbalance $|\vec l_\perp+\vec l_{q\perp}|$ dependence of the nuclear modification factor for fixed values of the transverse momenta $l_\perp$ and $l_{q\perp}$. For given equal values of the transverse momenta $l_\perp=l_{q\perp}$, the azimuthal angle $\Delta\phi$ varies from $\pi$ for $|\vec l_\perp+\vec l_{q\perp}|=0$ to 0 for $|\vec l_\perp+\vec l_{q\perp}|=2l_\perp$. Note that we require two jets have an angular separation $(y_l-y_{lq})^2+\Delta\phi^2>\Delta R^2$ and we set $\Delta R=1$ in the follow numerical calculations.  In Fig.~\ref{fig:RD_azAngle-distr}, we plot the azimuthal angle $\Delta\phi$ or transverse momentum imbalance $|\vec l_\perp+\vec l_{q\perp}|$  distribution of the dijet nuclear modification factor from double scattering $I_{eA}^D(l_{\perp},l_{q\perp},\Delta\phi,z)$  in $e$+Pb DIS for different values of
$[z,l_\perp ({\rm GeV}/c)]=(0.04,2), (0.1,3),(0.2,4),(0.4,4.9)$. These values are selected to satisfy the kinematic constraints: small longitudinal momentum transfer $ l_{\perp}^2 \leq (1-z)zQ^2/2 $, collinear approximation $Q^2\gg (\vec{l}_{\perp}+\vec{l}_{q\perp})^2$ and the dijet angular separation $(y_{l_q} - y_l)^2+\Delta\phi^2 > \Delta R^2$.

One can see that in the dijet spectrum from double scattering, the contribution containing ${\cal N}_g^{\rm qLM}$ is negligible since it is suppressed by both the color factor and the LPM interference.
The term containing ${\cal N}_g^{\rm gLPM}$ is the most dominant in which the gluon is emitted from the gluon propagator. The contribution from ${\cal N}_g^{\rm nonLPM}$ is small and finite, but its relative importance increases with large momentum fraction $z$ or small rapidity gap $|y_l -y_{l_q}|$. The magnitude of the dijet cross section from double scattering, however, decreases with the increase of $l_\perp$ when the transverse momenta of the initial quark and gluon become negligible and the dijet cross section from double scattering is power-suppressed relative to the LO cross section of single scattering.  We therefore have to limit ourselves to mini-jets if we want to observe the contributions from double scattering in the dijet spectrum.

\begin{widetext}

\bef
\includegraphics[width=8cm]{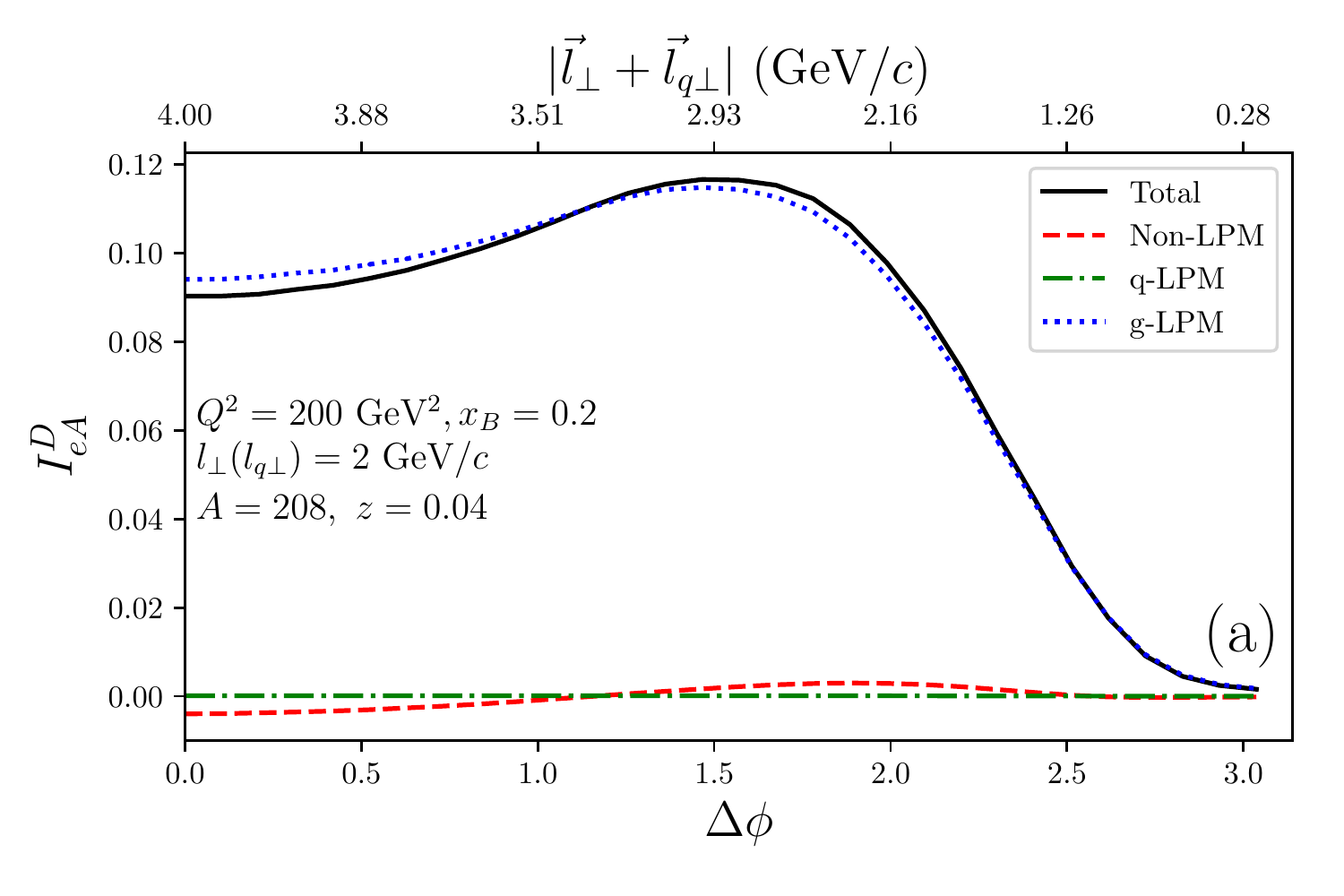}
\includegraphics[width=8cm]{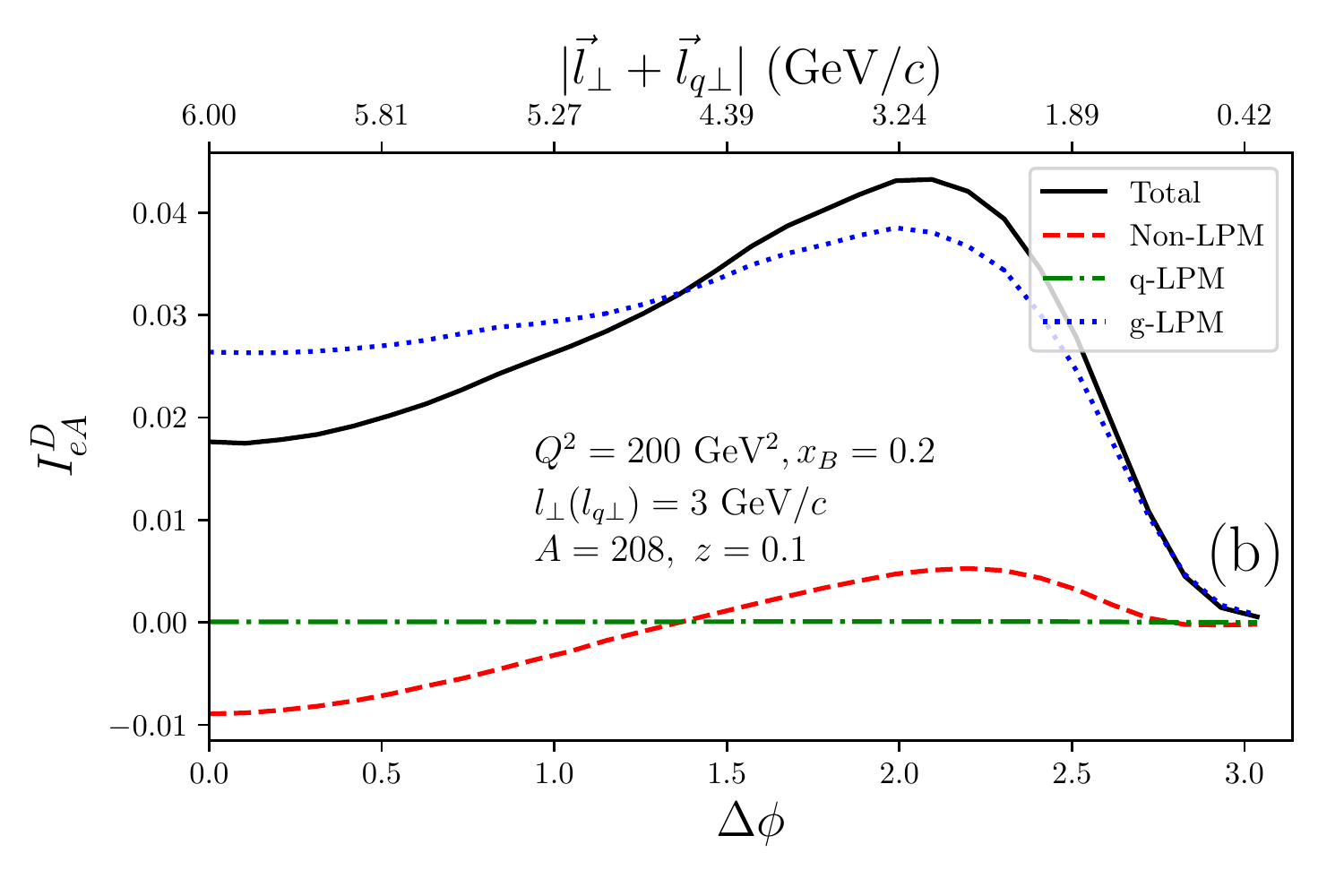}
\includegraphics[width=8cm]{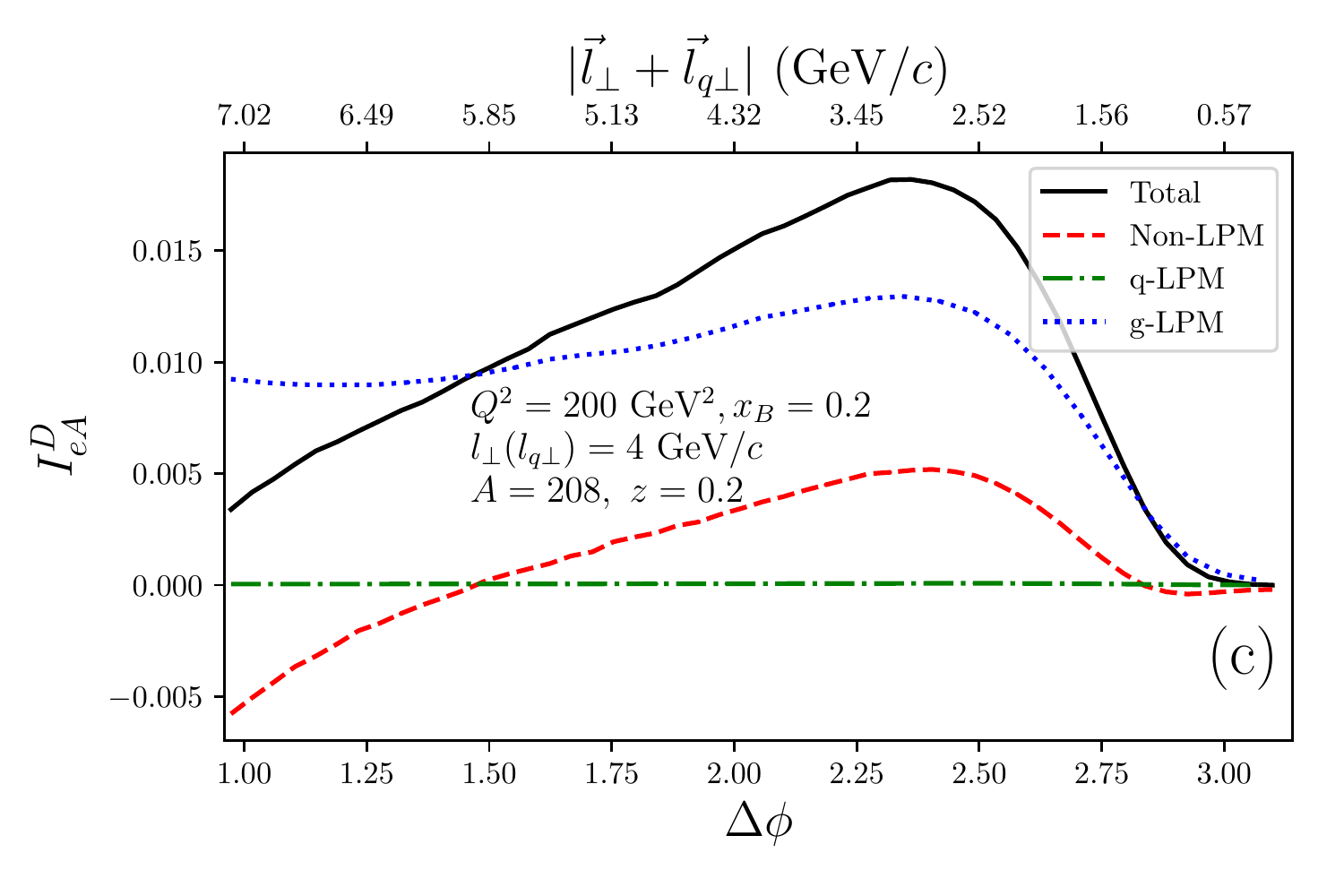}
\includegraphics[width=8cm]{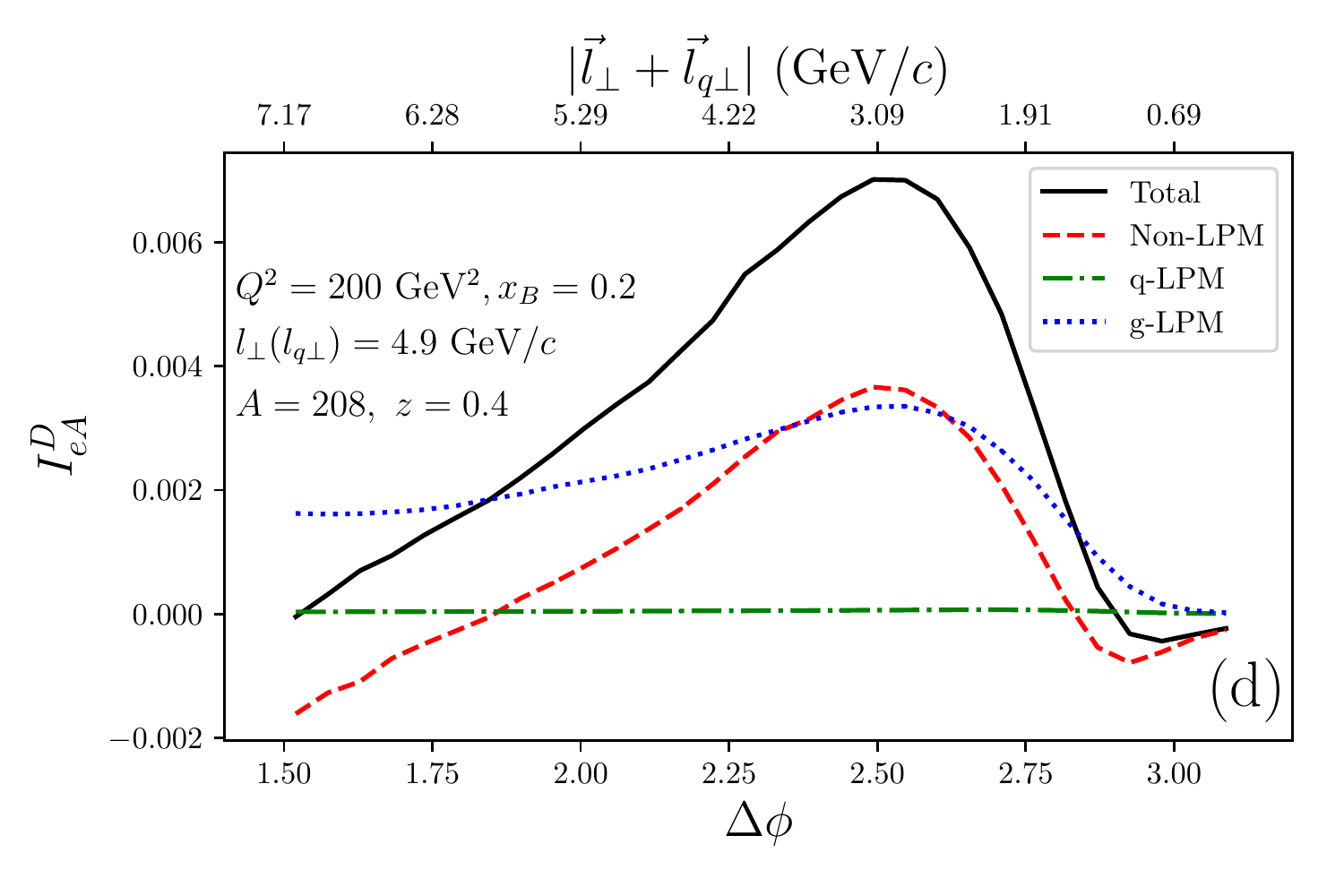}
\caption{The nuclear modification factor $I_{eA}^D(l_{\perp},l_{q\perp}=l_\perp,\Delta\phi,z)$ for the dijet cross section from double scattering as a function of the azimuthal angle $\Delta\phi$ or the transverse momentum imbalance $|\vec l_\perp+\vec l_{q\perp}|$ in $e$+Pb DIS with $x_B=0.2$ and  $Q^2 = 200$ GeV$^2$ for (a) [$z, l_\perp$(GeV/$c$)]=[0.04,2], (b) [0.1, 3], (c) [0.2, 4] and (d) [0.4, 4.9].}
\label{fig:RD_azAngle-distr}
\eef

\end{widetext}

We have included both the transverse momentum broadening in the initial quark TMD PDF and the gluon saturation in the medium gluon TMD PDF in the calculation of the dijet cross section from double scattering according to Eq.~(\ref{equ:nXsection}). Inside a Pb nucleus, the saturation scale in our simple model is $Q_s(0_\perp)=1.2$ GeV at zero impact-parameter. The corresponding total quark transverse momentum broadening squared is $\langle\Delta p^2_{\perp q}\rangle=(3/2)I_A\hat q_F^0=Q_s^2(0_\perp)/3=0.48$ GeV$^2$. The combined effect of the initial quark transverse momentum broadening and gluon saturation inside a nucleus leads to the peak structure in the azimuthal angle distribution of the nuclear modification factor from double scattering at $|\vec l_\perp+\vec l_{q\perp}|\approx 3$ GeV/$c$ as seen in Fig.~\ref{fig:RD_azAngle-distr}.

The transverse momentum broadening in quark TMD PDF inside a nucleus can also lead to nuclear modification of the dijet spectrum from single hard scattering in $e$+A DIS. Since single scattering dominates in the total dijet cross section, such a nuclear modification purely due to quark transverse momentum broadening also dominates the nuclear modification of the total dijet spectrum. To illustrate the relative importance of nuclear modification 
in single and double scattering, we show in  Fig. \ref{fig:Qs_azAngle-distr}(a) the nuclear modification factor of the dijet angular distribution from single (dashed line) and single + double (solid line) scattering with $l_\perp=2$ GeV/$c$ and $z=0.04$. The quark TMD PDF \cite{Hautmann:2017fcj} in a nucleon that we use has a Gaussian form at small transverse momentum and transits to a power-law form at large transverse momentum.  The transverse momentum broadening in a nucleus target according to Eq.~(\ref{eq:qpdf}) will therefore suppress the nuclear modification factor for the dijet spectrum from single scattering at $v_\perp=|\vec l_\perp+\vec l_{q\perp}|\approx 0$, but enhance the modification factor at intermediate value of $|\vec l_\perp+\vec l_{q\perp}|$ reaching a peak before falling back asymptotically to 1 at large $|\vec l_\perp+\vec l_{q\perp}|$. Such a nuclear modification of the dijet angular distribution in Fig. \ref{fig:Qs_azAngle-distr} is very similar to the typical Cronin effect of transverse momentum broadening in hadron spectra in $p$+A collisions \cite{Cronin:1974zm}.

\bef
\includegraphics[width=0.45\textwidth]{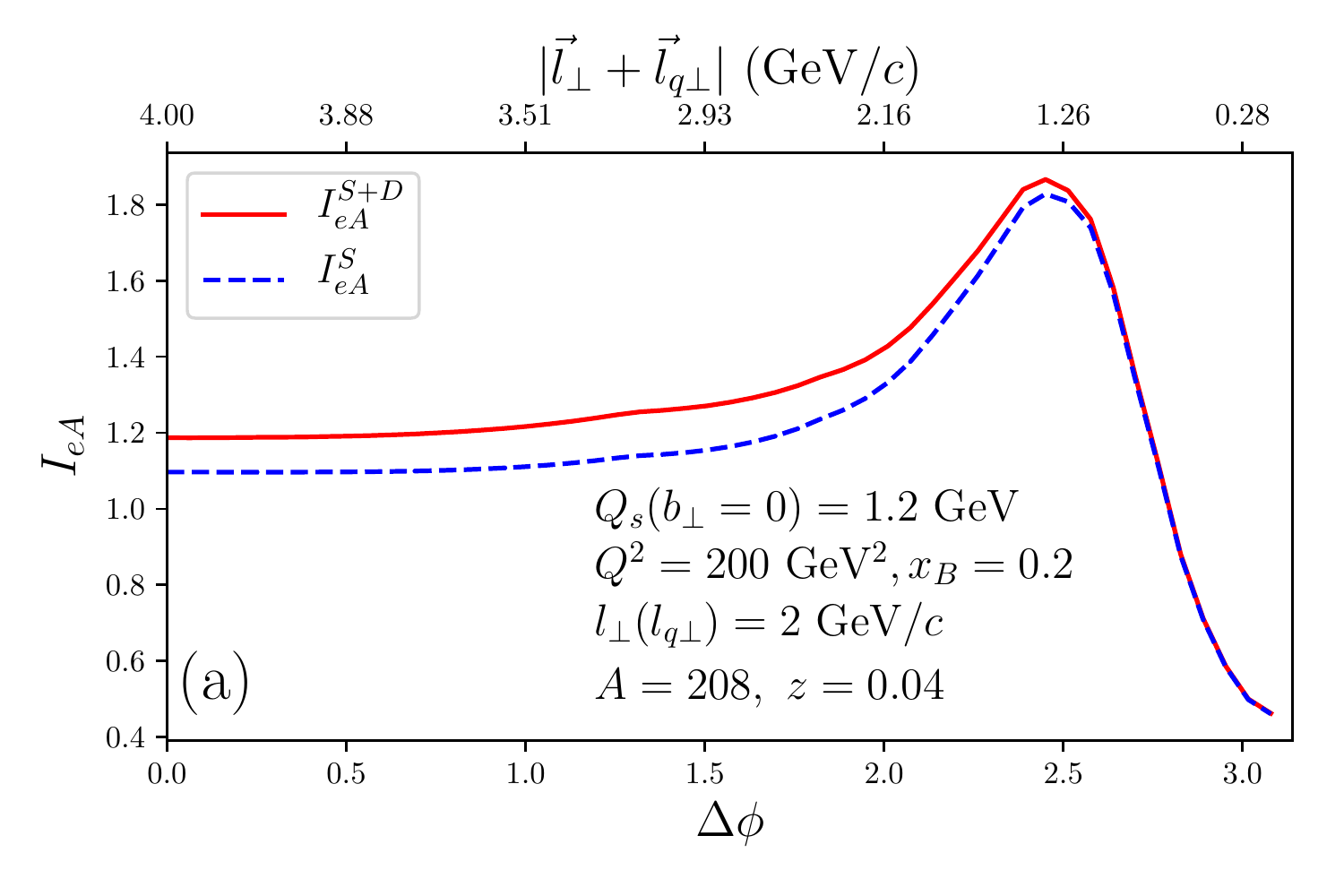}
\includegraphics[width=0.45\textwidth]{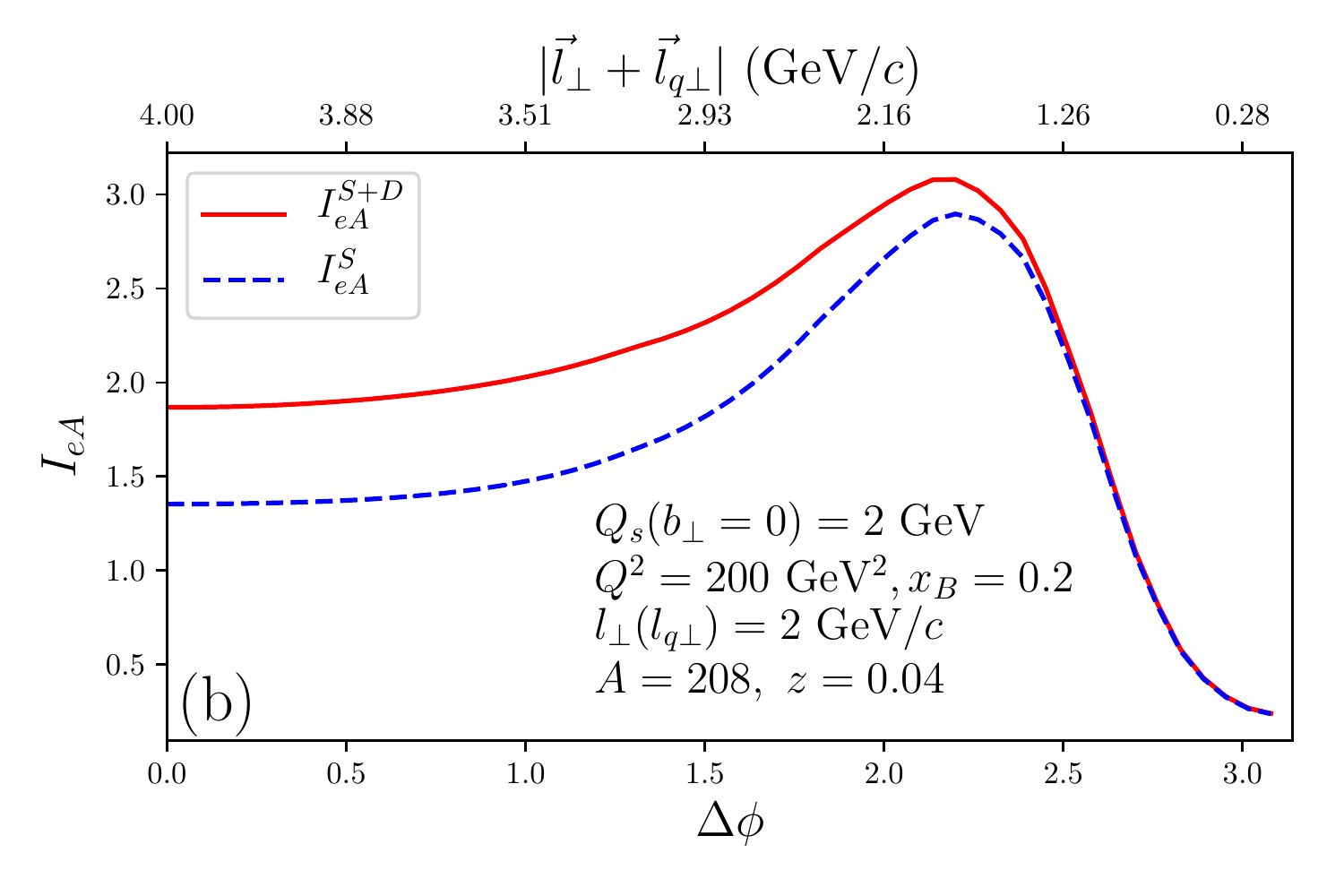}
\caption{The total nuclear modification factor $I_{eA}^{S+D}(l_{\perp},l_{q\perp}=l_\perp,\Delta\phi,z)$ and that from only single scattering $I_{eA}^{S}(l_{\perp},l_{q\perp}=l_\perp,\Delta\phi,z)$ as a function of the azimuthal angle $\Delta\phi$ or the transverse momentum imbalance $|\vec l_\perp+\vec l_{q\perp}|$ in $e$+Pb DIS with $x_B=0.2$ and  $Q^2 = 200$ GeV$^2$ for (a) $Q_s(0_\perp)=1.2$ GeV and  (b) $Q_s(0_\perp)=2$ GeV.}
\label{fig:Qs_azAngle-distr}
\eef

For $Q_s(0_\perp)=1.2$ GeV and $\langle\Delta p^2_{\perp q}\rangle=0.48$ GeV$^2$ in a Pb nucleus within our simple model for gluon saturation, the nuclear modification factor from single scattering peaks at $|\vec l_\perp+\vec l_{q\perp}|\approx 1.3$ GeV/$c$ as seen in Fig.~\ref{fig:Qs_azAngle-distr}(a) (dashed line). From Fig.~\ref{fig:RD_azAngle-distr}, we know that contributions from double scattering peak at $|\vec l_\perp+\vec l_{q\perp}|\approx 3$ GeV/$c$ in these ranges of kinematics. Contributions from double scattering, therefore, further enhance the total dijet spectrum at large $|\vec l_\perp+\vec l_{q\perp}|$ (small azimuthal angle $\Delta\phi$) as we see in Fig.~\ref{fig:Qs_azAngle-distr}(a) (solid line).

Since the peak of the total nuclear modification (enhancement) is caused mainly by the quark transverse momentum broadening which in turn is determined by the gluon saturation scale in our model, it will shift to a larger value of $|\vec l_\perp+\vec l_{q\perp}|$ (smaller angle $\Delta\phi$) when the saturation scale is increased to $Q_s(0_\perp)=2.0 $ GeV as shown in Fig.~\ref{fig:Qs_azAngle-distr}(b) (dashed line), which is the solution to the self-consistent equation in Eq.~(\ref{equ:Q2}) when the   medium gluon density is artificially increased  by a factor of 5.
Correspondingly, the contribution to the dijet spectrum from double scattering is also bigger with its peak moving to a higher value of $|\vec l_\perp+\vec l_{q\perp}|$ due to the increased quark $p_\perp$ broadening and gluon saturation scale $Q_s$ as shown in Fig.~\ref{fig:Qs_azAngle-distr_RD}. This would further enhance the total nuclear modification factor at large $|\vec l_\perp+\vec l_{q\perp}|$ as we see in Fig.~\ref{fig:Qs_azAngle-distr}(b) (solid line). 

\bef
\includegraphics[width=0.45\textwidth]{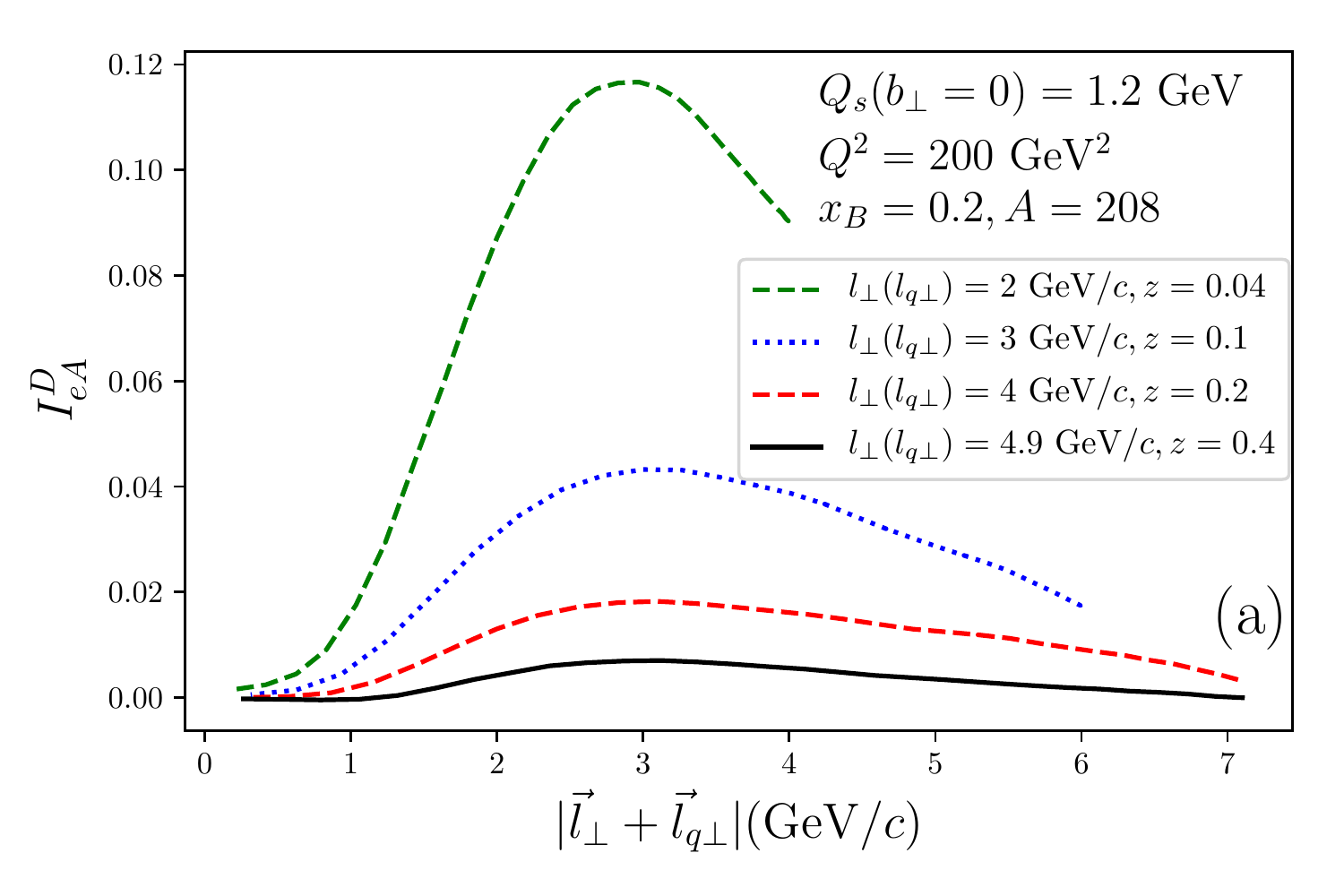}
\includegraphics[width=0.45\textwidth]{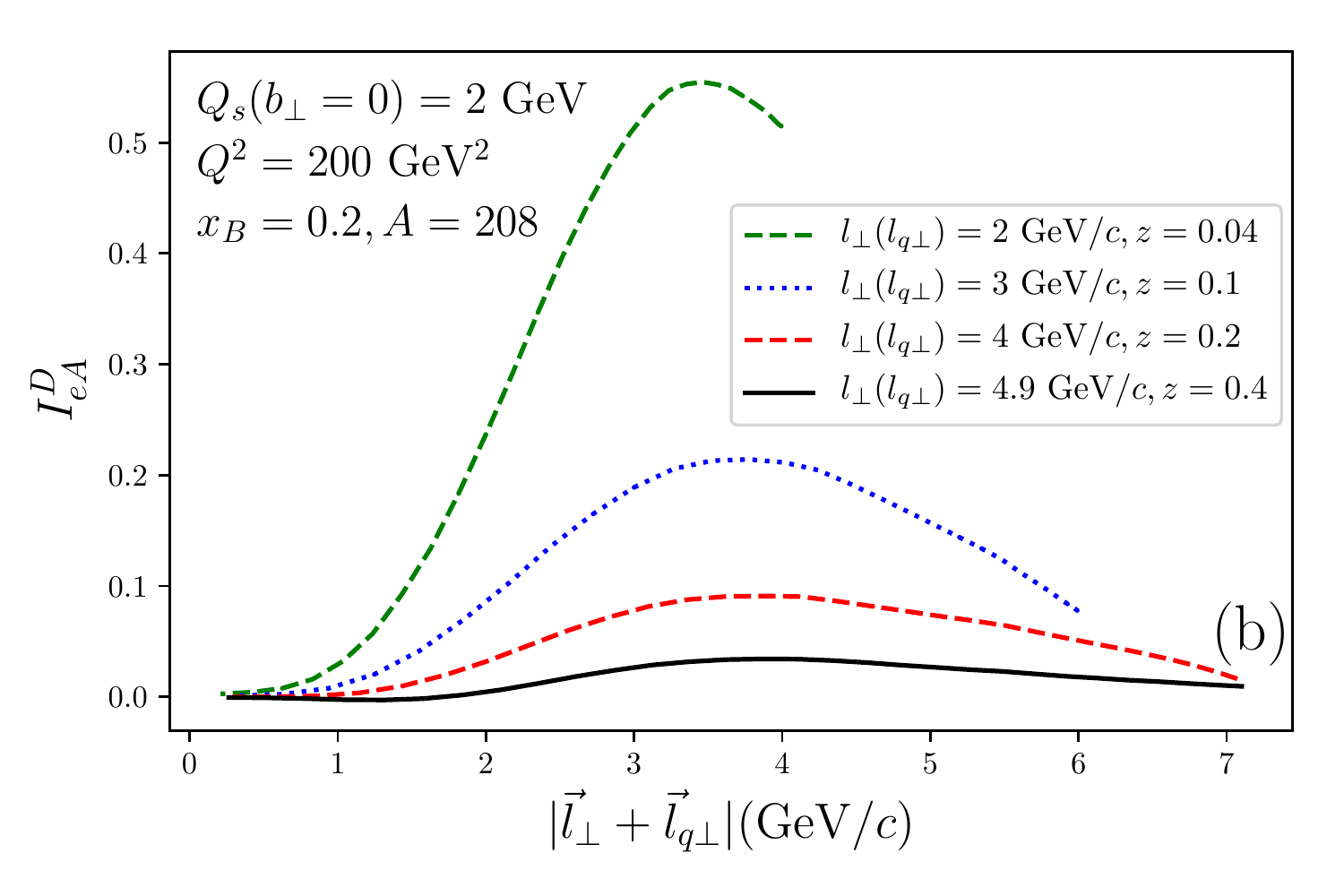}
\caption{The nuclear modification factor for the dijet spectrum from double scattering $I_{eA}^{D}(l_{\perp},l_{q\perp},\Delta\phi,z)$ as a function of the azimuthal angle $\Delta\phi$ or the transverse momentum imbalance $|\vec l_\perp+\vec l_{q\perp}|$ in $e$+Pb DIS for different dijet kinematics and (a) $Q_s(0_\perp)=1.2$ GeV and  (b) $Q_s(0_\perp)=2$ GeV.}
\label{fig:Qs_azAngle-distr_RD}
\eef

\subsubsection{Rapidity gap dependence}

In the Breit frame, the rapidity gap of the dijet is related to the longitudinal momentum fraction $z$ of the radiated gluon according to Eq.~(\ref{equ:dijet-rapidity}),
 \begin{equation}
 y_{l_q} - y_{l} = \ln (\frac{z}{1-z}).
 \end{equation}
 Shown in Fig.~\ref{fig:rapidity-distr} are (a) different contributions to the dijet nuclear modification factor from double scattering and (b) the nuclear modification factor for the dijet cross section with (solid) and without (dashed) double scattering in $e$+Pb DIS as a function of the rapidity gap $|y_{l_q} - y_{l}|$ or momentum fraction $z$. We have chosen $l_\perp=l_{q\perp}=2$ GeV/$c$ and $\Delta\phi=1$ in these calculations where double scattering has the maximum contribution. 
 
\bef
\includegraphics[width=0.45\textwidth]{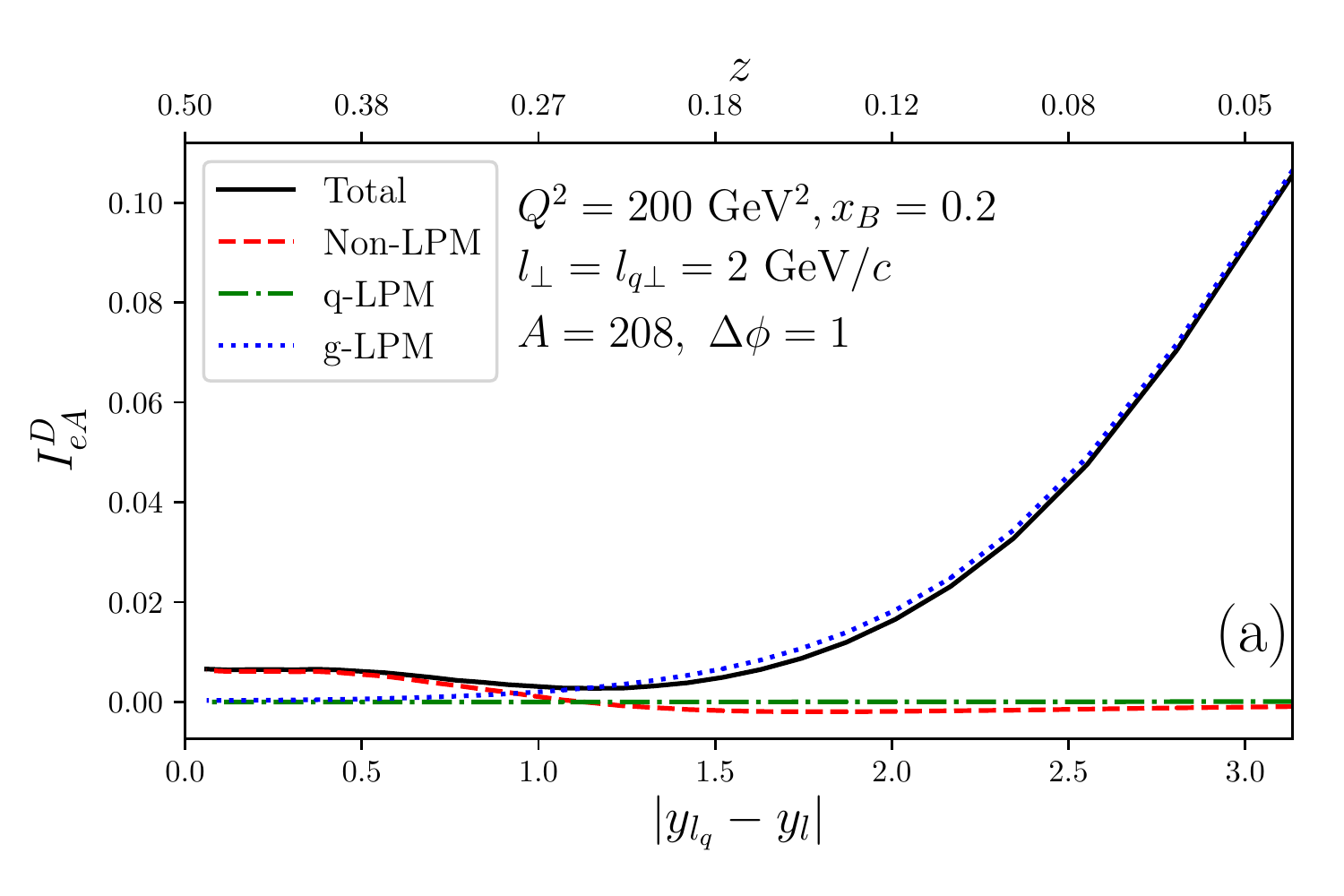}\\
\includegraphics[width=0.45\textwidth]{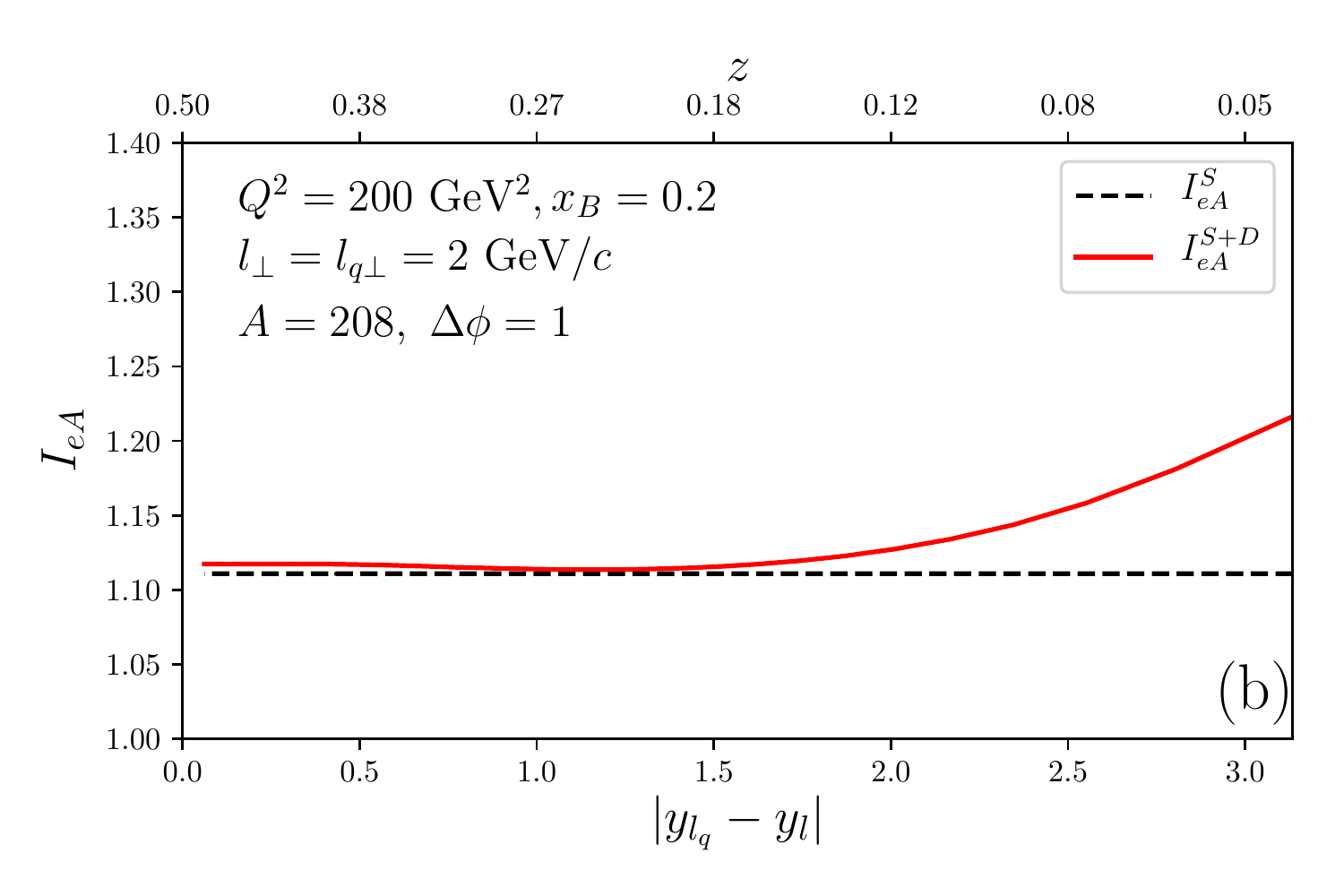}
\caption{The rapidity gap $|y_{l_q} - y_{l}|$ or momentum fraction $z$ dependence of (a) the nuclear modification factor from double scattering $I_{eA}^D(l_{\perp},l_{q\perp},\Delta\phi,z)$ (b) the nuclear modification factor with (solid) $I_{eA}^{S+D}(l_{\perp},l_{q\perp},\Delta\phi,z)$ and without  (dashed) double scattering $I_{eA}^{S}(l_{\perp},l_{q\perp},\Delta\phi,z)$ in $e$+Pb DIS. }
\label{fig:rapidity-distr}
\eef

 Since the nuclear modification factor of the dijet spectrum due to single scattering is mainly caused by the quark transverse momentum broadening, it will be approximately independent of the longitudinal momentum fraction $z$ or the rapidity gap $|y_{l_q} - y_{l}|$ as seen in Fig.~\ref{fig:rapidity-distr}(b) (dashed line). 

The contributions to the dijet spectrum from double scattering, however, have some unique $z$ or rapidity gap $|y_{l_q} - y_{l}|$ dependence. Among the three contributions from double scattering in Eqs.~(\ref{eq:hadronic-novp})-(\ref{eq:gLPM}), one can neglect ${\cal N}_g^{\rm qLPM}$ [dot-dashed line in Fig.~\ref{fig:rapidity-distr}(a)] since it is suppressed by both a color factor and the LPM interference. For given transverse momentum $l_\perp(l_{q\perp})$ and azimuthal angle $\Delta\phi$, the small but finite contribution ${\cal N}_g^{\rm nonLPM}$ [dashed line in Fig.\ref{fig:rapidity-distr} (a)] vanishes as $z\rightarrow 0$.  The most dominant contribution is from ${\cal N}_g^{\rm gLPM}$ [dotted line in Fig.~\ref{fig:rapidity-distr}(a)]  which contains the LPM interference factor,
\begin{equation}
   1-\cos\frac{y^-_{01}}{\tau_{gf}}=1-\cos\left[\frac{y^-_{01}(\vec l_\perp-(1-z)\vec v_\perp-\vec k_\perp)^2}{2q^-z(1-z)}\right],
\end{equation}
with $y_{01}^- = y_1^- - y_0^-$. As $z\rightarrow 0$ or at increasing rapidity gap, the formation time for the medium-induced gluon splitting becomes small such that the destructive LPM interference disappears, leading to an increased contribution due to incoherent dijet production induced by double scattering. This happens as the formation time $\tau_{gf}$ becomes small such that $2R_A/\tau_{gf}\stackrel{>}{\sim} 2\pi$, or
\begin{equation}
    \frac{2R_Al_\perp^2 m_Nx_B}{z(1-z)Q^2} \stackrel{>}{\sim} 2\pi.
\end{equation}
For the kinematics we use in Fig.~\ref{fig:rapidity-distr}, $Q^2=200$ GeV$^2$, $x_B=0.2$, $l_\perp=2$ GeV/$c$ and $R_A=6.6$ fm in a Pb nucleus, this leads to $z_{\rm incoh} \stackrel{<}{\sim} 0.04 $ when medium-induced dijet production becomes incoherent. This is consistent with what we observe in Fig.~\ref{fig:rapidity-distr}. One can therefore consider the increase of the nuclear modification factor for dijet cross section with the rapidity gap, as illustrated by the solid line in Fig.~\ref{fig:rapidity-distr}(b), as a unique feature caused by the LPM interference in medium-induced dijet production.

\subsubsection{Nuclei size dependence}

In the single scattering process, the dijet cross section depends on the nuclear quark TMD PDF which is proportional to the atomic number $A$ of the nucleus and the effective nucleon quark TMD PDF according to Eqs.~(\ref{eq:qpdfa}) and (\ref{eq:qpdf}). The corresponding nuclear modification will be determined by the quark transverse momentum broadening. At large transverse momentum $\vec v_\perp = \vec l_\perp+\vec l_{q\perp} $, if the dijet spectrum has an effective power-law form $1/v_\perp^{2n}$, and the $p_{\perp}$ broadening is given by Eq.~(\ref{eq:pT_BR}), the nuclear modification factor from single scattering has an enhancement,
\begin{equation}
    I_{e{\rm A}}^S \sim 1+\frac{3n}{2}\frac{R_A\hat q_F^0}{(\vec l_\perp+\vec l_{q\perp})^2},
\end{equation}
that is linear in the nuclear size $R_A$. The numerical result on the nuclear size dependence of $I_{e{\rm A}}^S$ in Fig.~\ref{fig:R_A-distr}(b) ( dashed line) indeed shows such an approximate linear dependence.

\bef
\centering
\includegraphics[width=0.45\textwidth]{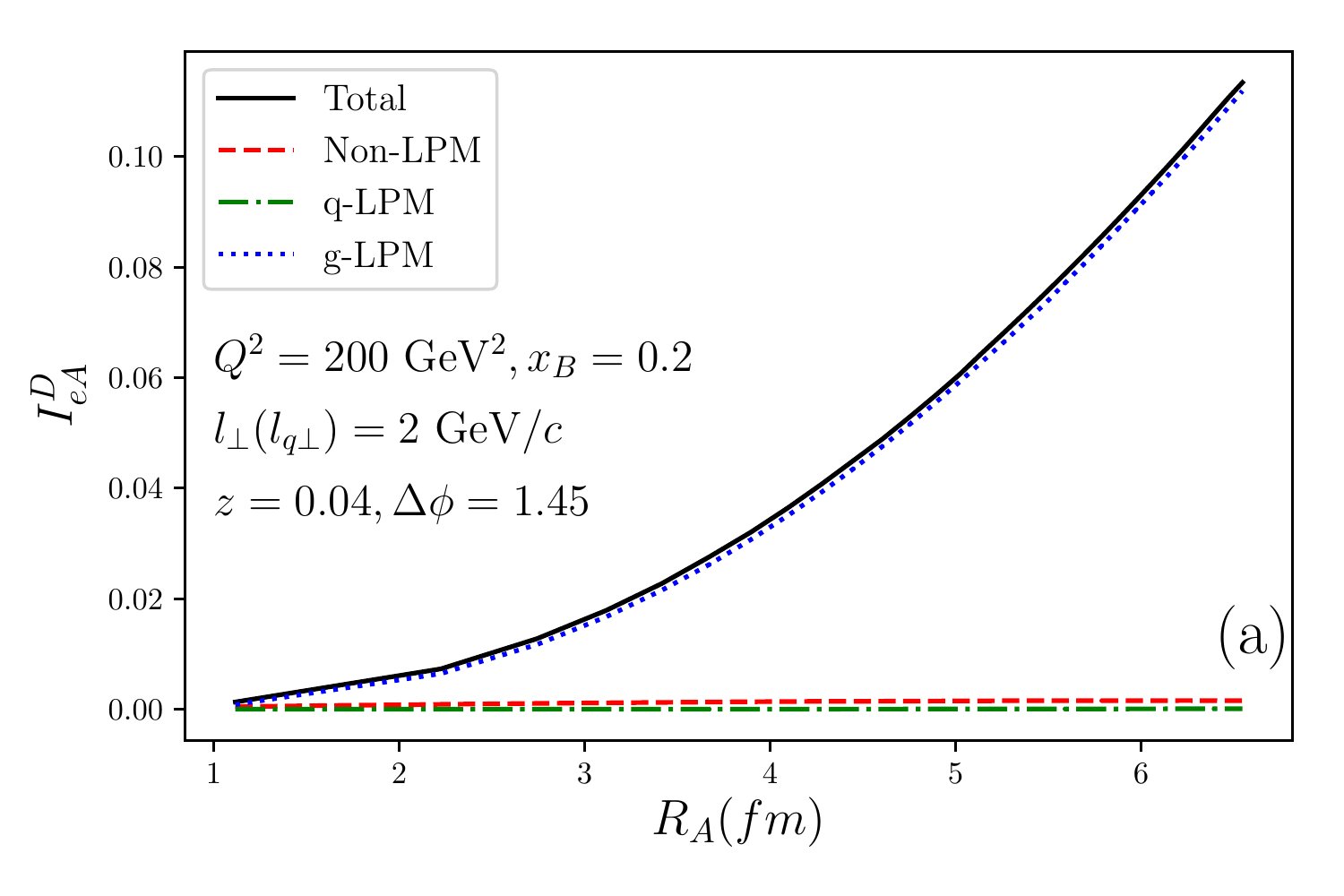}\\
\includegraphics[width=0.45\textwidth]{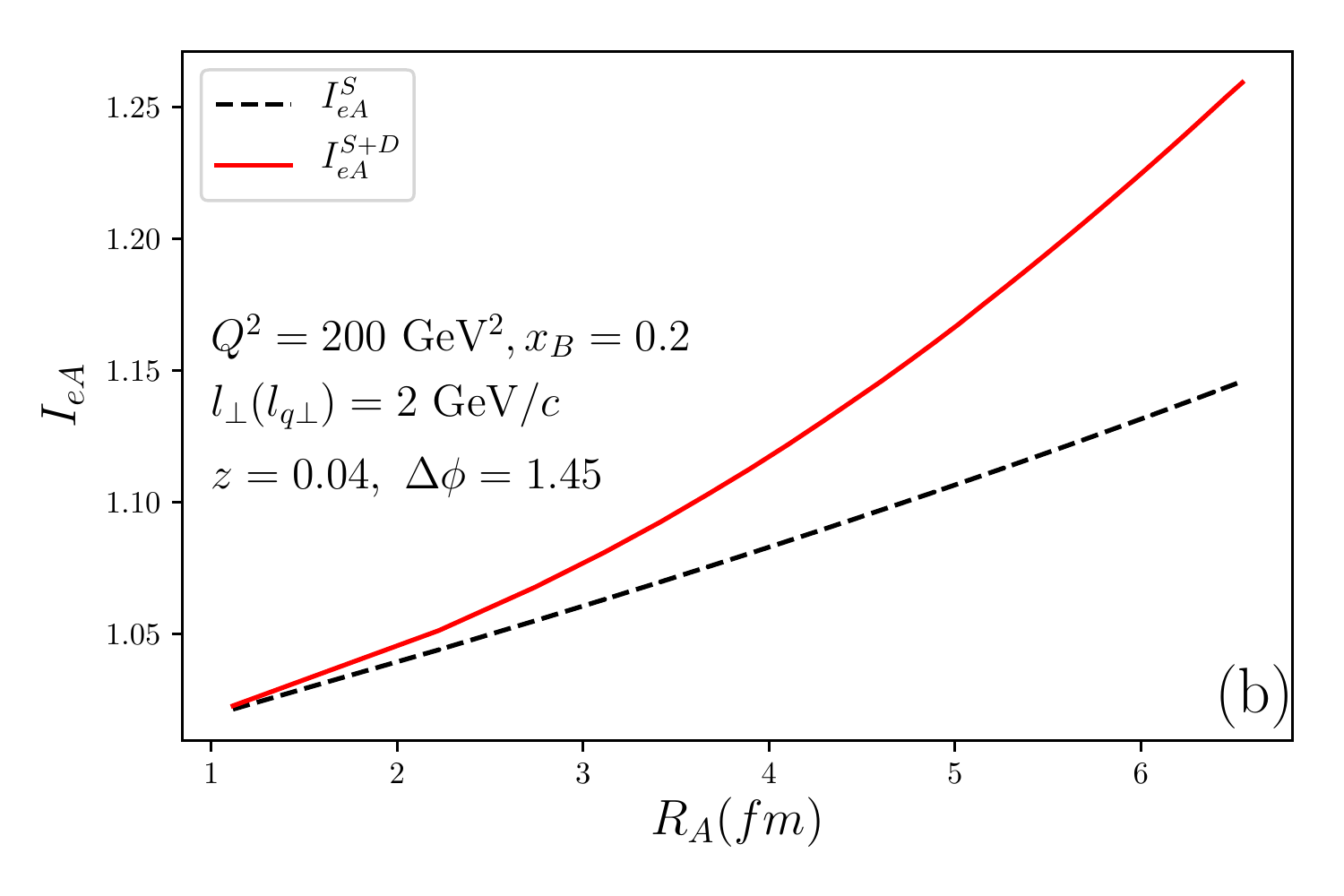}
\caption{The nuclear size $R_A$ dependence of the nuclear modification factor from (a) double  scattering $ I_{eA}^D(l_{\perp},l_{q\perp},  \Delta \phi,z)$ (b) single scattering (dashed) $ I_{eA}^S(l_{\perp},l_{q\perp},  \Delta \phi,z)$ and single+double scattering (solid) $ I_{eA}^{S+D}(l_{\perp},l_{q\perp},  \Delta \phi,z)$ in $e$+Pb DIS.}
\label{fig:R_A-distr}
\eef

In addition to the nuclear quark TMD PDF, the dijet spectrum induced by double scattering is also proportional to a path integral of the differential induced gluon splitting rate (per mean-free-path) over the total length of the quark propagation, according to Eq.~(\ref{equ:nXsection}). The nuclear modification factor for dijet spectrum due to double scattering should then have a linear nuclear size $R_A$ dependence,
\begin{eqnarray}
    I_{e{\rm A}}^D &\propto& \frac{1}{A} \int d^2b_\perp dy_0^- \rho_A(y_0^-,b_\perp) \int dy_1^- \rho_A(y_1^-,b_\perp) \nonumber \\
    &\sim& \frac{9R_A}{16\pi r_0^3},
\end{eqnarray}
from contribution $\mathcal{N}_g^{\rm{nonLPM}}$ that does not have LPM interference. This length dependence comes from the integration over the position of the nucleon $y_1$  involved in the secondary scattering.

For contributions ${\cal N}_g^{\rm qLPM}$ and ${\cal N}_g^{\rm gLPM}$ that have the LPM interference, the integration over the position of the nucleon involved in the secondary scattering has to be weighted with the LPM interference factor. In the case when the photon-quark scattering occurs at the center of the  nucleus, this leads to a nuclear size dependence,
\begin{eqnarray}
\int_0^{R_A} d y_1^- && \rho_A(y_1^-,0_\perp) [1-\cos(y_1^-/\tau_f)] \nonumber \\
&& = \frac{3}{4\pi r_0^3} R_A (1-\frac{\sin(R_A/\tau_f)}{R_A/\tau_f}),
\end{eqnarray}
 of the nuclear modification factor for the medium-induced dijet spectrum, which is approximately quadratic  when $R_A/\tau_f<1$.
Shown in Fig.~\ref{fig:R_A-distr}(a) are the nuclear modification factors of the dijet spectrum due to double scattering for $l_{\perp} = l_{q\perp} = 2$ GeV/$c$, $z=0.04$ at the azimuthal angle  $\Delta\phi = 1.45$ where the medium-induced dijet spectrum is the largest (see Fig.~\ref{fig:RD_azAngle-distr}). Again, the contribution from ${\cal N}_g^{\rm gLMP}$ dominates. 

Including contributions from both single and double scattering, the nuclear size dependence of the modification factor for the total dijet spectrum in $e$+Pb DIS, shown as the solid line in Fig.~\ref{fig:R_A-distr}(b), has a quadratic component due to double scattering on top of a linear dependence due to the transverse momentum broadening in the single scattering. Such a quadratic component in the nuclear size dependence of the nuclear modification factor is another unique feature due to the LPM interference in the dijet production induced by multiple scattering in the cold nuclear medium.

\section{Conclusion and Discussions}
\label{sec-conclusion}

In this study, we calculate the dijet spectrum at LO in pQCD within the framework of a generalized high-twist approach to multiple parton scattering in $e$+A DIS at EIC. We have specifically considered contributions to the dijet spectrum from both single and double scattering and examined in detail the dijet angular correlation which will be influenced by the transverse momentum broadening in the quark TMD PDF and saturation in the gluon TMD PDF in a large nucleus. We have employed a simple model in which we can determine the gluon saturation scale $Q_s^2$ and the quark transport coefficient $\hat q_F$ or transverse momentum broadening squared per unit length. 

In the single hard scattering, the quark transverse momentum broadening dominates the nuclear modification of the dijet angular correlation which resembles the Cronin effect  in the nuclear modification of the hadron transverse momentum spectra in $p$+A collisions. Corrections to the dijet cross section from double hard scattering is relatively small reaching to about 10\% at jet transverse momentum $l_\perp\approx 2$ GeV/$c$ and at the azimuthal angle where contributions from double scattering reach a peak value due to quark transverse momentum broadening and the saturation of medium gluon TMD PDF.  Therefore the nuclear modification of the di-minijet angular correlation is sensitive to the gluon saturation scale in cold nuclei.

We have also examined the dependence of the nuclear modification of the dijet angular correlation on the dijet rapidity gap and on the nuclear size. We found that the dijet correlation increases with the dijet rapidity gap. The nuclear size dependence has a quadratic component on top of a linear dependence. Both of these two features are unique consequences of the LPM interference in the gluon splitting induced by double scattering.

We should note that our calculations are in LO of pQCD. Though we have used TMD PDFs that include a power-law tail at high transverse momentum due to QCD evolution, these are not exactly higher order corrections and resummation of soft gluon radiations which will lead to higher order corrections to the dijet angular correlation in both $e+p$ and $e$+A DIS. Their nuclear modification and effects on the nuclear modification of the dijet spectra need more careful and quantitative investigations. Going into higher order corrections, one also needs to consider jet radius and jet algorithm dependence. 

In principle, gluon saturation occurs when its coherence length $1/x_Gp^+$ becomes larger than the nuclear size $2R_Am_N/p^+$ or $x_G<x_A\equiv 1/2m_NR_A$. In this saturation limit, the effective gluon distribution per nucleon in Eq.~(\ref{equ:saturated_phi}) should be bounded from below by $\phi^0_N(x_A,k_\perp,\mu^2)/A^{1/3}$. To estimate the effect of such gluon saturation due to large coherence length, we can restrict the region of integration over the phase space in the dijet spectra to require $x_G>x_A$, effectively setting the saturated gluon distribution per nucleon to be zero when the coherence length is larger than the nuclear size. As shown in Appendix \ref{extra-plots}, the modified dijet spectra are slightly smaller than that without the restriction $x_G>x_A$. Since the saturated gluon distribution below $x_G<x_A$ should be non-zero, what is estimated in Appendix \ref{extra-plots} is just a lower bound. For a more realistic estimate of this effect in the future, one should consider a gluon saturation model that takes into account of the coherence length of small-$x$ gluons. 

 As we have shown by our numerical calculations, effects of double scattering and the LPM interference on the nuclear modification of dijets are only significant and measurable for minijets at EIC. Identification and reconstruction of these minijets is, however, rather challenging if not impossible. It is more straightforward to measure the correlation of dihadrons with moderately high transverse momentum. We expect the nuclear modification of dijet correlation due to multiple scatterings and induced gluon splittings should also be applicable to dihdadron correlation. This can be calculated by convoluting the dijet cross section with TMD fragmentation functions. This will be our next step in a follow-up study.

\section*{acknowledgement}

We thank Z. B. Kang and F. Ringer for reading our manuscript and helpful comments. This work is supported in part by
the National Science Foundation of China under Grant Nos. 11935007, 11221504, 11861131009 and 11890714, by the Guangdong Major Project of Basic and Applied Basic Research No. 2020B0301030008, 
by the Director, Office of Energy Research, Office of High Energy and Nuclear Physics, Division of Nuclear Physics, of the U.S. Department of Energy under  Contract No. DE-AC02-05CH11231, by the US National Science Foundation under Grant No. ACI-1550228 within the JETSCAPE and OAC-2004571 within the X-SCAPE Collaboration. Computations are performed at the NSC3/CCNU.

\appendix

 \section{Hadronic tensors for dijet production from double scattering}
 
\label{append-HadronicTensor}

\begin{widetext}

In this appendix we list the hadronic tensors for all diagrams (Fig. \ref{fig:Central_11_22} to Fig. \ref{fig:Right_7}) of dijet production induced by double scattering. According to Eqs.~(\ref{eq:hadronic-novp}) and (\ref{eq:tqgcorr2}), these hadronic tensors can be expressed in the following form,
\begin{equation}
\frac{d W^{\mu\nu}}{dz d^2 {l}_{\perp} d^2 {l}_{q\perp} } = \int dx_0 \frac{ \alpha_s}{2\pi}\frac{1+z^{2}}{1-z}H_{(0)}^{\mu\nu}(x_0) \left[ \frac{2\pi \alpha_s}{ N_c}    \int d^2\vec{v}_{\perp}   \int \frac{d^2 \vec{k}_{\perp}}{(2\pi)^2} \int dy_0^- d^2b_\perp dy_1^- \rho(y_0^-, b_\perp)\rho(y_1^-, b_\perp) \mathcal{W} \right].
\end{equation}
In the following we list ${\cal W}$ according to the labeling of the corresponding cut diagrams. We also suppress the impact-parameter $b_\perp$ dependence of the effective quark and gluon TMD PDF inside ${\cal W}$. The definitions of momentum fractions $x_L, x_S, ...$ are given in Eq.~(\ref{equ:x-variables}).

\bef
\begin{center}
 \includegraphics[width=8cm]{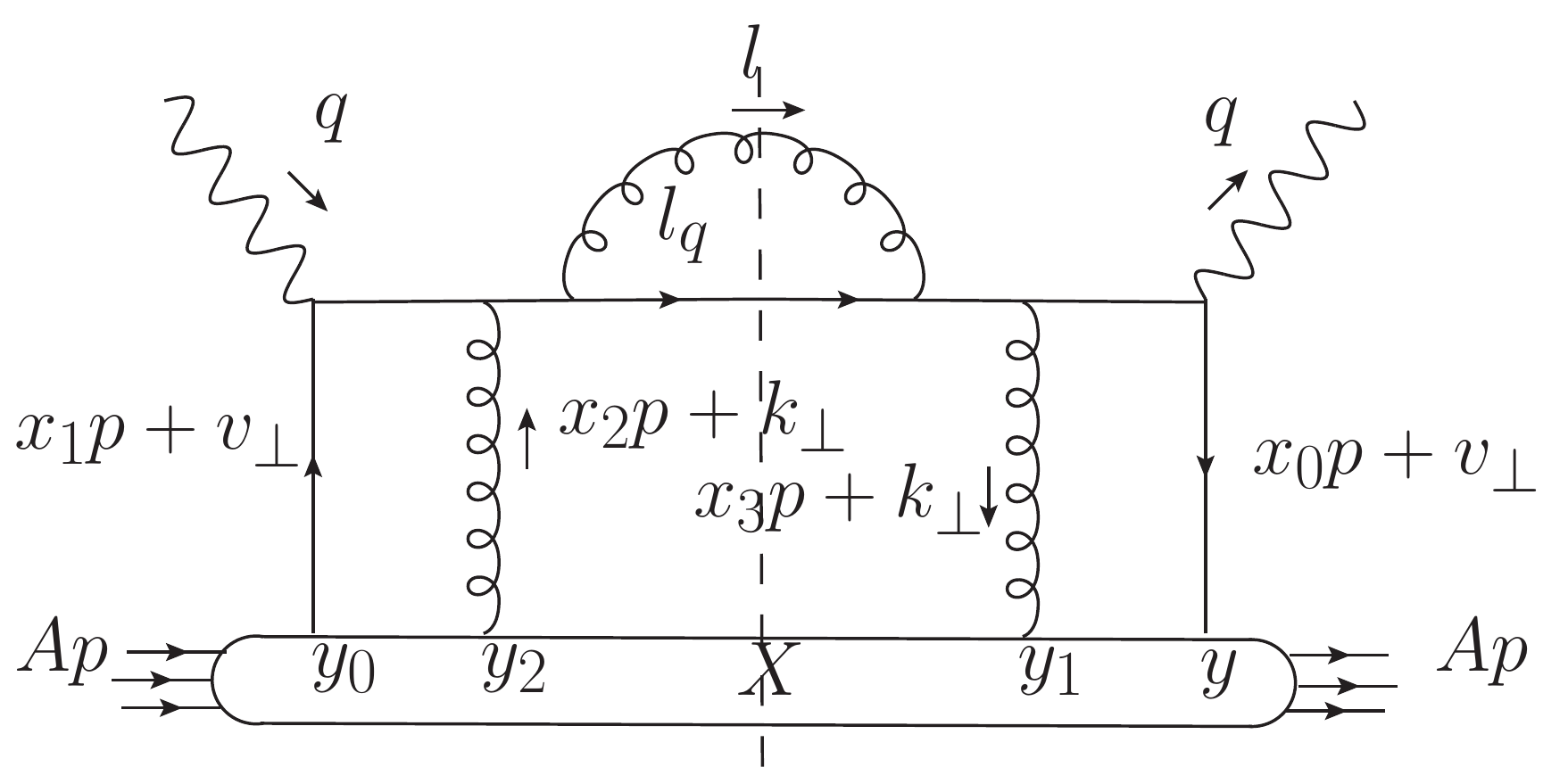} 
  \includegraphics[width=8cm]{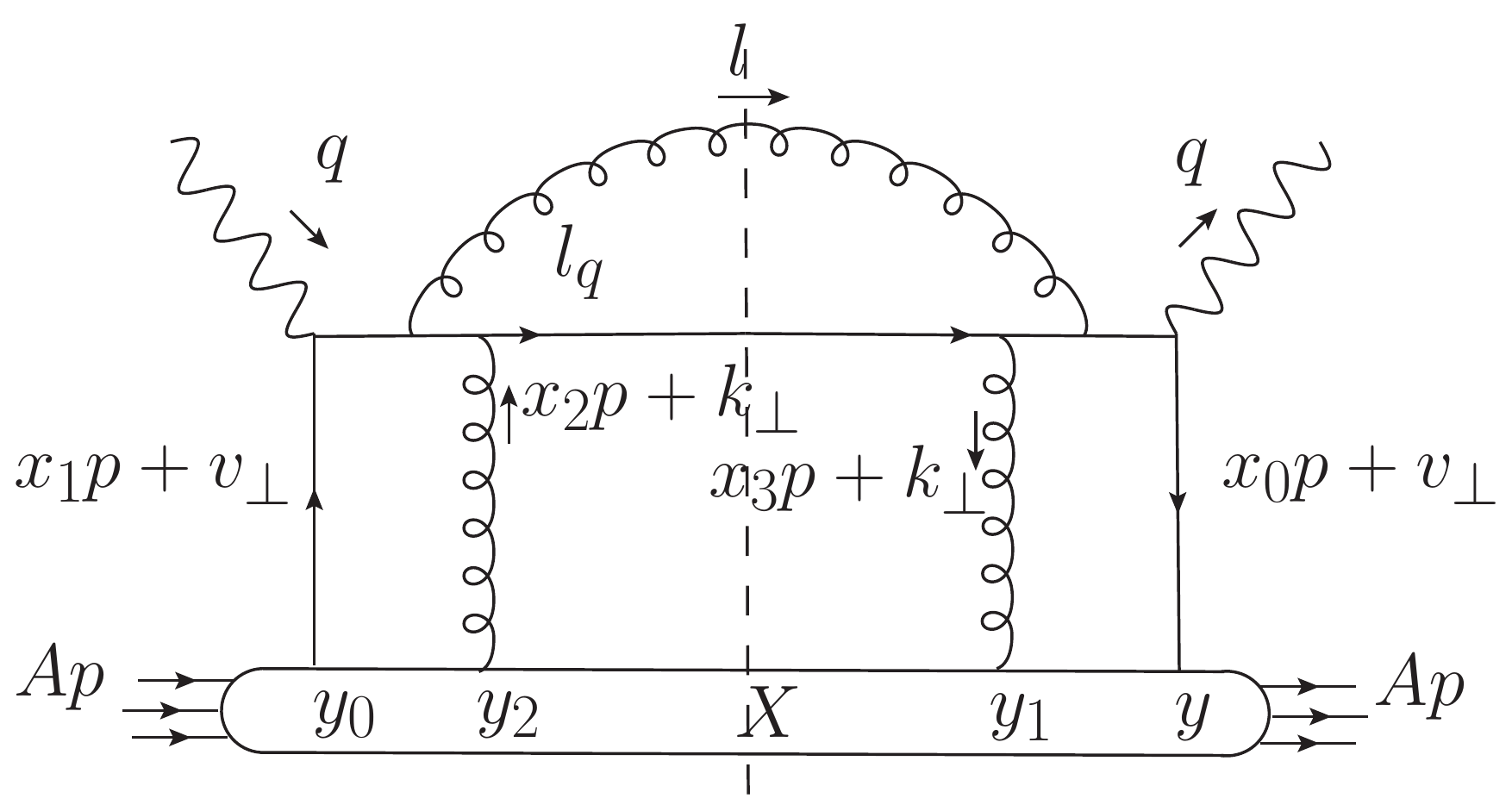} 
 \caption{Central Cut 11 and Central Cut 22}
 \label{fig:Central_11_22}
 \end{center}
\eef

\begin{align}
\begin{aligned}
   \mathcal{W}_{11} = &\frac{C_F}{[\vec{l}_{\perp} -(1- z)(\vec{l}_{\perp}+\vec{l}_{q\perp})]^2}  q_N(x+x_F, \vec{l}_{\perp}+ \vec{l}_{q\perp}-\vec{k}_{\perp}) \frac{  \phi_N(x_L+x_S- x_F,\vec{k}_{\perp})}{k_{\perp}^2} 
  \end{aligned}
  \end{align}
  \begin{align}
  \begin{aligned}
\mathcal{W}_{22} = 
  &  \frac{C_F}{[\vec{l}_{\perp} - (1-z)\vec{v}_{\perp}]^2}\left[q_N(x+x_L+x_E, \vec{v}_{\perp}) \frac{\phi_N(x_S- x_E,\vec{k}_{\perp})}{k_{\perp}^2}    \right.  \\
&   -q_N(x+x_L+x_E, \vec{v}_{\perp}) \frac{\phi_N (x_L+x_S-x_{F},\vec{k}_{\perp})}{k_{\perp}^2}e^{i(x_L+x_E- x_F)p^+y_1^-}   \\
 &- q_N(x+x_{F})\frac{\phi_N(x_S- x_E,\vec{k}_{\perp})}{k_{\perp}^2}e^{-i(x_L+x_E-x_{F})p^+y_1^-}  \\
& \left. + q_N(x+x_{F})\frac{\phi_N(x_L+x_S-x_{F},k_{\perp})}{k_{\perp}^2}\right] 
\end{aligned}
\end{align}

  \bef
 \centering
 \includegraphics[width=8cm]{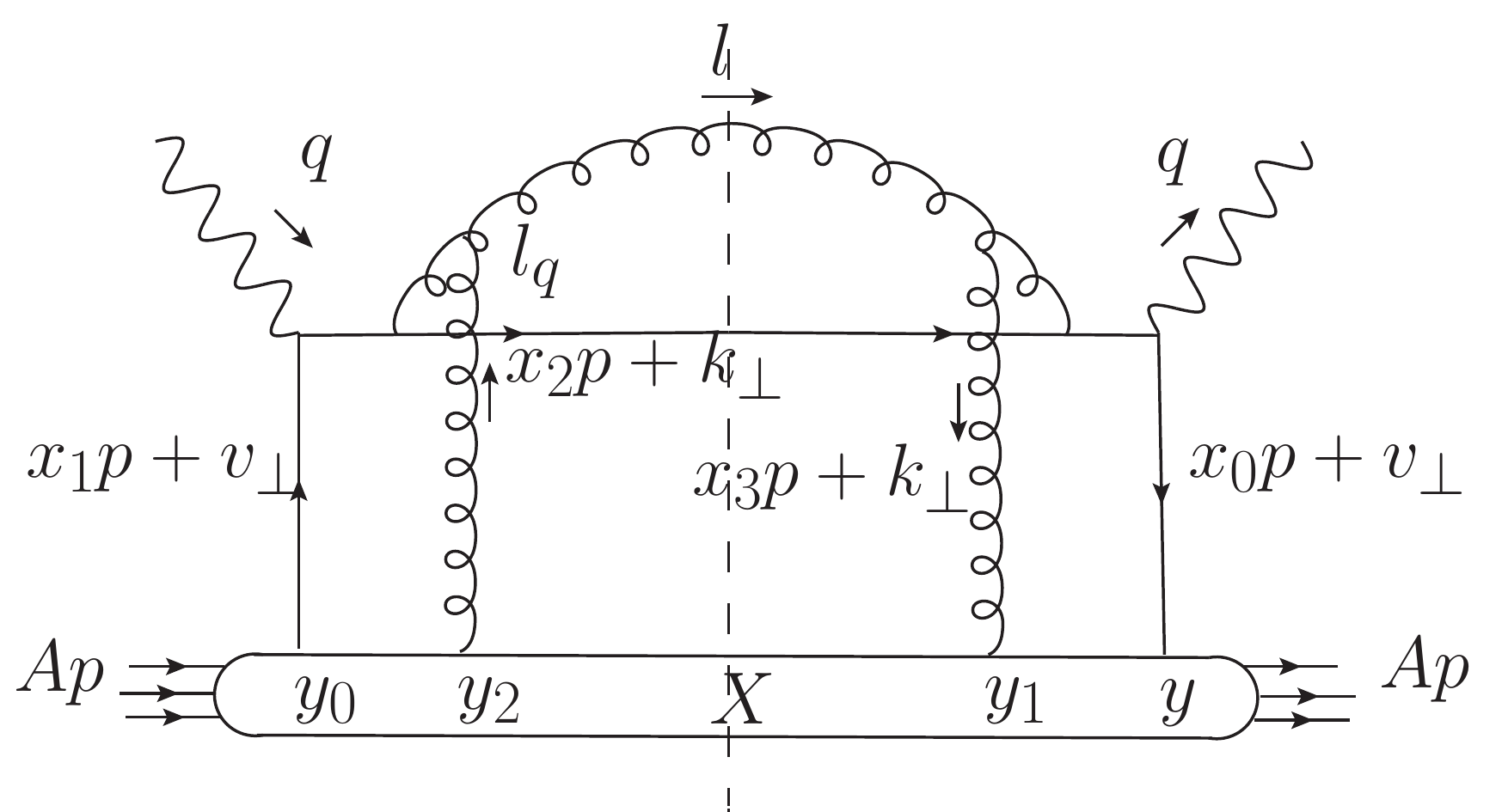}
 \caption{Central Cut 33}
 \label{fig:Central_33}
\eef

\begin{equation}
  \begin{aligned}
\mathcal{W}_{33} = & \frac{C_A}{[\vec{l}_{\perp} -(1- z)\vec{v}_{\perp}-\vec{k}_{\perp}]^2} \left[q_N(x+x_L+x_S+\frac{z}{1-z}x_D, \vec{v}_{\perp})\frac{\phi_N(-\frac{z}{1-z}x_D,\vec{k}_{\perp})}{k_{\perp}^2}  \right.  \\
 &  -q_N(x+x_L+\frac{z}{1-z}x_D+x_S, \vec{v}_{\perp}) \frac{ \phi_N (x_L+x_S-x_{F},\vec{k}_{\perp})}{k_{\perp}^2}e^{i(x_L+x_S+\frac{z}{1-z}x_D - x_F)p^+y_1^-}\\
 &  - q_N(x+x_{F})\frac{\phi_N(-\frac{z}{1-z}x_D,\vec{k}_{\perp})}{k_{\perp}^2}e^{-i(x_L+x_S+\frac{z}{1-z}x_D-x_{F})p^+y_1^-}   \\
& \left. + q_N(x+x_{F},\vec{v}_{\perp})\frac{\phi_N(x_L+x_S-x_{F},\vec{k}_{\perp})}{k_{\perp}^2} \right]  \\
  \end{aligned}
  \end{equation}

  \bef
 \centering
 \includegraphics[width=8cm]{twist4_NLO12_new.pdf} 
  \includegraphics[width=8cm]{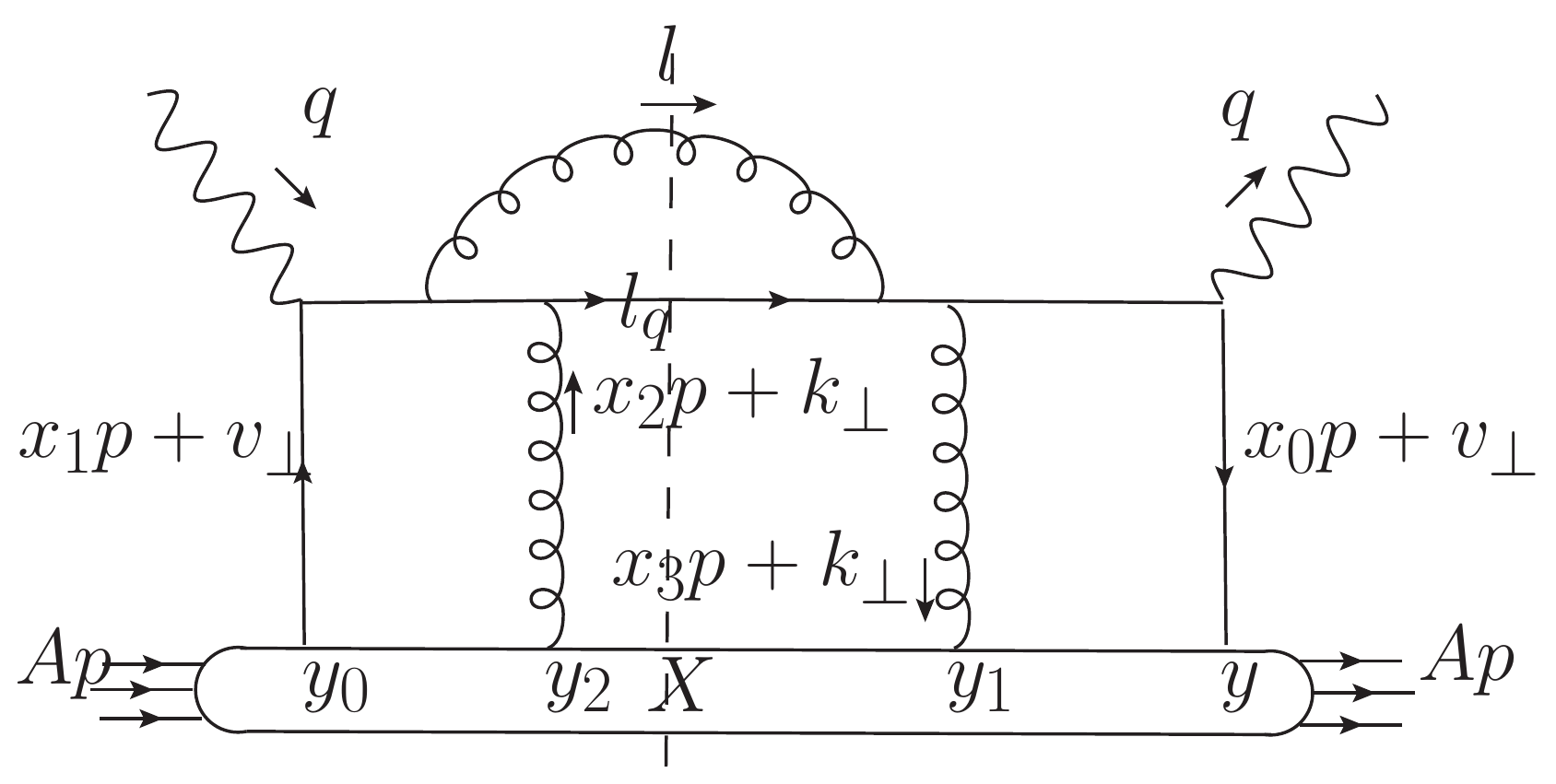} 
 \caption{NLO: Central Cut 12 and Central Cut 21}
 \label{fig:Central_12_21}
\eef

\begin{equation}
  \begin{aligned}
   \mathcal{W}_{12} = & \frac{1}{2N_c} \frac{[\vec{l}_{\perp} -(1- z)\vec{v}_{\perp}]\cdot [\vec{l}_{\perp} -(1- z)(\vec{v}_{\perp}+\vec{k}_{\perp})]}{[\vec{l}_{\perp} -(1- z)\vec{v}_{\perp}]^2[\vec{l}_{\perp} -(1- z)(\vec{v}_{\perp}+\vec{k}_{\perp})]^2}  \left[q_N(x+x_F, \vec{v}_{\perp})\frac{\phi_N(x_L+x_S-x_F,\vec{k}_{\perp})}{k_{\perp}^2} \right. \\
 &\left.  - q_N(x+x_{L}+x_E,\vec{v}_{\perp})\frac{\phi_N(x_L+x_S-x_F,\vec{k}_{\perp})}{k_{\perp}^2}e^{i(x_L+x_E-x_F)p^+y_1^-} \right] \\
  \end{aligned}
  \end{equation}
  
  \begin{equation}
    \begin{aligned}
 \mathcal{W}_{21} = &\frac{1}{2N_c} \frac{[\vec{l}_{\perp} -(1- z)\vec{v}_{\perp}]\cdot [\vec{l}_{\perp} -(1- z)(\vec{v}_{\perp}+\vec{k}_{\perp})]}{[\vec{l}_{\perp} -(1- z)\vec{v}_{\perp}]^2[\vec{l}_{\perp} -(1- z)(\vec{v}_{\perp}+\vec{k}_{\perp})]^2} \left[q_N(x+x_F, \vec{v}_{\perp})\frac{\phi_N(x_L+x_S-x_F,\vec{k}_{\perp})}{k_{\perp}^2}  \right.\\
 &\left. - q_N(x+x_{F},\vec{v}_{\perp})\frac{\phi_N(x_S-x_{E},\vec{k}_{\perp})}{k_{\perp}^2}e^{-i(x_L+x_E-x_F)p^+y_1^-} \right]  \\
   \end{aligned}
  \end{equation}
  
   \bef
 \centering
 \includegraphics[width=8cm]{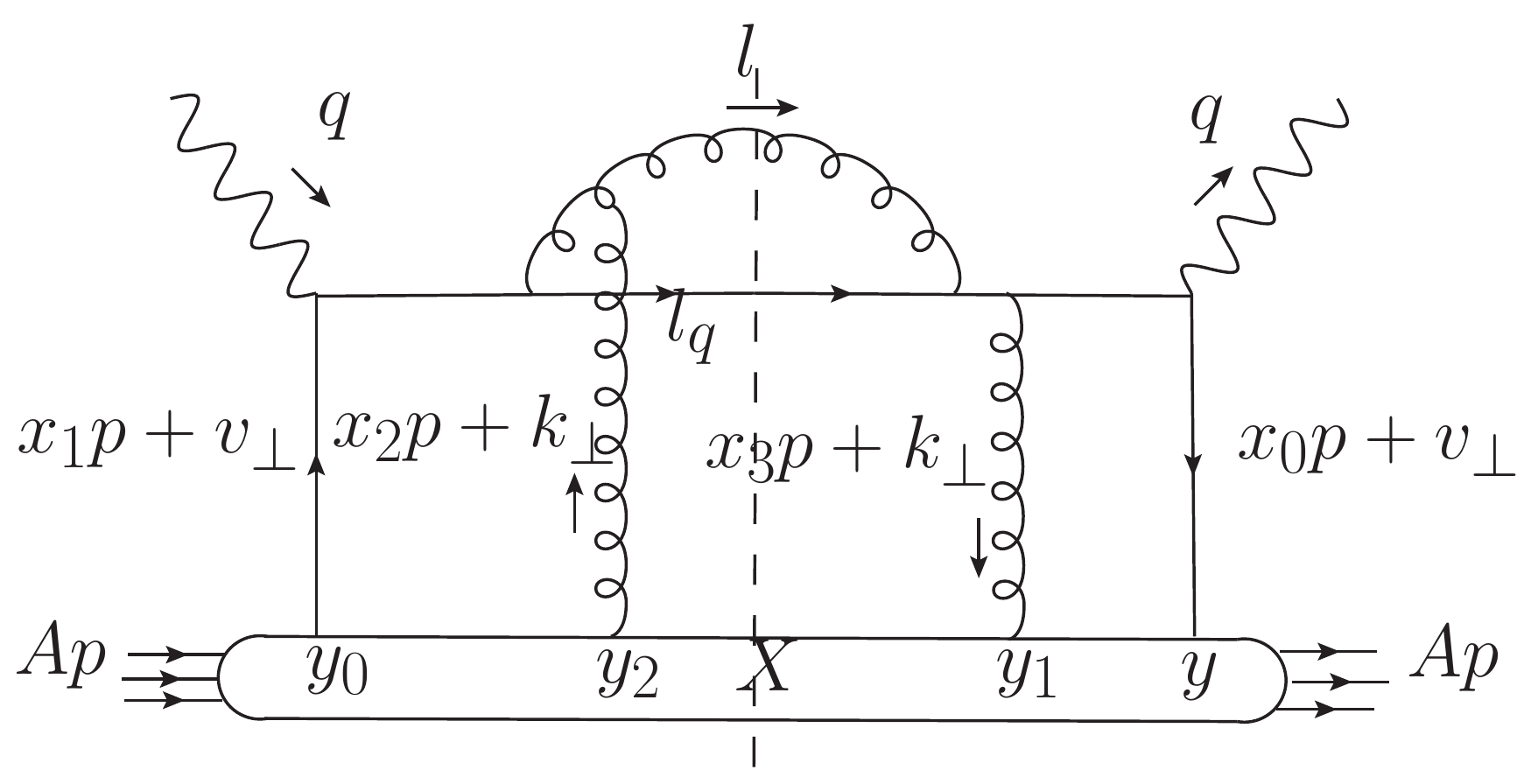} 
  \includegraphics[width=8cm]{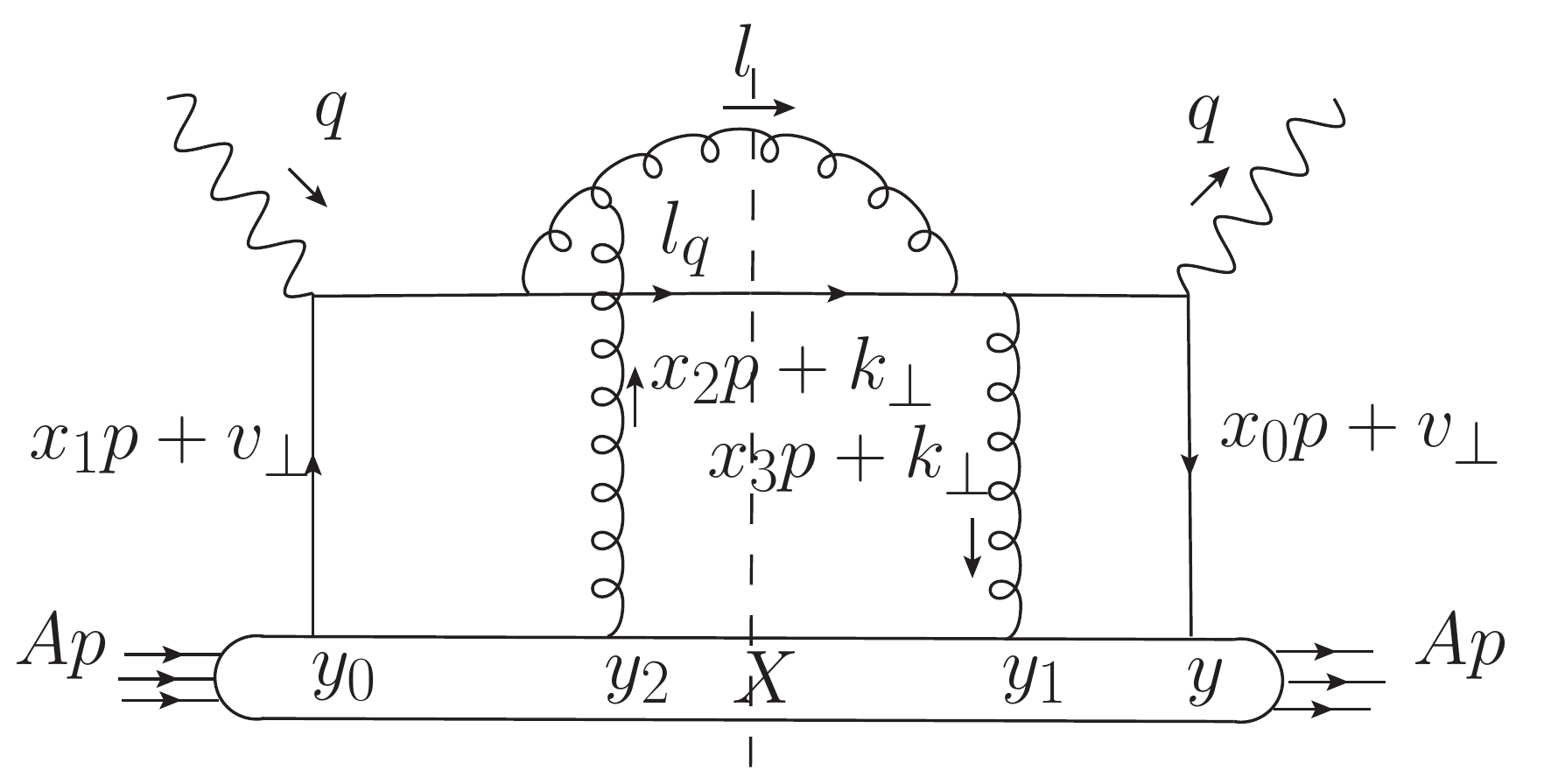} 
 \caption{NLO: Central Cut 13 and Central Cut 31}
  \label{fig:Central_13_31}
\eef

\begin{equation}
  \begin{aligned}
 \mathcal{W}_{13} =& \frac{C_A}{2} \frac{[\vec{l}_{\perp} - (1-z)\vec{v}_{\perp}-\vec{k}_{\perp}]\cdot [\vec{l}_{\perp} - (1-z)(\vec{v}_{\perp}+\vec{k}_{\perp})] }{[\vec{l}_{\perp} - (1-z)\vec{v}_{\perp}-\vec{k}_{\perp}]^2[\vec{l}_{\perp} - (1-z)(\vec{v}_{\perp}+\vec{k}_{\perp})]^2} \\
 &\times \left[ q_N(x+x_L+\frac{z}{1-z}x_D+x_S, \vec{v}_{\perp}) \frac{ \phi_N(x_L+x_S-x_F,\vec{k}_{\perp})}{k_{\perp}^2} e^{i(x_L+\frac{z}{1-z}x_D+x_S  -x_F)p^+y_1^-} \right. \\
 &\left. -  q_N(x+x_F, \vec{v}_{\perp})\frac{\phi_N(x_L+x_S-x_F, \vec{k}_{\perp})}{k_{\perp}^2}  \right] \\
  \end{aligned}
  \end{equation}

\begin{equation}
  \begin{aligned}
 \mathcal{W}_{31} =& \frac{C_A}{2} \frac{[\vec{l}_{\perp} - (1-z)\vec{v}_{\perp}-\vec{k}_{\perp}]\cdot [\vec{l}_{\perp} - (1-z)(\vec{v}_{\perp}+\vec{k}_{\perp})] }{[\vec{l}_{\perp} - (1-z)\vec{v}_{\perp}-\vec{k}_{\perp}]^2[\vec{l}_{\perp} - (1-z)(\vec{v}_{\perp}+\vec{k}_{\perp})]^2} \\
  & \times\left[  q_N(x+x_F, \vec{v}_{\perp}) \frac{ \phi_N(-\frac{z}{1-z}x_D,\vec{k}_{\perp})}{k_{\perp}^2} e^{-i(x_L+\frac{z}{1-z}x_D+x_S  -x_F)p^+y_1^-} \right. \\
  &\left. -  q_N(x+x_F, \vec{v}_{\perp})\frac{\phi_N(x_L+x_S-x_F, \vec{k}_{\perp})}{k_{\perp}^2}  \right] \\
  \end{aligned}
  \end{equation}
 
\bef
 \centering
 \includegraphics[width=8cm]{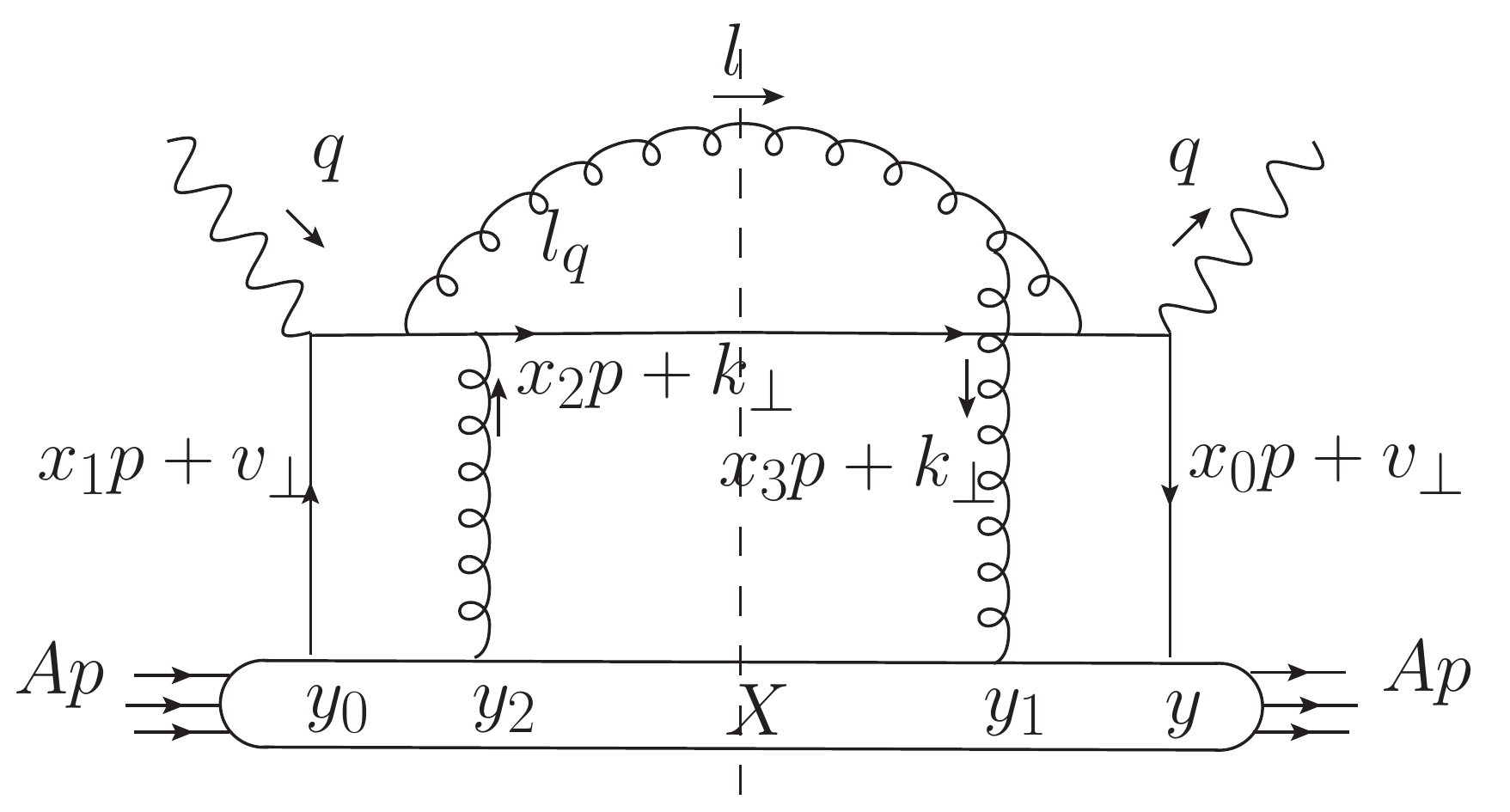}
  \includegraphics[width=8cm]{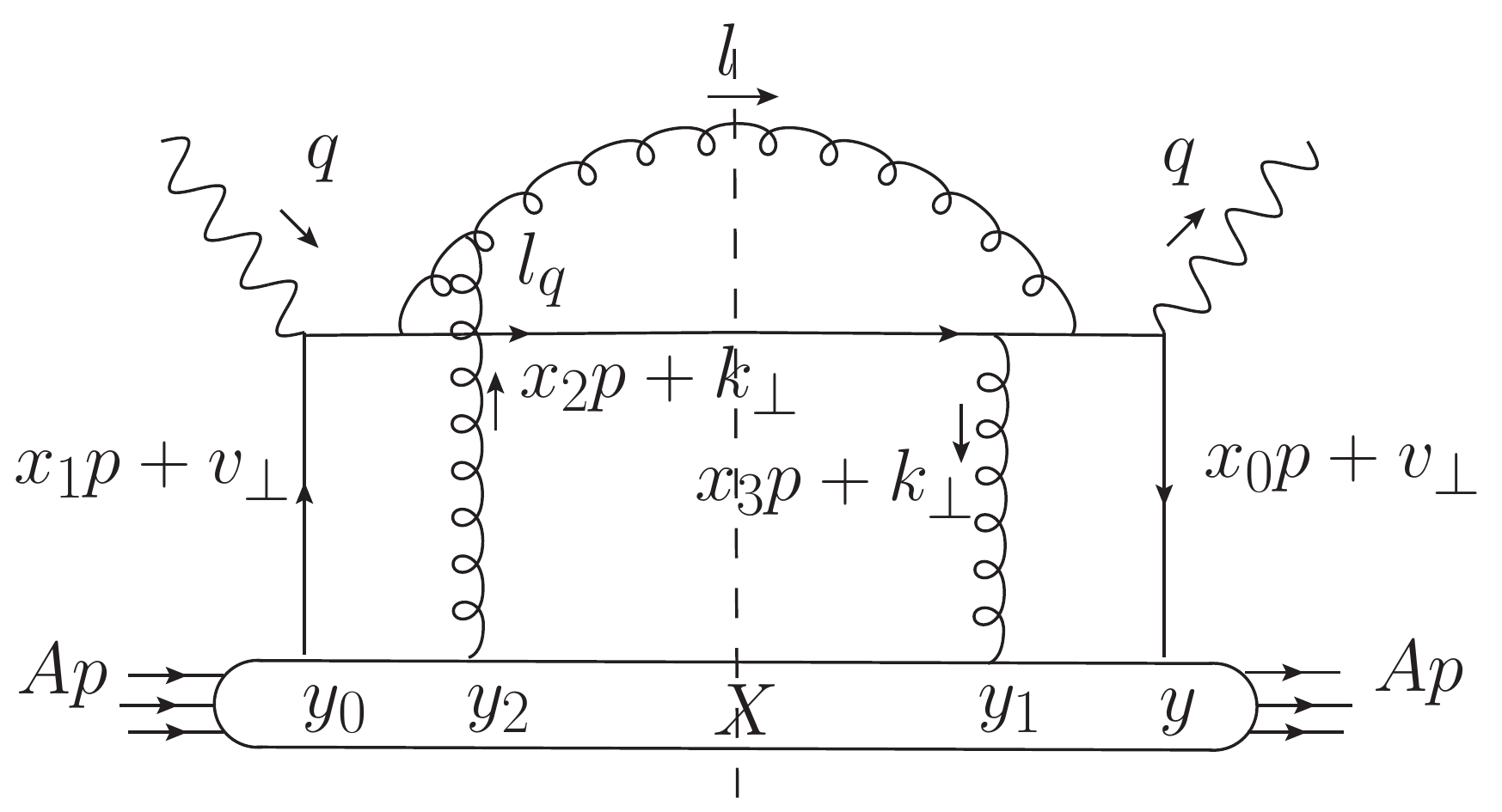}  
 \caption{NLO: Central Cut 23 and Central Cut 32}
  \label{fig:Central_23_32}
\eef

\begin{align}
  \begin{aligned}
\mathcal{W}_{23} =  & -\frac{C_A}{2} \frac{[\vec{l}_{\perp} - (1-z)\vec{v}_{\perp}]\cdot [\vec{l}_{\perp} - (1-z)\vec{v}_{\perp}-\vec{k}_{\perp}]}{[\vec{l}_{\perp} - (1-z)\vec{v}_{\perp}]^2[\vec{l}_{\perp} - (1-z)\vec{v}_{\perp}-\vec{k}_{\perp}]^2}  \\
&\times \left[ q_N(x+x_L+x_S+\frac{z}{1-z}x_D, \vec{v}_{\perp})\frac{ \phi_N(x_S-x_E,\vec{k}_{\perp})}{k_{\perp}^2} e^{i(\frac{z}{1-z}x_D +x_S -x_E)p^+y_1^-} \right.\\ 
 & -q_N(x+x_L+\frac{z}{1-z}x_D+x_S, \vec{v}_{\perp} )\frac{\phi_N(x_L+x_S-x_F, \vec{k}_{\perp} )}{k_{\perp}^2}  e^{i(x_L+x_S+\frac{z}{1-z}x_D-x_F)p^+y_1^-} \\
 &\left.  -  q_N(x+x_F, \vec{v}_{\perp})\frac{\phi_N(x_S-x_E, \vec{k}_{\perp} )}{k_{\perp}^2}e^{-i(x_L+x_E-x_F)p^+y_1^-} +  q_N(x+x_F, \vec{v}_{\perp})\frac{\phi_N(x_L+x_S-x_F)}{k_{
 \perp}^2}  \right] \\
   \end{aligned}
  \end{align}
  
  \begin{equation}
  \begin{aligned}
\mathcal{W}_{32} =&- \frac{C_A}{2} \frac{[\vec{l}_{\perp} - (1-z)\vec{v}_{\perp}]\cdot [\vec{l}_{\perp} - (1-z)\vec{v}_{\perp}-\vec{k}_{\perp}]}{[\vec{l}_{\perp} - (1-z)\vec{v}_{\perp}]^2[\vec{l}_{\perp} - (1-z)\vec{v}_{\perp}-\vec{k}_{\perp}]^2} \\
  & \left[   q_N(x+x_L+x_E, \vec{v}_{\perp})\frac{\phi_N(-\frac{z}{1-z}x_D, \vec{k}_{\perp} )}{k_{\perp}^2}e^{-i(\frac{z}{1-z}x_D+x_S-x_E)p^+y_1^-}  \right.\\
  & - q_N(x+x_L+x_E, \vec{v}_{\perp}) \frac{\phi_N(x_L+x_S-x_F,\vec{k}_{\perp})}{k_{\perp}^2} e^{i(x_L+x_E -x_F)p^+y_1^-} \\
  &  - q_N(x+x_F, \vec{v}_{\perp})\frac{\phi_N(-\frac{z}{1-z}x_D, \vec{k}_{\perp})}{k_{\perp}^2} e^{-i(x_L+\frac{z}{1-z}x_D+x_S-x_F)p^+y_1^-}  \\ 
 &\left. + q_N(x+x_F, \vec{v}_{\perp} )\frac{\phi_N(x_L+x_S-x_F, \vec{k}_{\perp} )}{k_{\perp}^2}\right]   \\
  \end{aligned}
  \end{equation}

\bef
 \centering
 \includegraphics[width=8cm]{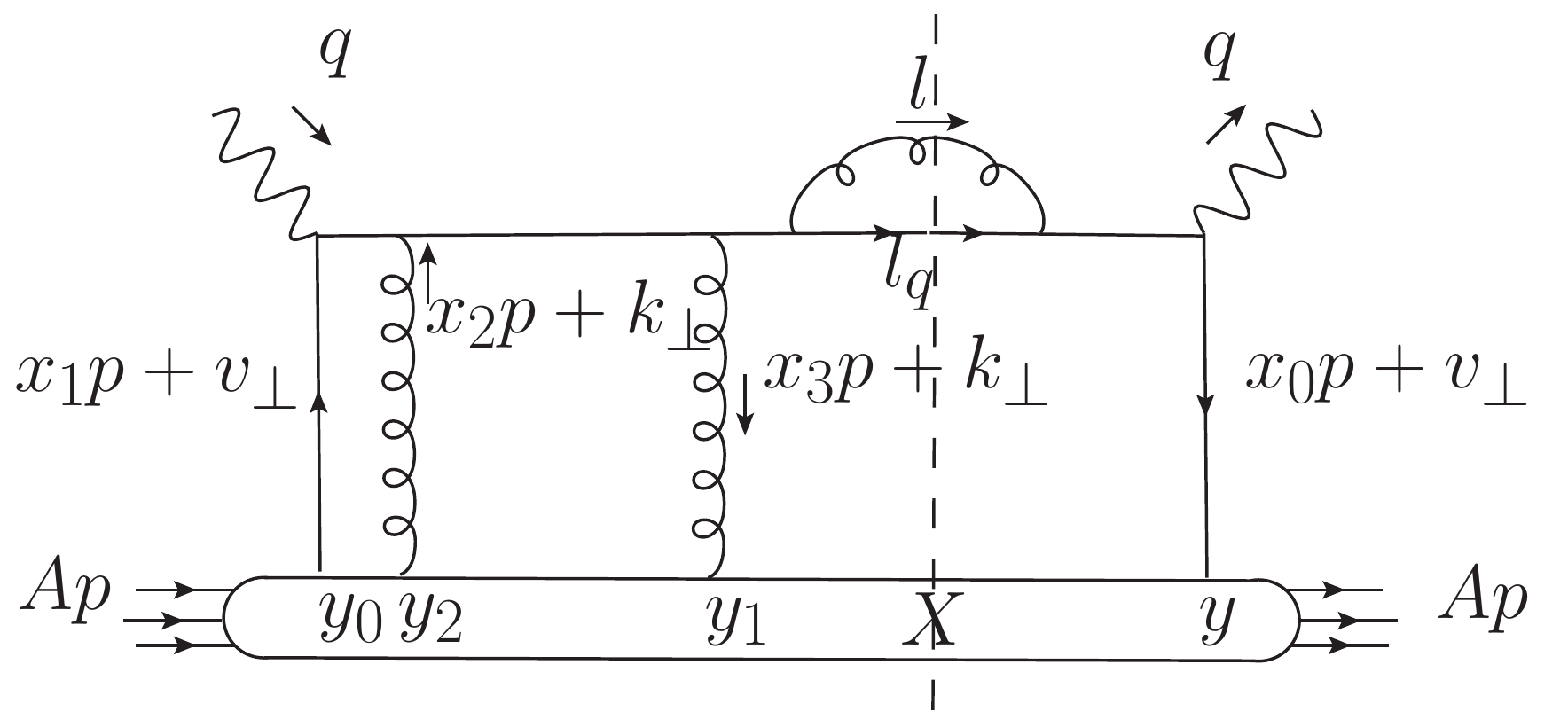} 
 \includegraphics[width=8cm]{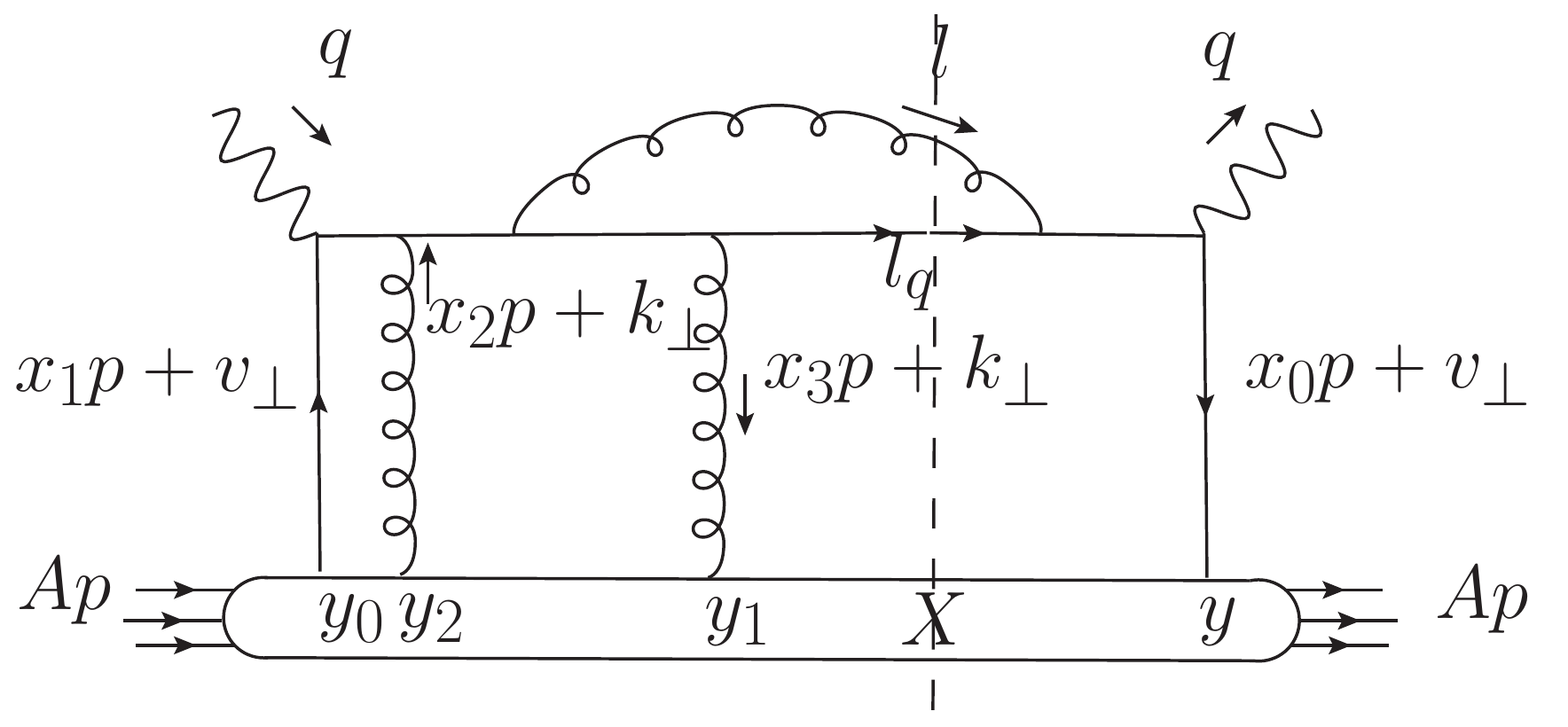} 
 \caption{Right Cut 1 and Right Cut 2}
 \label{fig:Right_1_2}
\eef

\begin{equation}
  \begin{aligned}
 \mathcal{W}_{R1} =&-\frac{C_F}{[\vec{l}_{\perp} - (1-z)\vec{v}_{\perp}]^2 }  q_N(x+x_L+x_E, \vec{v}_{\perp})\frac{ \phi_N(x_H-x_F,\vec{k}_{\perp})}{k_{\perp}^2} e^{i(x_L+x_E-x_F)p^+y_1^-}  \\ 
  \end{aligned}
  \end{equation}
  
  \begin{equation}
  \begin{aligned}
 \mathcal{W}_{L1} =&- \frac{C_F}{[\vec{l}_{\perp} - (1-z)\vec{v}_{\perp}]^2 }q_N(x+x_F, \vec{v}_{\perp})\frac{ \phi_N(x_H-x_E-x_L,\vec{k}_{\perp})}{k_{\perp}^2} e^{-i(x_L+x_E-x_F)p^+y_1^-}  \\ 
  \end{aligned}
  \end{equation}

 \begin{align}
  \begin{aligned}
\mathcal{W}_{R2} =& \frac{1}{2N_c}  \frac{[\vec{l}_{\perp} - (1-z)\vec{v}_{\perp}]\cdot [\vec{l}_{\perp} - (1-z)(\vec{v}_{\perp}+\vec{k}_{\perp})]}{[\vec{l}_{\perp} - (1-z)\vec{v}_{\perp}]^2 [\vec{l}_{\perp} - (1-z)(\vec{v}_{\perp}+\vec{k}_{\perp})]^2}\\
  \end{aligned}
\end{align}  

 \begin{equation}
  \begin{aligned}
\mathcal{W}_{L2} =&  \frac{1}{2N_c}  \frac{[\vec{l}_{\perp} - (1-z)\vec{v}_{\perp}]\cdot [\vec{l}_{\perp} - (1-z)(\vec{v}_{\perp}+\vec{k}_{\perp})]}{[\vec{l}_{\perp} - (1-z)\vec{v}_{\perp}]^2 [\vec{l}_{\perp} - (1-z)(\vec{v}_{\perp}+\vec{k}_{\perp})]^2}\\
&\times\left[ q_N(x+x_F, \vec{v}_{\perp}) \phi_N(x_S-x_E,\vec{k}_{\perp}) e^{-i(x_L+x_E-x_F)p^+y_1^-} \right.  \\
& \left. - q_N(x+x_F, \vec{v}_{\perp}) \phi_N(x_H-x_F,\vec{k}_{\perp}) e^{-i(x_L+x_E-x_F)p^+y_1^-}    \right] \\
  \end{aligned}
\end{equation}  

  \bef
 \centering
 \includegraphics[width=8cm]{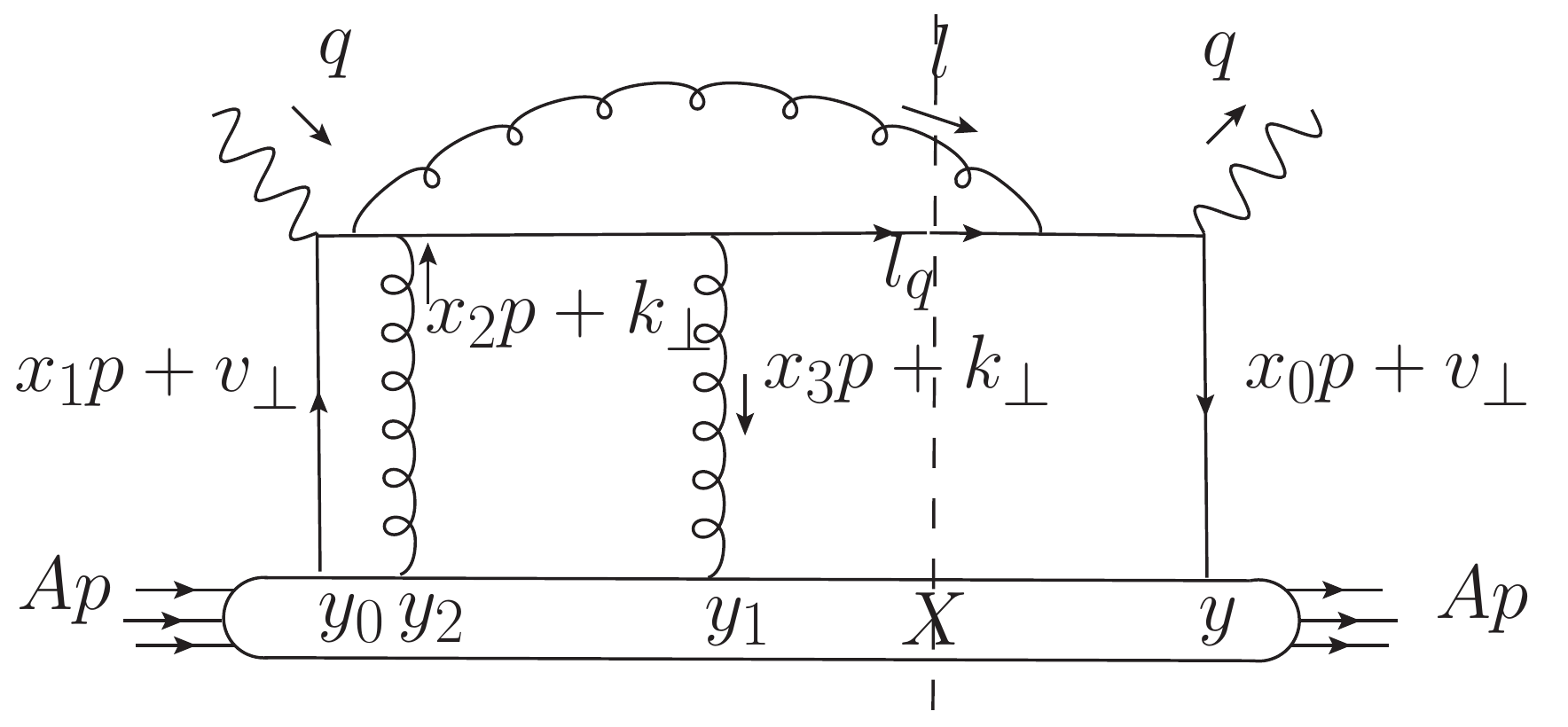} 
  \includegraphics[width=8cm]{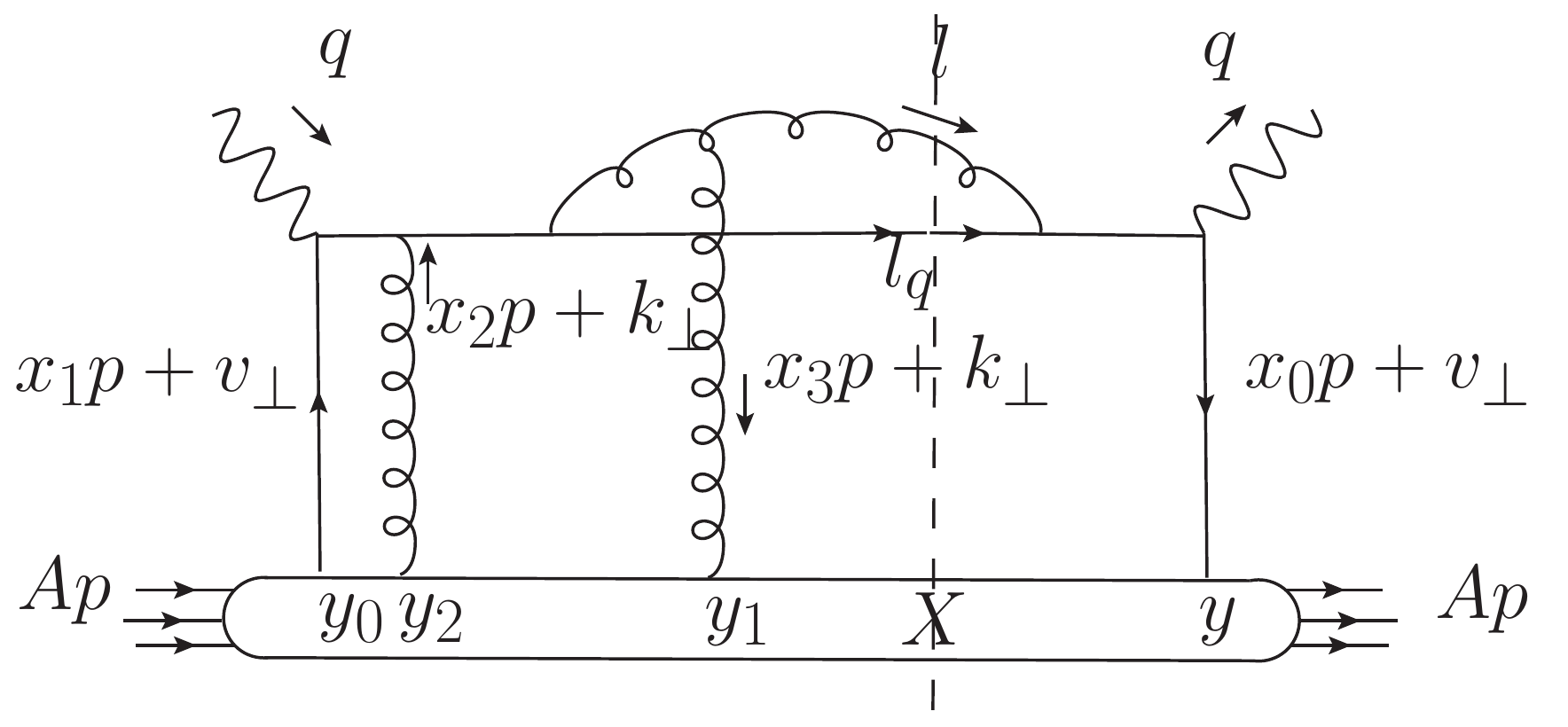} 
 \caption{Right Cut 3 and Right Cut 4}
  \label{fig:Right_3_4}
\eef

 \begin{equation}
  \begin{aligned}
\mathcal{W}_{R3} =& \frac{C_F}{[\vec{l}_{\perp} - (1-z)\vec{v}_{\perp}]^2 }\left[ q_N(x+x_L+x_E, \vec{v}_{\perp}) \phi_N(x_L+x_S-x_F,\vec{k}_{\perp}) e^{i(x_L+x_E-x_F)p^+y_1^-} \right. \\
&\left. - q_N(x+x_L+x_E, \vec{v}_{\perp}) \phi_N(x_S-x_E,\vec{k}_{\perp})    \right] \\
  \end{aligned}
\end{equation}  

 \begin{equation}
  \begin{aligned}
\mathcal{W}_{L3} =&\frac{C_F}{[\vec{l}_{\perp} - (1-z)\vec{v}_{\perp}]^2 } \left[   q_N(x+x_F, \vec{v}_{\perp}) \phi_N(x_S-x_E,\vec{k}_{\perp})e^{-i(x_L+x_E-x_F)p^+y_1^-} \right. \\
&\left. - q_N(x+x_L+x_E, \vec{v}_{\perp}) \phi_N(x_S-x_E,\vec{k}_{\perp})      \right]  \\  
\end{aligned}
\end{equation}  

 \begin{equation}
  \begin{aligned}
\mathcal{W}_{R4} =&\frac{C_A}{2}  \frac{[\vec{l}_{\perp} - (1-z)\vec{v}_{\perp}]\cdot [\vec{l}_{\perp} - (1-z)\vec{v}_{\perp}-z\vec{k}_{\perp}]}{[\vec{l}_{\perp} - (1-z)\vec{v}_{\perp}]^2 [\vec{l}_{\perp} - (1-z)\vec{v}_{\perp}-z\vec{k}_{\perp}]^2} \\
&\left[q_N(x+x_L+x_E, \vec{v}_{\perp}) \phi_N(x_H-x_F,\vec{k}_{\perp}) e^{i(x_L+x_E-x_F)p^+y_1^-} \right. \\
&\left. - q_N(x+x_L+x_E, \vec{v}_{\perp}) \phi_N(x_L+\frac{z}{1-z}x_D+x_E-x_F,\vec{k}_{\perp})e^{i(x_L+x_E-x_F)p^+y_1^-}    \right] \\
  \end{aligned}
\end{equation}  

 \begin{equation}
  \begin{aligned}
\mathcal{W}_{L4} =&\frac{C_A}{2}  \frac{[\vec{l}_{\perp} - (1-z)\vec{v}_{\perp}]\cdot [\vec{l}_{\perp} - (1-z)\vec{v}_{\perp}-z\vec{k}_{\perp}]}{[\vec{l}_{\perp} - (1-z)\vec{v}_{\perp}]^2 [\vec{l}_{\perp} - (1-z)\vec{v}_{\perp}-z\vec{k}_{\perp}]^2}\\
&\times \left[q_N(x+x_F, \vec{v}_{\perp}) \phi_N(x_H-x_L-x_E,\vec{k}_{\perp}) e^{-i(x_L+x_E-x_F)p^+y_1^-} \right. \\
&\left. - q_N(x+x_F, \vec{v}_{\perp}) \phi_N(\frac{z}{1-z}x_D,\vec{k}_{\perp})e^{-i(x_L+x_E-x_F)p^+y_1^-}     \right] \\
  \end{aligned}
\end{equation}

 \bef
 \centering
 \includegraphics[width=8cm]{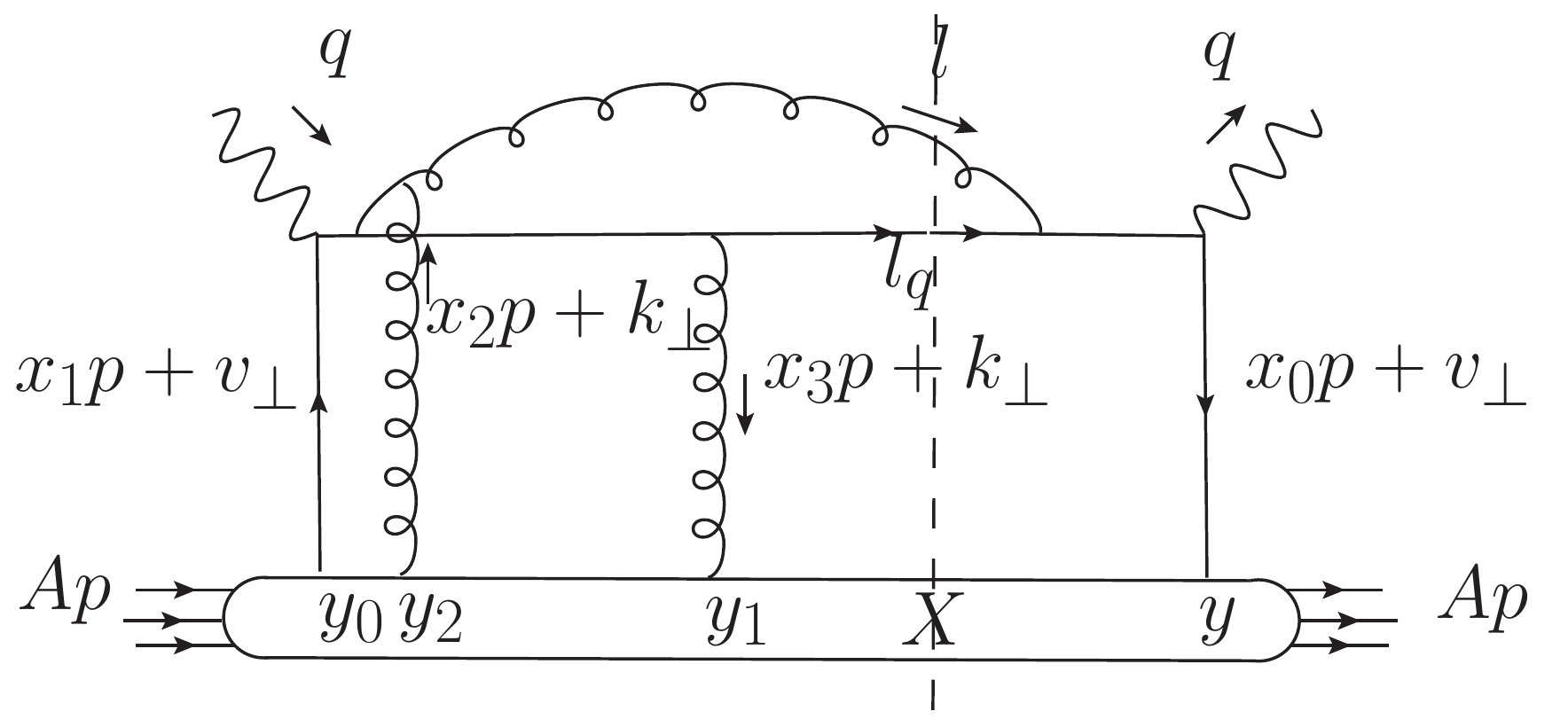} 
 \includegraphics[width=8cm]{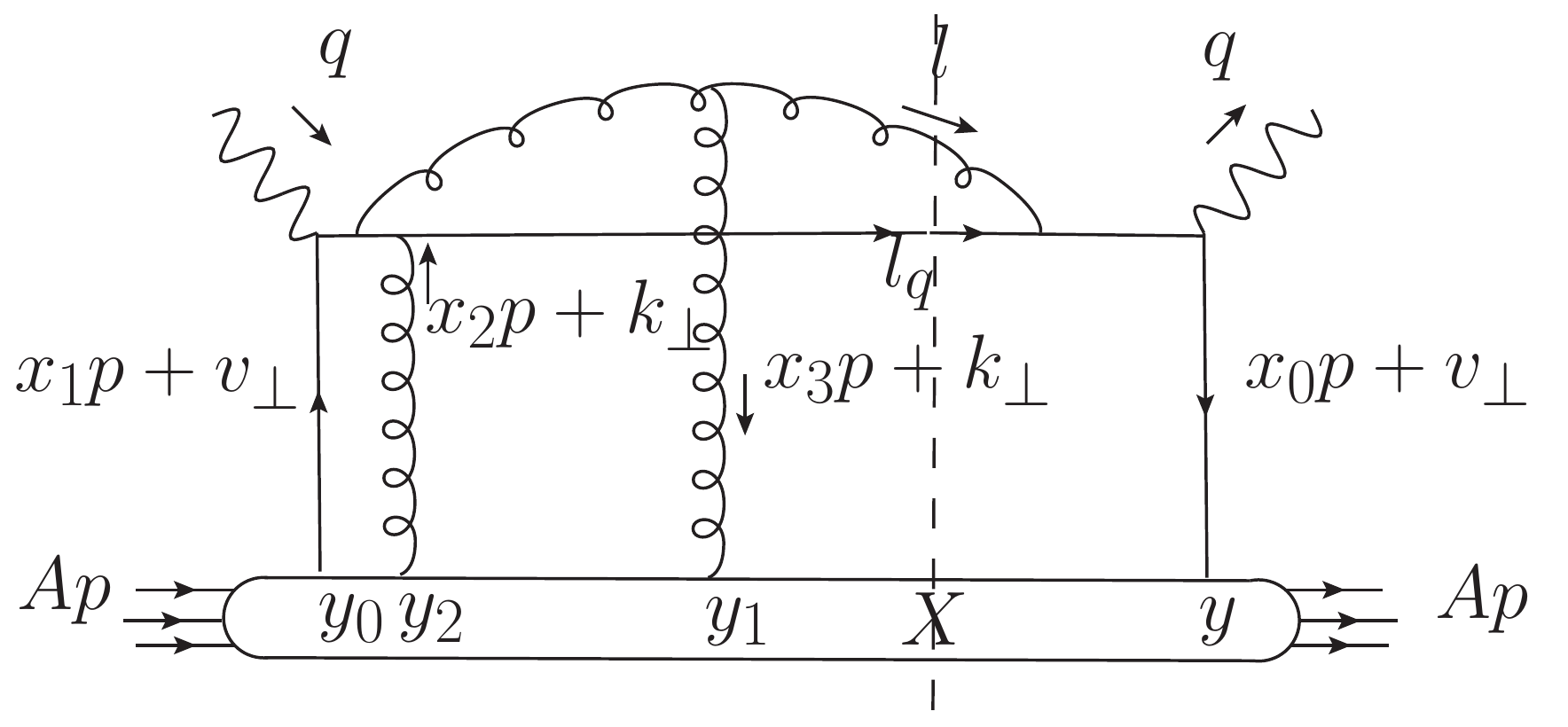} 
 \caption{Right Cut 5 and Right Cut 6}
   \label{fig:Right_5_6}
\eef

 \begin{equation}
  \begin{aligned}
\mathcal{W}_{R5} =&\frac{C_A}{2}  \frac{[\vec{l}_{\perp} - (1-z)\vec{v}_{\perp}]\cdot [\vec{l}_{\perp} - (1-z)\vec{v}_{\perp}-\vec{k}_{\perp}]}{[\vec{l}_{\perp} - (1-z)\vec{v}_{\perp}]^2 [\vec{l}_{\perp} - (1-z)\vec{v}_{\perp}-\vec{k}_{\perp}]^2 }\\
&\times \left[ q_N(x+x_L+x_E, \vec{v}_{\perp}) \phi_N(-\frac{z}{1-z}x_D,\vec{k}_{\perp}) e^{-i(\frac{z}{1-z}x_D+x_S-x_E)p^+y_1^-} \right.  \\
& \left.  - q_N(x+x_L+x_E, \vec{v}_{\perp}) \phi_N(x_L+x_S-x_F,\vec{k}_{\perp})e^{i(x_L+x_E-x_F)p^+y_1^-}   \right]\\
  \end{aligned}
\end{equation}  

 \begin{equation}
  \begin{aligned}
\mathcal{W}_{L5} =& \frac{C_A}{2}  \frac{[\vec{l}_{\perp} - (1-z)\vec{v}_{\perp}]\cdot [\vec{l}_{\perp} - (1-z)\vec{v}_{\perp}-\vec{k}_{\perp}]}{[\vec{l}_{\perp} - (1-z)\vec{v}_{\perp}]^2 [\vec{l}_{\perp} - (1-z)\vec{v}_{\perp}-\vec{k}_{\perp}]^2 }\\
&\times  \left[ q_N(x+x_L+x_S+\frac{z}{1-z}x_D, \vec{v}_{\perp}) \phi_N(x_S-x_E,\vec{k}_{\perp}) e^{i(x_S+\frac{z}{1-z}x_D-x_E)p^+y_1^-} \right. \\
& \left.  - q_N(x+x_F, \vec{v}_{\perp}) \phi_N(x_S-x_E,\vec{k}_{\perp})e^{-i(x_L+x_E-x_F)p^+y_1^-}   \right]   \\
  \end{aligned}
\end{equation}

 \begin{equation}
  \begin{aligned}
\mathcal{W}_{R6} =&\frac{C_A}{2}  \frac{[\vec{l}_{\perp} - (1-z)\vec{v}_{\perp}]\cdot [\vec{l}_{\perp} - (1-z)\vec{v}_{\perp}-\vec{k}_{\perp}]}{[\vec{l}_{\perp} - (1-z)\vec{v}_{\perp}]^2 [\vec{l}_{\perp} - (1-z)\vec{v}_{\perp}-\vec{k}_{\perp}]^2 }\\
&\times \left[ q_N(x+x_L+x_E, \vec{v}_{\perp}) \phi_N(x_E-x_S,\vec{k}_{\perp}) e^{-i(\frac{z}{1-z}x_D+x_S-x_E)p^+y_1^-} \right.   \\
&\left.  - q_N(x+x_L+x_E, \vec{v}_{\perp}) \phi_N(x_L+x_S-x_F+\frac{z}{1-z}x_D,\vec{k}_{\perp})e^{i(x_L+x_E-x_F)p^+y_1^-}   \right]  \\
  \end{aligned}
\end{equation}  

 \begin{equation}
  \begin{aligned}
\mathcal{W}_{L6} =&\frac{C_A}{2}  \frac{[\vec{l}_{\perp} - (1-z)\vec{v}_{\perp}]\cdot [\vec{l}_{\perp} - (1-z)\vec{v}_{\perp}-\vec{k}_{\perp}]}{[\vec{l}_{\perp} - (1-z)\vec{v}_{\perp}]^2 [\vec{l}_{\perp} - (1-z)\vec{v}_{\perp}-\vec{k}_{\perp}]^2 }\\
&\times \left[ q_N(x+x_L+\frac{z}{1-z}x_D+x_S, \vec{v}_{\perp}) \phi_N(\frac{z}{1-z}x_D,\vec{k}_{\perp}) e^{i(\frac{z}{1-z}x_D+x_S-x_E)p^+y_1^-} \right.   \\
&\left.  - q_N(x+x_F, \vec{v}_{\perp}) \phi_N(\frac{z}{1-z}x_D,\vec{k}_{\perp})e^{-i(x_L+x_E-x_F)p^+y_1^-}   \right] \\
  \end{aligned}
\end{equation}  

\bef
 \centering
 \includegraphics[width=8cm]{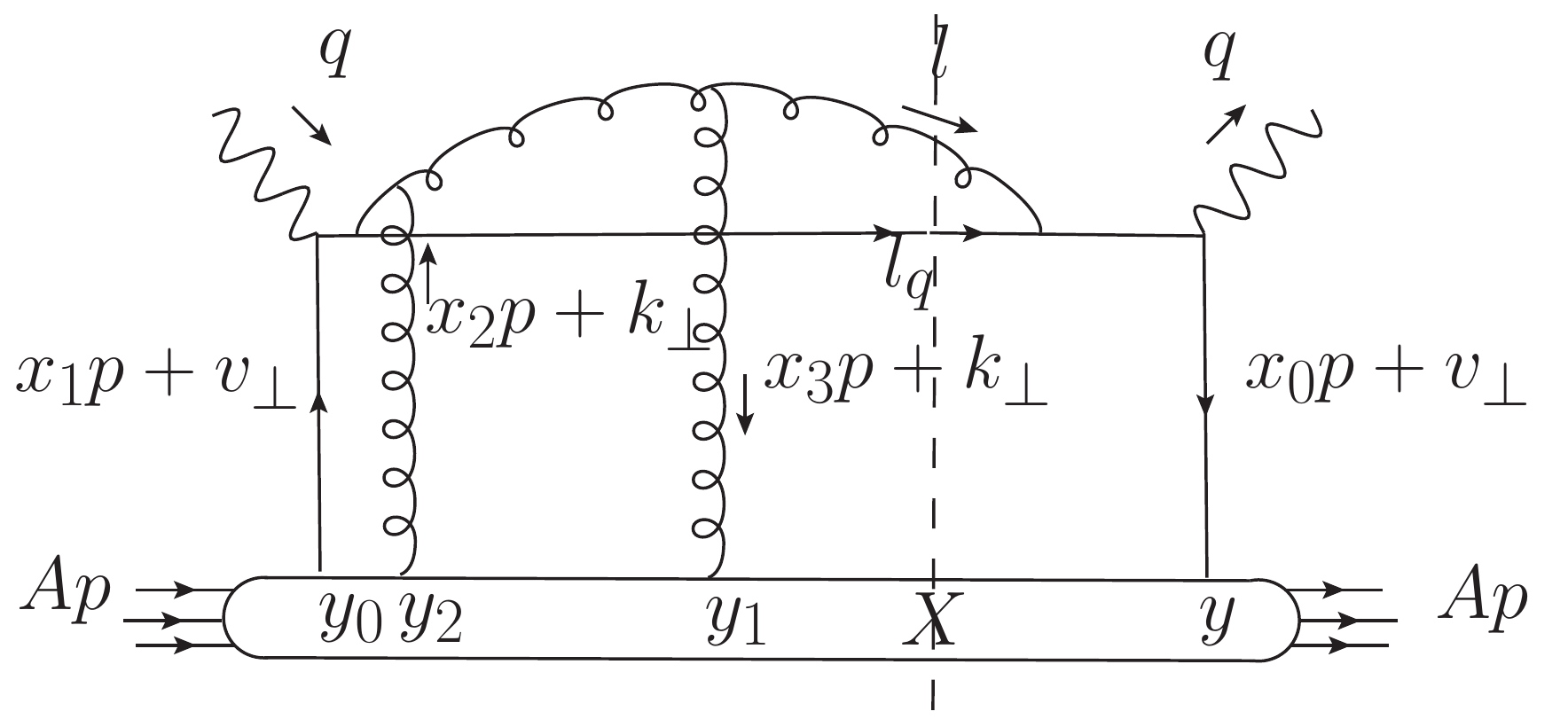} 
 \caption{Right Cut 7}
 \label{fig:Right_7}
\eef

 \begin{equation}
  \begin{aligned}
\mathcal{W}_{R7} =& C_A  \frac{1}{[\vec{l}_{\perp} - (1-z)\vec{v}_{\perp}]^2  } \left[ - q_N(x+x_L+x_E, \vec{v}_{\perp}) \phi_N(\frac{z}{1-z}x_D,\vec{k}_{\perp})   
\right. \\
&\left.+q_N(x+x_L+x_E, \vec{v}_{\perp}) \phi_N(x_L+x_E+\frac{z}{1-z}x_D-x_F,\vec{k}_{\perp})e^{i(x_L+x_E-x_F)p^+y_1^-}  \right] \\
  \end{aligned}
\end{equation}  

  \begin{equation}
  \begin{aligned}
\mathcal{W}_{L7} =&C_A  \frac{1}{[\vec{l}_{\perp} - (1-z)\vec{v}_{\perp}]^2  }\left[  -q_N(x+x_L+x_E, \vec{v}_{\perp}) \phi_N(\frac{z}{1-z}x_D,\vec{k}_{\perp})  \right.\\
&\left. + q_N(x+x_F, \vec{v}_{\perp}) \phi_N(\frac{z}{1-z}x_D,\vec{k}_{\perp})e^{-i(x_L+x_E-x_F)p^+y_1^-}  \right]  \\
  \end{aligned}
\end{equation}  
\end{widetext}

 \section{Effect of coherence of small-$x$ gluons}
 \label{extra-plots}
 For small-$x$ gluons, saturation will occur when the coherence length becomes larger than the nuclear size,
 \begin{equation}
    \frac{1}{x_Gp^+} > \frac{m_N}{p^+}2R_A 
 \end{equation}
 or $x_G<x_A$, $x_A=1/(2m_NR_A)$.. In this limit, the low bound of the saturated gluon distribution per nucleon should be $\phi^0_N(x_A,k_\perp,\mu^2)/A^{1/3}$ when the saturated gluon density of nucleons along the trajectory of the struck quark becomes equivalent to that of a single nucleon.  To estimate the effect of this small-$x$ gluon coherence, we can require $x_G>x_A$ in the calculation of the dijet cross section. These results are shown in Figs.~\ref{fig:RD_azAngle-distr_APD}(b) - \ref{fig:R_A-distr_APPD} (b) in comparison to that without the restriction $x_G>x_A$ in Figs.~\ref{fig:RD_azAngle-distr_APD}(a) - \ref{fig:R_A-distr_APPD}(a). We have chosen the kinematics $Q^2=25$ GeV$^2$, $ l_\perp=  1.41$ GeV/$c$, $x_B=0.2$ for these calculations.
 
 From the azimuthal angle dependence in Fig.~\ref{fig:RD_azAngle-distr_APD}, the rapidity gap in Fig.~\ref{fig:rapidity-distr_APPD} and nuclei size dependence in Fig.~\ref{fig:R_A-distr_APPD}, we see that the restriction $x_G>x_A$ reduces the dijet cross section from double parton scattering slightly. However, the rapidity gap dependence (Fig.~\ref{fig:rapidity-distr_APPD}) and nuclear size dependence (Fig.~\ref{fig:R_A-distr_APPD}) show the same pattern when such coherence is not taken into account. The results here only illustrate the upper bound on the effect of the coherence length of small-$x$ gluons since the saturated gluon distribution per nucleon should be non-zero, lying between the $\phi^0_N(x_A,k_\perp,\mu^2)$ and the lower bound $\phi^0_N(x_A,k_\perp,\mu^2)/A^{1/3}$.
 
 \begin{widetext}

\bef
\includegraphics[width=8cm]{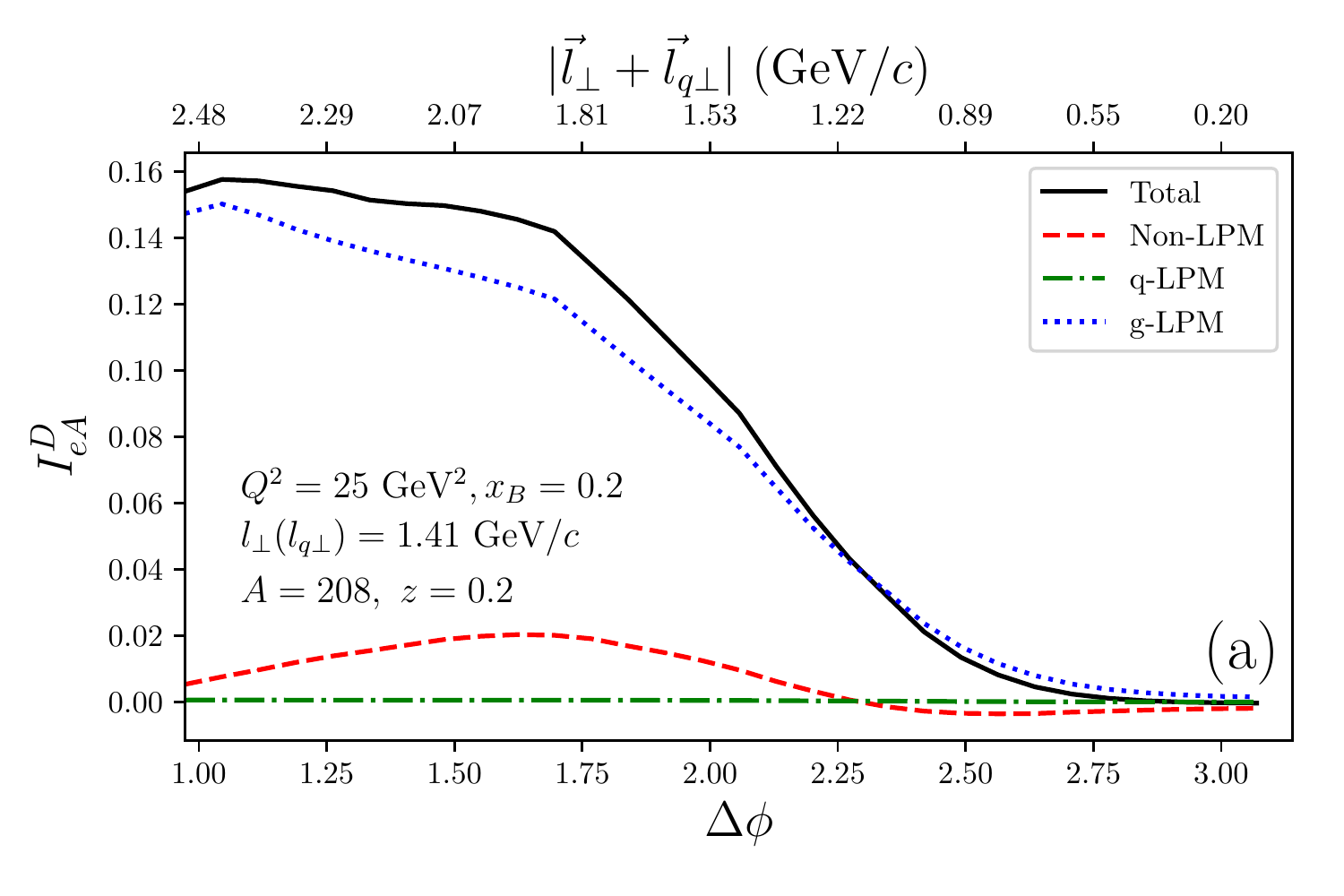}
\includegraphics[width=8cm]{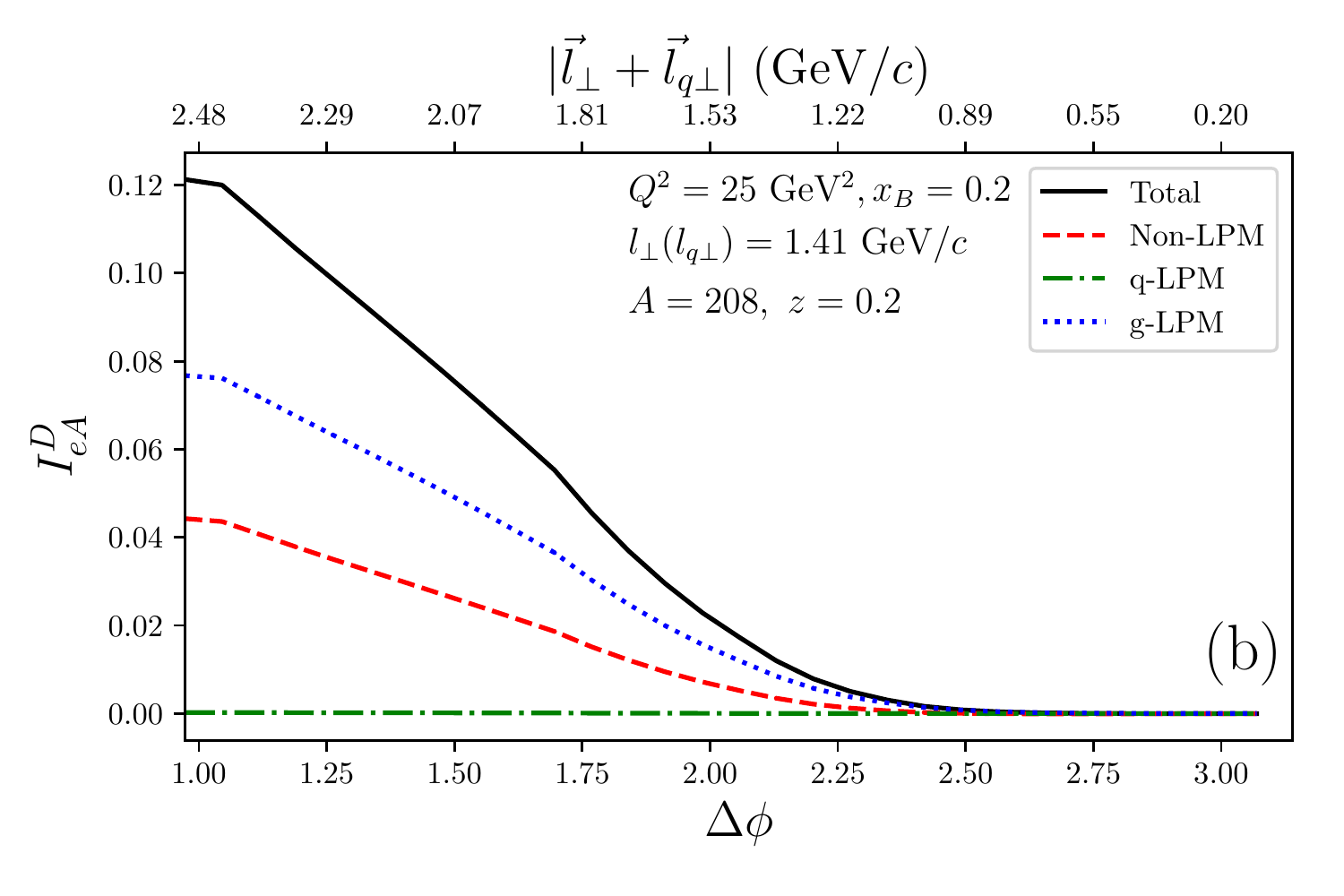}
\caption{The nuclear modification factor $I_{eA}^D(l_{\perp},l_{q\perp}=l_\perp,\Delta\phi,z)$ for the dijet cross section from double scattering as a function of the azimuthal angle $\Delta\phi$ or the transverse momentum imbalance $|\vec l_\perp+\vec l_{q\perp}|$ in $e$+Pb DIS with $x_B=0.2$, $ z= 0.2$, $ l_\perp$=1.41 GeV/$c$ and $Q^2$=25 GeV$^2$  (a) without (b) and with the constraint $x_G>x_A$ in the dijet cross section.}
\label{fig:RD_azAngle-distr_APD}
\eef
 
\bef
\includegraphics[width=8cm]{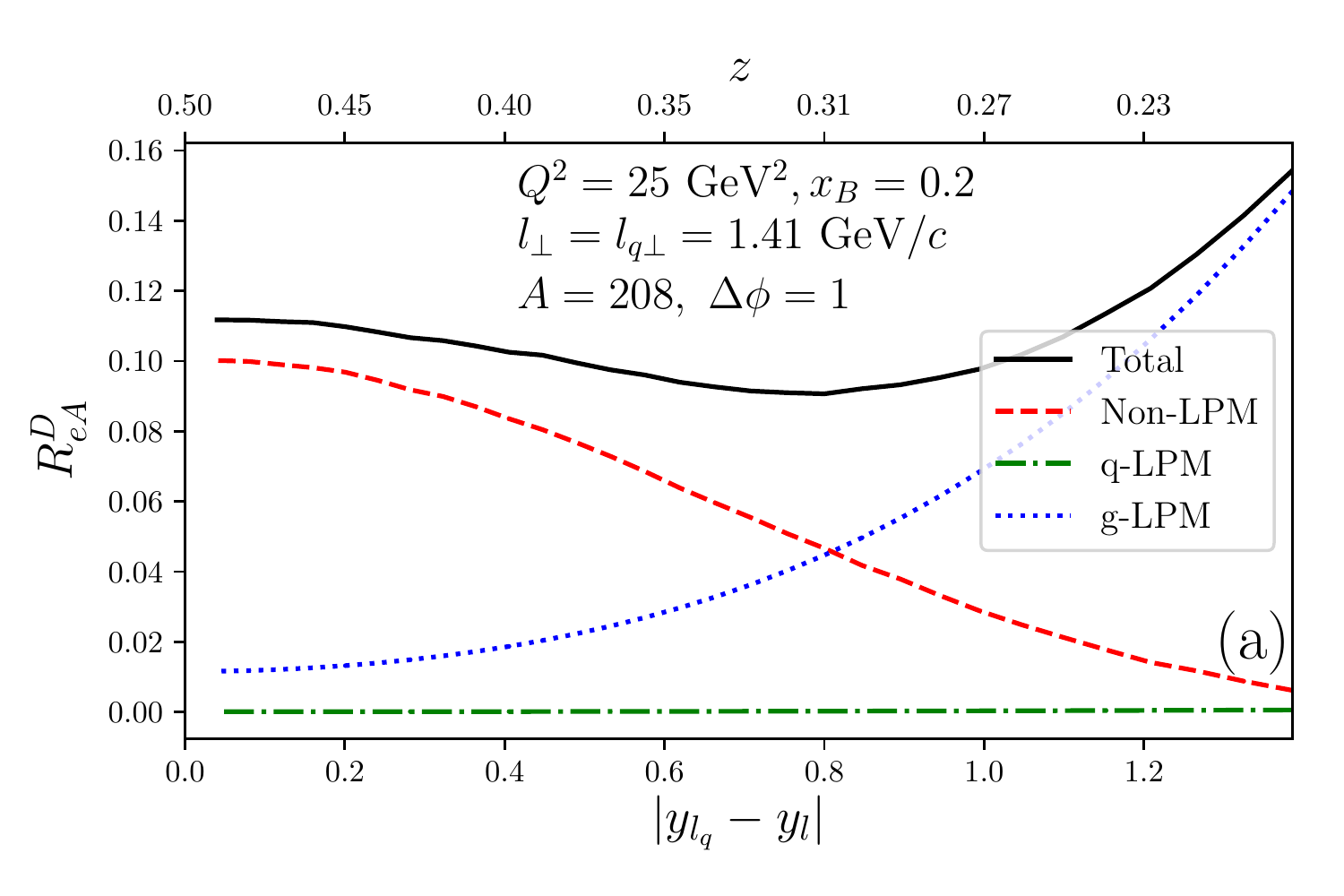}
\includegraphics[width=8cm]{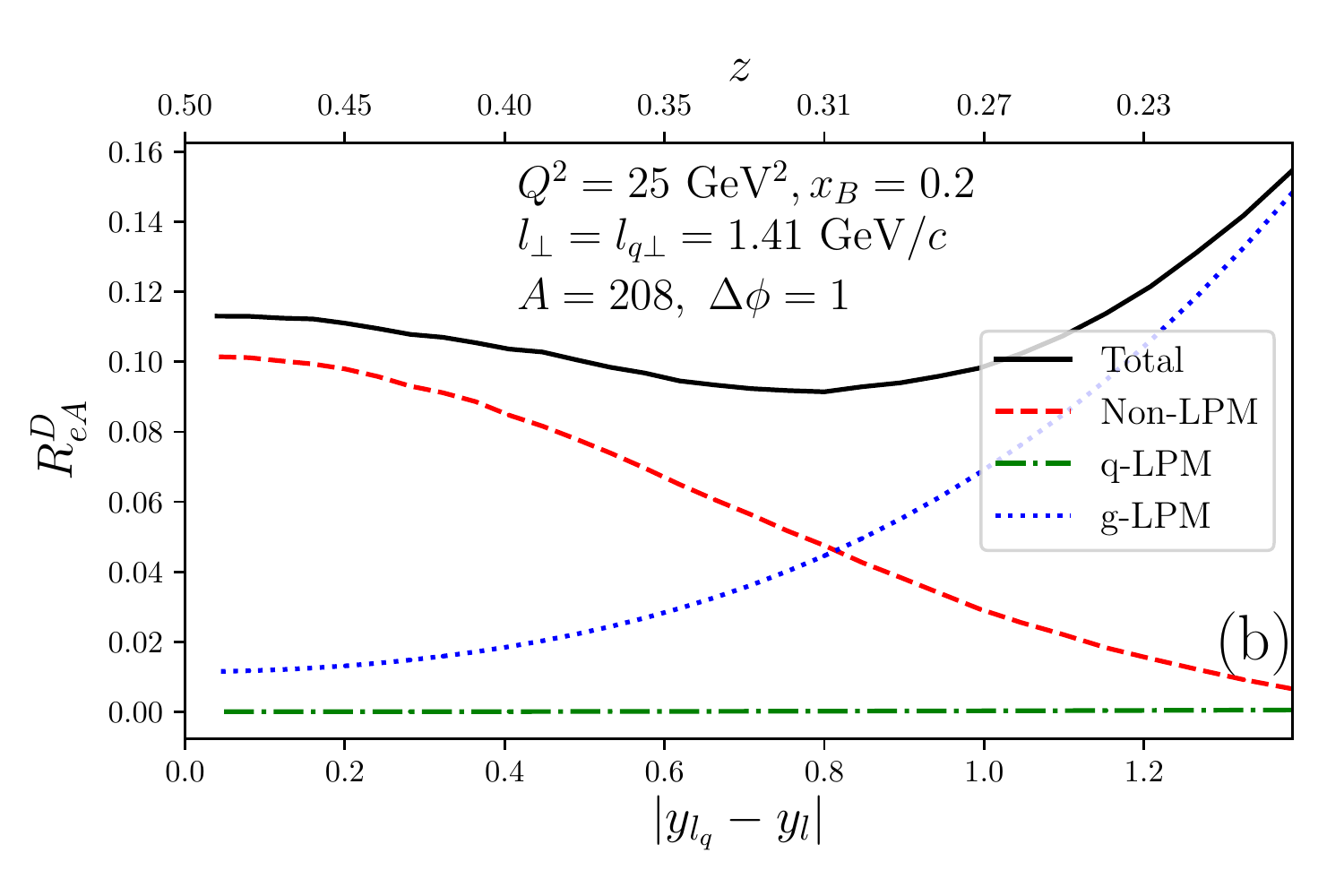}
\includegraphics[width=8cm]{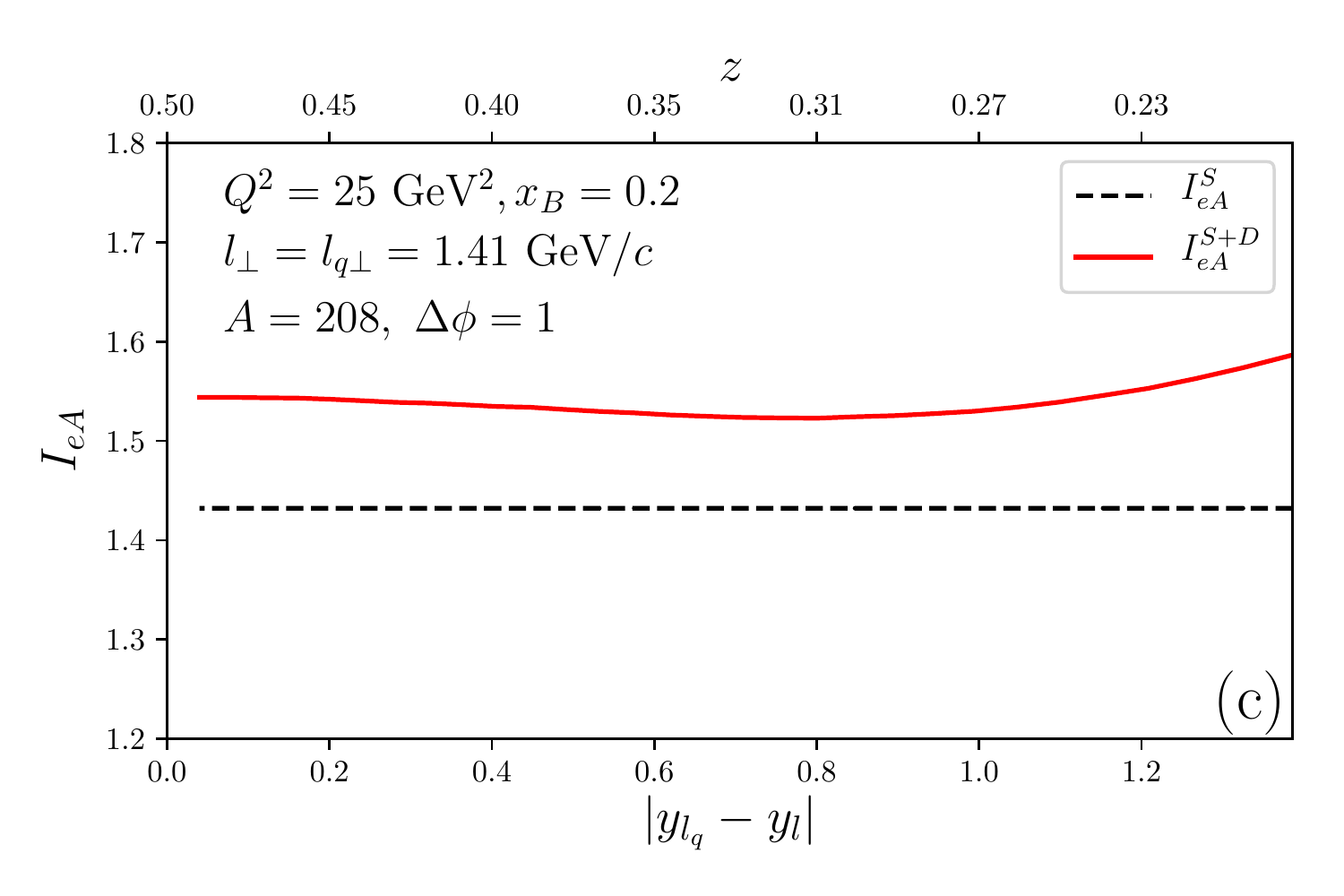}
\includegraphics[width=8cm]{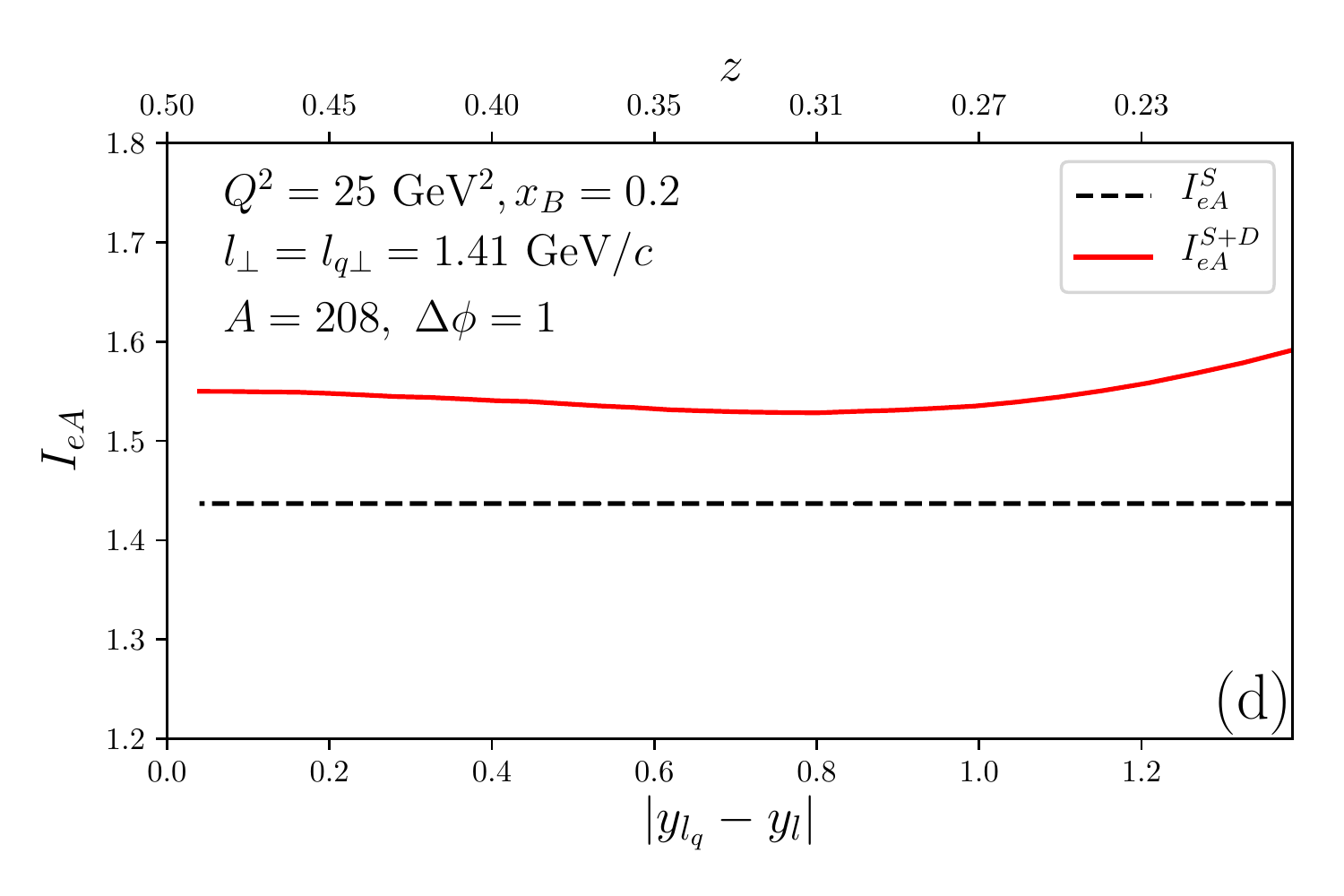}
\caption{The rapidity gap $|y_{l_q} - y_{l}|$ dependence of the nuclear modification factor from double scattering $I_{eA}^D(l_{\perp},l_{q\perp},\Delta\phi,z)$ (a) without (b) and with the constraint $x_G>x_A$ ,  the nuclear modification factor $I_{eA}^{S+D}(l_{\perp},l_{q\perp},\Delta\phi,z)$ with (solid) and $I_{eA}^{S}(l_{\perp},l_{q\perp},\Delta\phi,z)$ without  (dashed) double scattering  (c) without (d) and with constraint $x_G>x_A$ in $e$+Pb DIS.}
\label{fig:rapidity-distr_APPD}
\eef 

\bef
\centering
\includegraphics[width=0.45\textwidth]{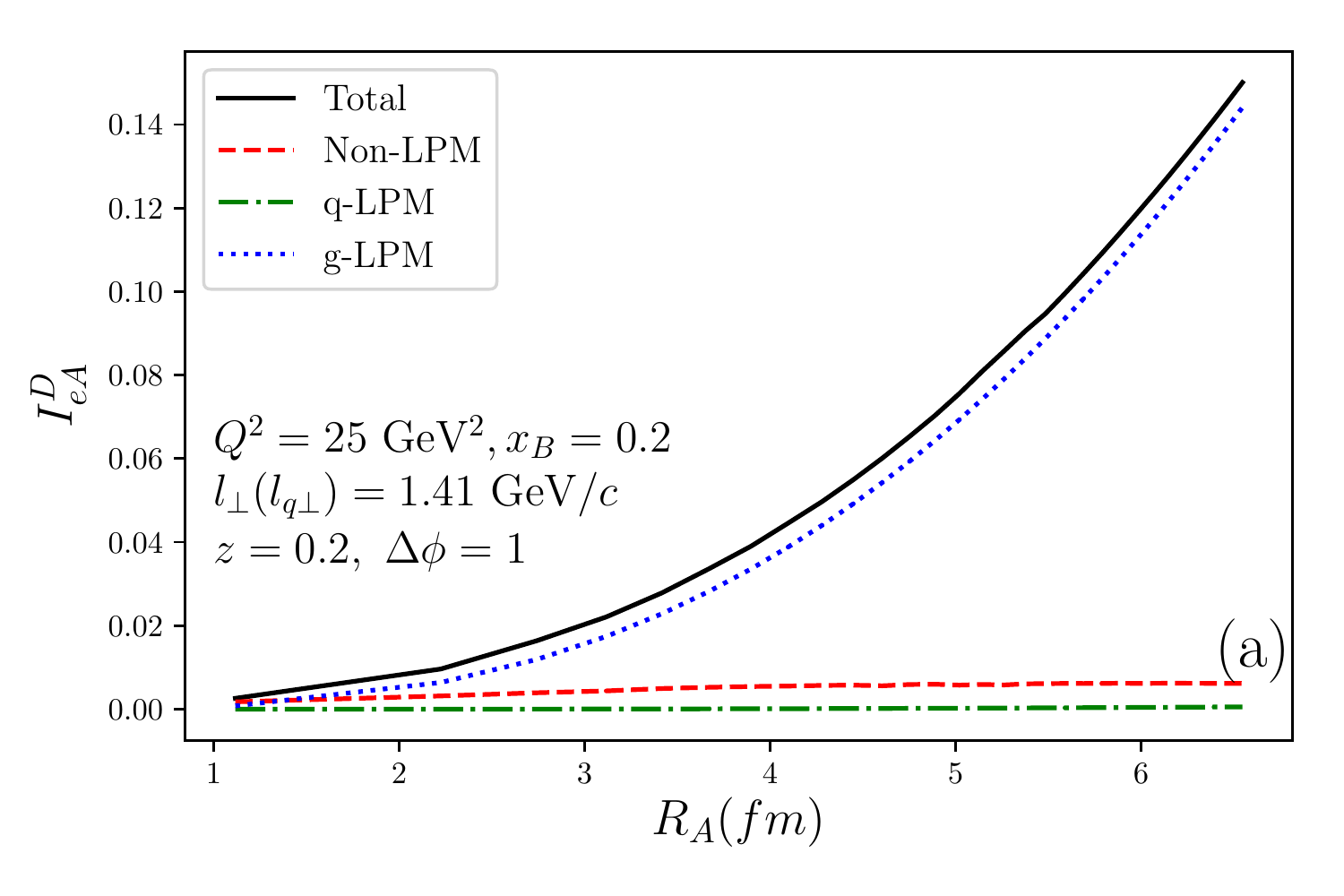}
\includegraphics[width=0.45\textwidth]{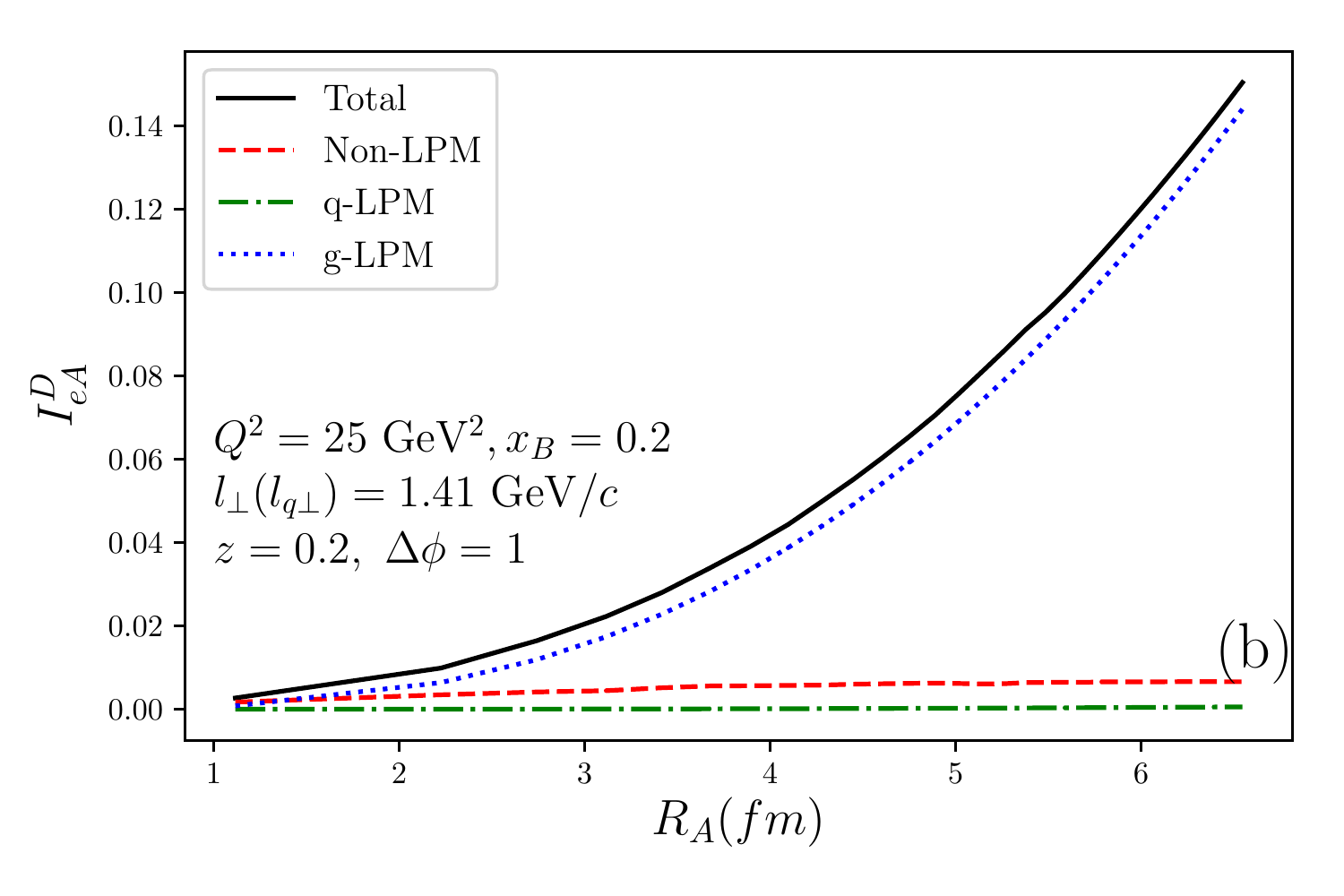}
\includegraphics[width=0.45\textwidth]{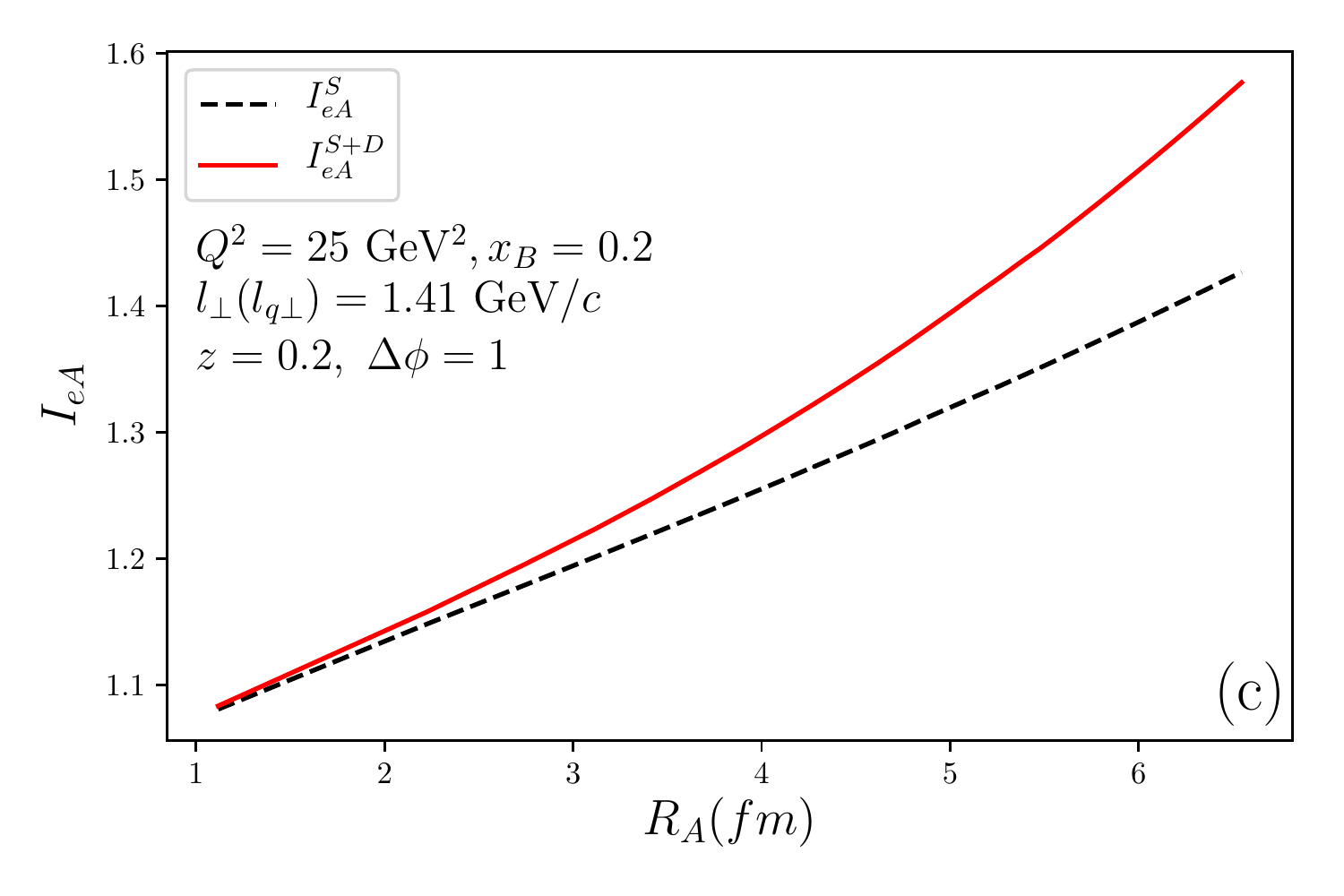}
\includegraphics[width=0.45\textwidth]{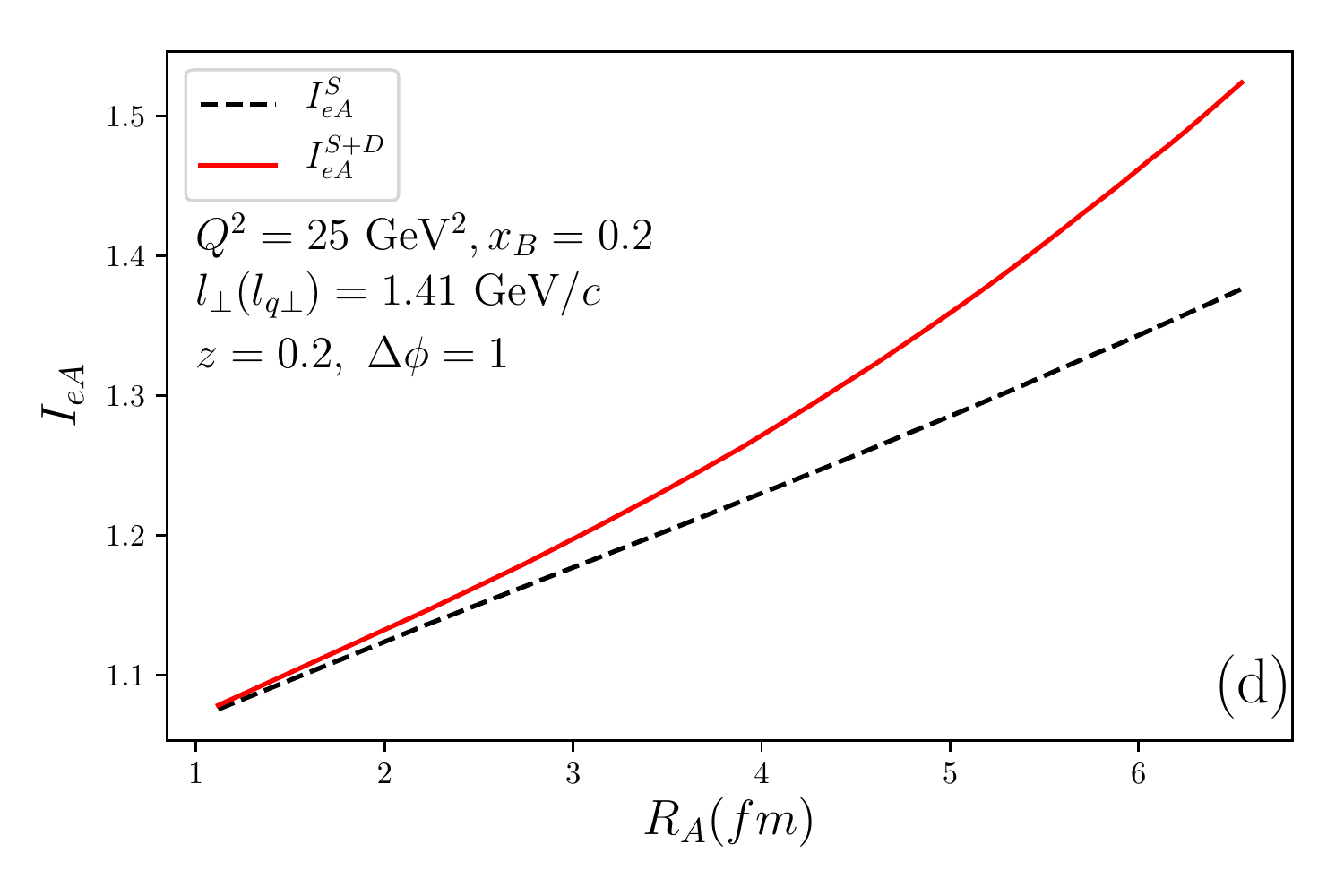}
\caption{The nuclear size $R_A$ dependence of the nuclear modification factor from double  scattering $ I_{eA}^D(l_{\perp},l_{q\perp},  \Delta \phi,z)$ (a) with (b) and without  (b) constraint $x_G>x_A$, the nuclear modification factor $I_{eA}^{S+D}(l_{\perp},l_{q\perp},  \Delta\phi,z)$ with (solid) and without (dashed) double scattering $I_{eA}^S(l_{\perp},l_{q\perp},\Delta \phi,z)$ without (c) and with (d) the constraint $x_G>x_A$  in $e$+Pb DIS.}
\label{fig:R_A-distr_APPD}
\eef

 \end{widetext}

\bibliographystyle{apsrev4-1}
\bibliography{Paper_Dijet}

\begin{thebibliography}{92}%
\makeatletter
\providecommand \@ifxundefined [1]{%
 \@ifx{#1\undefined}
}%
\providecommand \@ifnum [1]{%
 \ifnum #1\expandafter \@firstoftwo
 \else \expandafter \@secondoftwo
 \fi
}%
\providecommand \@ifx [1]{%
 \ifx #1\expandafter \@firstoftwo
 \else \expandafter \@secondoftwo
 \fi
}%
\providecommand \natexlab [1]{#1}%
\providecommand \enquote  [1]{``#1''}%
\providecommand \bibnamefont  [1]{#1}%
\providecommand \bibfnamefont [1]{#1}%
\providecommand \citenamefont [1]{#1}%
\providecommand \href@noop [0]{\@secondoftwo}%
\providecommand \href [0]{\begingroup \@sanitize@url \@href}%
\providecommand \@href[1]{\@@startlink{#1}\@@href}%
\providecommand \@@href[1]{\endgroup#1\@@endlink}%
\providecommand \@sanitize@url [0]{\catcode `\\12\catcode `\$12\catcode
  `\&12\catcode `\#12\catcode `\^12\catcode `\_12\catcode `\%12\relax}%
\providecommand \@@startlink[1]{}%
\providecommand \@@endlink[0]{}%
\providecommand \url  [0]{\begingroup\@sanitize@url \@url }%
\providecommand \@url [1]{\endgroup\@href {#1}{\urlprefix }}%
\providecommand \urlprefix  [0]{URL }%
\providecommand \Eprint [0]{\href }%
\providecommand \doibase [0]{http://dx.doi.org/}%
\providecommand \selectlanguage [0]{\@gobble}%
\providecommand \bibinfo  [0]{\@secondoftwo}%
\providecommand \bibfield  [0]{\@secondoftwo}%
\providecommand \translation [1]{[#1]}%
\providecommand \BibitemOpen [0]{}%
\providecommand \bibitemStop [0]{}%
\providecommand \bibitemNoStop [0]{.\EOS\space}%
\providecommand \EOS [0]{\spacefactor3000\relax}%
\providecommand \BibitemShut  [1]{\csname bibitem#1\endcsname}%
\let\auto@bib@innerbib\@empty
\bibitem [{\citenamefont {Gyulassy}\ \emph {et~al.}(2004)\citenamefont
  {Gyulassy}, \citenamefont {Vitev}, \citenamefont {Wang},\ and\ \citenamefont
  {Zhang}}]{Gyulassy:2003mc}%
  \BibitemOpen
  \bibfield  {author} {\bibinfo {author} {\bibfnamefont {M.}~\bibnamefont
  {Gyulassy}}, \bibinfo {author} {\bibfnamefont {I.}~\bibnamefont {Vitev}},
  \bibinfo {author} {\bibfnamefont {X.-N.}\ \bibnamefont {Wang}}, \ and\
  \bibinfo {author} {\bibfnamefont {B.-W.}\ \bibnamefont {Zhang}},\ }\href
  {\doibase 10.1142/9789812795533_0003} {\ ,\ \bibinfo {pages} {123} (\bibinfo
  {year} {2004})},\ \Eprint {http://arxiv.org/abs/nucl-th/0302077}
  {arXiv:nucl-th/0302077} \BibitemShut {NoStop}%
\bibitem [{\citenamefont {Wang}(2005)}]{Wang:2004dn}%
  \BibitemOpen
  \bibfield  {author} {\bibinfo {author} {\bibfnamefont {X.-N.}\ \bibnamefont
  {Wang}},\ }\href {\doibase 10.1016/j.nuclphysa.2004.12.037} {\bibfield
  {journal} {\bibinfo  {journal} {Nucl. Phys. A}\ }\textbf {\bibinfo {volume}
  {750}},\ \bibinfo {pages} {98} (\bibinfo {year} {2005})},\ \Eprint
  {http://arxiv.org/abs/nucl-th/0405017} {arXiv:nucl-th/0405017} \BibitemShut
  {NoStop}%
\bibitem [{\citenamefont {Majumder}\ and\ \citenamefont
  {Van~Leeuwen}(2011)}]{Majumder:2010qh}%
  \BibitemOpen
  \bibfield  {author} {\bibinfo {author} {\bibfnamefont {A.}~\bibnamefont
  {Majumder}}\ and\ \bibinfo {author} {\bibfnamefont {M.}~\bibnamefont
  {Van~Leeuwen}},\ }\href {\doibase 10.1016/j.ppnp.2010.09.001} {\bibfield
  {journal} {\bibinfo  {journal} {Prog. Part. Nucl. Phys.}\ }\textbf {\bibinfo
  {volume} {66}},\ \bibinfo {pages} {41} (\bibinfo {year} {2011})},\ \Eprint
  {http://arxiv.org/abs/1002.2206} {arXiv:1002.2206 [hep-ph]} \BibitemShut
  {NoStop}%
\bibitem [{\citenamefont {Muller}\ \emph {et~al.}(2012)\citenamefont {Muller},
  \citenamefont {Schukraft},\ and\ \citenamefont {Wyslouch}}]{Muller:2012zq}%
  \BibitemOpen
  \bibfield  {author} {\bibinfo {author} {\bibfnamefont {B.}~\bibnamefont
  {Muller}}, \bibinfo {author} {\bibfnamefont {J.}~\bibnamefont {Schukraft}}, \
  and\ \bibinfo {author} {\bibfnamefont {B.}~\bibnamefont {Wyslouch}},\ }\href
  {\doibase 10.1146/annurev-nucl-102711-094910} {\bibfield  {journal} {\bibinfo
   {journal} {Ann. Rev. Nucl. Part. Sci.}\ }\textbf {\bibinfo {volume} {62}},\
  \bibinfo {pages} {361} (\bibinfo {year} {2012})},\ \Eprint
  {http://arxiv.org/abs/1202.3233} {arXiv:1202.3233 [hep-ex]} \BibitemShut
  {NoStop}%
\bibitem [{\citenamefont {Mehtar-Tani}\ \emph {et~al.}(2013)\citenamefont
  {Mehtar-Tani}, \citenamefont {Milhano},\ and\ \citenamefont
  {Tywoniuk}}]{Mehtar-Tani:2013pia}%
  \BibitemOpen
  \bibfield  {author} {\bibinfo {author} {\bibfnamefont {Y.}~\bibnamefont
  {Mehtar-Tani}}, \bibinfo {author} {\bibfnamefont {J.~G.}\ \bibnamefont
  {Milhano}}, \ and\ \bibinfo {author} {\bibfnamefont {K.}~\bibnamefont
  {Tywoniuk}},\ }\href {\doibase 10.1142/S0217751X13400137} {\bibfield
  {journal} {\bibinfo  {journal} {Int. J. Mod. Phys. A}\ }\textbf {\bibinfo
  {volume} {28}},\ \bibinfo {pages} {1340013} (\bibinfo {year} {2013})},\
  \Eprint {http://arxiv.org/abs/1302.2579} {arXiv:1302.2579 [hep-ph]}
  \BibitemShut {NoStop}%
\bibitem [{\citenamefont {Qin}\ and\ \citenamefont {Wang}(2015)}]{Qin:2015srf}%
  \BibitemOpen
  \bibfield  {author} {\bibinfo {author} {\bibfnamefont {G.-Y.}\ \bibnamefont
  {Qin}}\ and\ \bibinfo {author} {\bibfnamefont {X.-N.}\ \bibnamefont {Wang}},\
  }\href {\doibase 10.1142/S0218301315300143} {\bibfield  {journal} {\bibinfo
  {journal} {Int. J. Mod. Phys. E}\ }\textbf {\bibinfo {volume} {24}},\
  \bibinfo {pages} {1530014} (\bibinfo {year} {2015})},\ \Eprint
  {http://arxiv.org/abs/1511.00790} {arXiv:1511.00790 [hep-ph]} \BibitemShut
  {NoStop}%
\bibitem [{\citenamefont {Cao}\ and\ \citenamefont {Wang}(2021)}]{Cao:2020wlm}%
  \BibitemOpen
  \bibfield  {author} {\bibinfo {author} {\bibfnamefont {S.}~\bibnamefont
  {Cao}}\ and\ \bibinfo {author} {\bibfnamefont {X.-N.}\ \bibnamefont {Wang}},\
  }\href {\doibase 10.1088/1361-6633/abc22b} {\bibfield  {journal} {\bibinfo
  {journal} {Rept. Prog. Phys.}\ }\textbf {\bibinfo {volume} {84}},\ \bibinfo
  {pages} {024301} (\bibinfo {year} {2021})},\ \Eprint
  {http://arxiv.org/abs/2002.04028} {arXiv:2002.04028 [hep-ph]} \BibitemShut
  {NoStop}%
\bibitem [{\citenamefont {Baier}\ \emph
  {et~al.}(1997{\natexlab{a}})\citenamefont {Baier}, \citenamefont
  {Dokshitzer}, \citenamefont {Mueller}, \citenamefont {Peigne},\ and\
  \citenamefont {Schiff}}]{Baier:1996kr}%
  \BibitemOpen
  \bibfield  {author} {\bibinfo {author} {\bibfnamefont {R.}~\bibnamefont
  {Baier}}, \bibinfo {author} {\bibfnamefont {Y.~L.}\ \bibnamefont
  {Dokshitzer}}, \bibinfo {author} {\bibfnamefont {A.~H.}\ \bibnamefont
  {Mueller}}, \bibinfo {author} {\bibfnamefont {S.}~\bibnamefont {Peigne}}, \
  and\ \bibinfo {author} {\bibfnamefont {D.}~\bibnamefont {Schiff}},\ }\href
  {\doibase 10.1016/S0550-3213(96)00553-6} {\bibfield  {journal} {\bibinfo
  {journal} {Nucl. Phys. B}\ }\textbf {\bibinfo {volume} {483}},\ \bibinfo
  {pages} {291} (\bibinfo {year} {1997}{\natexlab{a}})},\ \Eprint
  {http://arxiv.org/abs/hep-ph/9607355} {arXiv:hep-ph/9607355} \BibitemShut
  {NoStop}%
\bibitem [{\citenamefont {Baier}\ \emph
  {et~al.}(1997{\natexlab{b}})\citenamefont {Baier}, \citenamefont
  {Dokshitzer}, \citenamefont {Mueller}, \citenamefont {Peigne},\ and\
  \citenamefont {Schiff}}]{Baier:1996sk}%
  \BibitemOpen
  \bibfield  {author} {\bibinfo {author} {\bibfnamefont {R.}~\bibnamefont
  {Baier}}, \bibinfo {author} {\bibfnamefont {Y.~L.}\ \bibnamefont
  {Dokshitzer}}, \bibinfo {author} {\bibfnamefont {A.~H.}\ \bibnamefont
  {Mueller}}, \bibinfo {author} {\bibfnamefont {S.}~\bibnamefont {Peigne}}, \
  and\ \bibinfo {author} {\bibfnamefont {D.}~\bibnamefont {Schiff}},\ }\href
  {\doibase 10.1016/S0550-3213(96)00581-0} {\bibfield  {journal} {\bibinfo
  {journal} {Nucl. Phys. B}\ }\textbf {\bibinfo {volume} {484}},\ \bibinfo
  {pages} {265} (\bibinfo {year} {1997}{\natexlab{b}})},\ \Eprint
  {http://arxiv.org/abs/hep-ph/9608322} {arXiv:hep-ph/9608322} \BibitemShut
  {NoStop}%
\bibitem [{\citenamefont {Casalderrey-Solana}\ and\ \citenamefont
  {Wang}(2008)}]{CasalderreySolana:2007sw}%
  \BibitemOpen
  \bibfield  {author} {\bibinfo {author} {\bibfnamefont {J.}~\bibnamefont
  {Casalderrey-Solana}}\ and\ \bibinfo {author} {\bibfnamefont {X.-N.}\
  \bibnamefont {Wang}},\ }\href {\doibase 10.1103/PhysRevC.77.024902}
  {\bibfield  {journal} {\bibinfo  {journal} {Phys. Rev. C}\ }\textbf {\bibinfo
  {volume} {77}},\ \bibinfo {pages} {024902} (\bibinfo {year} {2008})},\
  \Eprint {http://arxiv.org/abs/0705.1352} {arXiv:0705.1352 [hep-ph]}
  \BibitemShut {NoStop}%
\bibitem [{\citenamefont {Liang}\ \emph {et~al.}(2008)\citenamefont {Liang},
  \citenamefont {Wang},\ and\ \citenamefont {Zhou}}]{Liang:2008vz}%
  \BibitemOpen
  \bibfield  {author} {\bibinfo {author} {\bibfnamefont {Z.-T.}\ \bibnamefont
  {Liang}}, \bibinfo {author} {\bibfnamefont {X.-N.}\ \bibnamefont {Wang}}, \
  and\ \bibinfo {author} {\bibfnamefont {J.}~\bibnamefont {Zhou}},\ }\href
  {\doibase 10.1103/PhysRevD.77.125010} {\bibfield  {journal} {\bibinfo
  {journal} {Phys. Rev. D}\ }\textbf {\bibinfo {volume} {77}},\ \bibinfo
  {pages} {125010} (\bibinfo {year} {2008})},\ \Eprint
  {http://arxiv.org/abs/0801.0434} {arXiv:0801.0434 [hep-ph]} \BibitemShut
  {NoStop}%
\bibitem [{\citenamefont {Chen}\ \emph {et~al.}(2010)\citenamefont {Chen},
  \citenamefont {Greiner}, \citenamefont {Wang}, \citenamefont {Wang},\ and\
  \citenamefont {Xu}}]{Chen:2010te}%
  \BibitemOpen
  \bibfield  {author} {\bibinfo {author} {\bibfnamefont {X.-F.}\ \bibnamefont
  {Chen}}, \bibinfo {author} {\bibfnamefont {C.}~\bibnamefont {Greiner}},
  \bibinfo {author} {\bibfnamefont {E.}~\bibnamefont {Wang}}, \bibinfo {author}
  {\bibfnamefont {X.-N.}\ \bibnamefont {Wang}}, \ and\ \bibinfo {author}
  {\bibfnamefont {Z.}~\bibnamefont {Xu}},\ }\href {\doibase
  10.1103/PhysRevC.81.064908} {\bibfield  {journal} {\bibinfo  {journal} {Phys.
  Rev. C}\ }\textbf {\bibinfo {volume} {81}},\ \bibinfo {pages} {064908}
  (\bibinfo {year} {2010})},\ \Eprint {http://arxiv.org/abs/1002.1165}
  {arXiv:1002.1165 [nucl-th]} \BibitemShut {NoStop}%
\bibitem [{\citenamefont {Burke}\ \emph {et~al.}(2014)\citenamefont {Burke}
  \emph {et~al.}}]{Burke:2013yra}%
  \BibitemOpen
  \bibfield  {author} {\bibinfo {author} {\bibfnamefont {K.~M.}\ \bibnamefont
  {Burke}} \emph {et~al.} (\bibinfo {collaboration} {JET}),\ }\href {\doibase
  10.1103/PhysRevC.90.014909} {\bibfield  {journal} {\bibinfo  {journal} {Phys.
  Rev. C}\ }\textbf {\bibinfo {volume} {90}},\ \bibinfo {pages} {014909}
  (\bibinfo {year} {2014})},\ \Eprint {http://arxiv.org/abs/1312.5003}
  {arXiv:1312.5003 [nucl-th]} \BibitemShut {NoStop}%
\bibitem [{\citenamefont {Xie}\ \emph {et~al.}(2019)\citenamefont {Xie},
  \citenamefont {Wei}, \citenamefont {Qin},\ and\ \citenamefont
  {Zhang}}]{Xie:2019oxg}%
  \BibitemOpen
  \bibfield  {author} {\bibinfo {author} {\bibfnamefont {M.}~\bibnamefont
  {Xie}}, \bibinfo {author} {\bibfnamefont {S.-Y.}\ \bibnamefont {Wei}},
  \bibinfo {author} {\bibfnamefont {G.-Y.}\ \bibnamefont {Qin}}, \ and\
  \bibinfo {author} {\bibfnamefont {H.-Z.}\ \bibnamefont {Zhang}},\ }\href
  {\doibase 10.1140/epjc/s10052-019-7100-1} {\bibfield  {journal} {\bibinfo
  {journal} {Eur. Phys. J. C}\ }\textbf {\bibinfo {volume} {79}},\ \bibinfo
  {pages} {589} (\bibinfo {year} {2019})},\ \Eprint
  {http://arxiv.org/abs/1901.04155} {arXiv:1901.04155 [hep-ph]} \BibitemShut
  {NoStop}%
\bibitem [{\citenamefont {Feal}\ \emph {et~al.}(2021)\citenamefont {Feal},
  \citenamefont {Salgado},\ and\ \citenamefont {Vazquez}}]{Feal:2019xfl}%
  \BibitemOpen
  \bibfield  {author} {\bibinfo {author} {\bibfnamefont {X.}~\bibnamefont
  {Feal}}, \bibinfo {author} {\bibfnamefont {C.~A.}\ \bibnamefont {Salgado}}, \
  and\ \bibinfo {author} {\bibfnamefont {R.~A.}\ \bibnamefont {Vazquez}},\
  }\href {\doibase 10.1016/j.physletb.2021.136251} {\bibfield  {journal}
  {\bibinfo  {journal} {Phys. Lett. B}\ }\textbf {\bibinfo {volume} {816}},\
  \bibinfo {pages} {136251} (\bibinfo {year} {2021})},\ \Eprint
  {http://arxiv.org/abs/1911.01309} {arXiv:1911.01309 [hep-ph]} \BibitemShut
  {NoStop}%
\bibitem [{\citenamefont {Kumar}\ \emph {et~al.}(2017)\citenamefont {Kumar},
  \citenamefont {Bianchi}, \citenamefont {Elledge}, \citenamefont {Majumder},
  \citenamefont {Qin},\ and\ \citenamefont {Shen}}]{Kumar:2017des}%
  \BibitemOpen
  \bibfield  {author} {\bibinfo {author} {\bibfnamefont {A.}~\bibnamefont
  {Kumar}}, \bibinfo {author} {\bibfnamefont {E.}~\bibnamefont {Bianchi}},
  \bibinfo {author} {\bibfnamefont {J.}~\bibnamefont {Elledge}}, \bibinfo
  {author} {\bibfnamefont {A.}~\bibnamefont {Majumder}}, \bibinfo {author}
  {\bibfnamefont {G.-Y.}\ \bibnamefont {Qin}}, \ and\ \bibinfo {author}
  {\bibfnamefont {C.}~\bibnamefont {Shen}},\ }\href {\doibase
  10.1016/j.nuclphysa.2017.05.015} {\bibfield  {journal} {\bibinfo  {journal}
  {Nucl. Phys. A}\ }\textbf {\bibinfo {volume} {967}},\ \bibinfo {pages} {536}
  (\bibinfo {year} {2017})},\ \Eprint {http://arxiv.org/abs/1706.07547}
  {arXiv:1706.07547 [nucl-th]} \BibitemShut {NoStop}%
\bibitem [{\citenamefont {Kumar}\ \emph {et~al.}(2020)\citenamefont {Kumar},
  \citenamefont {Majumder},\ and\ \citenamefont {Shen}}]{Kumar:2019uvu}%
  \BibitemOpen
  \bibfield  {author} {\bibinfo {author} {\bibfnamefont {A.}~\bibnamefont
  {Kumar}}, \bibinfo {author} {\bibfnamefont {A.}~\bibnamefont {Majumder}}, \
  and\ \bibinfo {author} {\bibfnamefont {C.}~\bibnamefont {Shen}},\ }\href
  {\doibase 10.1103/PhysRevC.101.034908} {\bibfield  {journal} {\bibinfo
  {journal} {Phys. Rev. C}\ }\textbf {\bibinfo {volume} {101}},\ \bibinfo
  {pages} {034908} (\bibinfo {year} {2020})},\ \Eprint
  {http://arxiv.org/abs/1909.03178} {arXiv:1909.03178 [nucl-th]} \BibitemShut
  {NoStop}%
\bibitem [{\citenamefont {Soltz}(2019)}]{Soltz:2019aea}%
  \BibitemOpen
  \bibfield  {author} {\bibinfo {author} {\bibfnamefont {R.}~\bibnamefont
  {Soltz}} (\bibinfo {collaboration} {Jetscape}),\ }\href {\doibase
  10.22323/1.345.0048} {\bibfield  {journal} {\bibinfo  {journal} {PoS}\
  }\textbf {\bibinfo {volume} {HardProbes2018}},\ \bibinfo {pages} {048}
  (\bibinfo {year} {2019})}\BibitemShut {NoStop}%
\bibitem [{\citenamefont {Cao}\ \emph {et~al.}(2021)\citenamefont {Cao} \emph
  {et~al.}}]{JETSCAPE:2021ehl}%
  \BibitemOpen
  \bibfield  {author} {\bibinfo {author} {\bibfnamefont {S.}~\bibnamefont
  {Cao}} \emph {et~al.} (\bibinfo {collaboration} {JETSCAPE}),\ }\href
  {\doibase 10.1103/PhysRevC.104.024905} {\bibfield  {journal} {\bibinfo
  {journal} {Phys. Rev. C}\ }\textbf {\bibinfo {volume} {104}},\ \bibinfo
  {pages} {024905} (\bibinfo {year} {2021})},\ \Eprint
  {http://arxiv.org/abs/2102.11337} {arXiv:2102.11337 [nucl-th]} \BibitemShut
  {NoStop}%
\bibitem [{\citenamefont {Guo}\ and\ \citenamefont {Wang}(2000)}]{Guo:2000nz}%
  \BibitemOpen
  \bibfield  {author} {\bibinfo {author} {\bibfnamefont {X.-F.}\ \bibnamefont
  {Guo}}\ and\ \bibinfo {author} {\bibfnamefont {X.-N.}\ \bibnamefont {Wang}},\
  }\href {\doibase 10.1103/PhysRevLett.85.3591} {\bibfield  {journal} {\bibinfo
   {journal} {Phys. Rev. Lett.}\ }\textbf {\bibinfo {volume} {85}},\ \bibinfo
  {pages} {3591} (\bibinfo {year} {2000})},\ \Eprint
  {http://arxiv.org/abs/hep-ph/0005044} {arXiv:hep-ph/0005044} \BibitemShut
  {NoStop}%
\bibitem [{\citenamefont {Wang}\ and\ \citenamefont
  {Guo}(2001)}]{Wang:2001ifa}%
  \BibitemOpen
  \bibfield  {author} {\bibinfo {author} {\bibfnamefont {X.-N.}\ \bibnamefont
  {Wang}}\ and\ \bibinfo {author} {\bibfnamefont {X.-f.}\ \bibnamefont {Guo}},\
  }\href {\doibase 10.1016/S0375-9474(01)01130-7} {\bibfield  {journal}
  {\bibinfo  {journal} {Nucl. Phys. A}\ }\textbf {\bibinfo {volume} {696}},\
  \bibinfo {pages} {788} (\bibinfo {year} {2001})},\ \Eprint
  {http://arxiv.org/abs/hep-ph/0102230} {arXiv:hep-ph/0102230} \BibitemShut
  {NoStop}%
\bibitem [{\citenamefont {Wang}\ and\ \citenamefont
  {Wang}(2002)}]{Wang:2002ri}%
  \BibitemOpen
  \bibfield  {author} {\bibinfo {author} {\bibfnamefont {E.}~\bibnamefont
  {Wang}}\ and\ \bibinfo {author} {\bibfnamefont {X.-N.}\ \bibnamefont
  {Wang}},\ }\href {\doibase 10.1103/PhysRevLett.89.162301} {\bibfield
  {journal} {\bibinfo  {journal} {Phys. Rev. Lett.}\ }\textbf {\bibinfo
  {volume} {89}},\ \bibinfo {pages} {162301} (\bibinfo {year} {2002})},\
  \Eprint {http://arxiv.org/abs/hep-ph/0202105} {arXiv:hep-ph/0202105}
  \BibitemShut {NoStop}%
\bibitem [{\citenamefont {Zhang}\ \emph {et~al.}(2004)\citenamefont {Zhang},
  \citenamefont {Wang},\ and\ \citenamefont {Wang}}]{Zhang:2003wk}%
  \BibitemOpen
  \bibfield  {author} {\bibinfo {author} {\bibfnamefont {B.-W.}\ \bibnamefont
  {Zhang}}, \bibinfo {author} {\bibfnamefont {E.}~\bibnamefont {Wang}}, \ and\
  \bibinfo {author} {\bibfnamefont {X.-N.}\ \bibnamefont {Wang}},\ }\href
  {\doibase 10.1103/PhysRevLett.93.072301} {\bibfield  {journal} {\bibinfo
  {journal} {Phys. Rev. Lett.}\ }\textbf {\bibinfo {volume} {93}},\ \bibinfo
  {pages} {072301} (\bibinfo {year} {2004})},\ \Eprint
  {http://arxiv.org/abs/nucl-th/0309040} {arXiv:nucl-th/0309040} \BibitemShut
  {NoStop}%
\bibitem [{\citenamefont {Zhang}\ and\ \citenamefont
  {Wang}(2003)}]{Zhang:2003yn}%
  \BibitemOpen
  \bibfield  {author} {\bibinfo {author} {\bibfnamefont {B.-W.}\ \bibnamefont
  {Zhang}}\ and\ \bibinfo {author} {\bibfnamefont {X.-N.}\ \bibnamefont
  {Wang}},\ }\href {\doibase 10.1016/S0375-9474(03)01003-0} {\bibfield
  {journal} {\bibinfo  {journal} {Nucl. Phys. A}\ }\textbf {\bibinfo {volume}
  {720}},\ \bibinfo {pages} {429} (\bibinfo {year} {2003})},\ \Eprint
  {http://arxiv.org/abs/hep-ph/0301195} {arXiv:hep-ph/0301195} \BibitemShut
  {NoStop}%
\bibitem [{\citenamefont {Arleo}(2003)}]{Arleo:2003jz}%
  \BibitemOpen
  \bibfield  {author} {\bibinfo {author} {\bibfnamefont {F.}~\bibnamefont
  {Arleo}},\ }\href {\doibase 10.1140/epjc/s2003-01289-x} {\bibfield  {journal}
  {\bibinfo  {journal} {Eur. Phys. J. C}\ }\textbf {\bibinfo {volume} {30}},\
  \bibinfo {pages} {213} (\bibinfo {year} {2003})},\ \Eprint
  {http://arxiv.org/abs/hep-ph/0306235} {arXiv:hep-ph/0306235} \BibitemShut
  {NoStop}%
\bibitem [{\citenamefont {Majumder}(2012)}]{Majumder:2009ge}%
  \BibitemOpen
  \bibfield  {author} {\bibinfo {author} {\bibfnamefont {A.}~\bibnamefont
  {Majumder}},\ }\href {\doibase 10.1103/PhysRevD.85.014023} {\bibfield
  {journal} {\bibinfo  {journal} {Phys. Rev. D}\ }\textbf {\bibinfo {volume}
  {85}},\ \bibinfo {pages} {014023} (\bibinfo {year} {2012})},\ \Eprint
  {http://arxiv.org/abs/0912.2987} {arXiv:0912.2987 [nucl-th]} \BibitemShut
  {NoStop}%
\bibitem [{\citenamefont {Chang}\ \emph {et~al.}(2014)\citenamefont {Chang},
  \citenamefont {Deng},\ and\ \citenamefont {Wang}}]{Chang:2014fba}%
  \BibitemOpen
  \bibfield  {author} {\bibinfo {author} {\bibfnamefont {N.-B.}\ \bibnamefont
  {Chang}}, \bibinfo {author} {\bibfnamefont {W.-T.}\ \bibnamefont {Deng}}, \
  and\ \bibinfo {author} {\bibfnamefont {X.-N.}\ \bibnamefont {Wang}},\ }\href
  {\doibase 10.1103/PhysRevC.89.034911} {\bibfield  {journal} {\bibinfo
  {journal} {Phys. Rev. C}\ }\textbf {\bibinfo {volume} {89}},\ \bibinfo
  {pages} {034911} (\bibinfo {year} {2014})},\ \Eprint
  {http://arxiv.org/abs/1401.5109} {arXiv:1401.5109 [nucl-th]} \BibitemShut
  {NoStop}%
\bibitem [{\citenamefont {Li}\ \emph {et~al.}(2021)\citenamefont {Li},
  \citenamefont {Liu},\ and\ \citenamefont {Vitev}}]{Li:2020zbk}%
  \BibitemOpen
  \bibfield  {author} {\bibinfo {author} {\bibfnamefont {H.~T.}\ \bibnamefont
  {Li}}, \bibinfo {author} {\bibfnamefont {Z.~L.}\ \bibnamefont {Liu}}, \ and\
  \bibinfo {author} {\bibfnamefont {I.}~\bibnamefont {Vitev}},\ }\href
  {\doibase 10.1016/j.physletb.2021.136261} {\bibfield  {journal} {\bibinfo
  {journal} {Phys. Lett. B}\ }\textbf {\bibinfo {volume} {816}},\ \bibinfo
  {pages} {136261} (\bibinfo {year} {2021})},\ \Eprint
  {http://arxiv.org/abs/2007.10994} {arXiv:2007.10994 [hep-ph]} \BibitemShut
  {NoStop}%
\bibitem [{\citenamefont {Ru}\ \emph {et~al.}(2021)\citenamefont {Ru},
  \citenamefont {Kang}, \citenamefont {Wang}, \citenamefont {Xing},\ and\
  \citenamefont {Zhang}}]{Ru:2019qvz}%
  \BibitemOpen
  \bibfield  {author} {\bibinfo {author} {\bibfnamefont {P.}~\bibnamefont
  {Ru}}, \bibinfo {author} {\bibfnamefont {Z.-B.}\ \bibnamefont {Kang}},
  \bibinfo {author} {\bibfnamefont {E.}~\bibnamefont {Wang}}, \bibinfo {author}
  {\bibfnamefont {H.}~\bibnamefont {Xing}}, \ and\ \bibinfo {author}
  {\bibfnamefont {B.-W.}\ \bibnamefont {Zhang}},\ }\href {\doibase
  10.1103/PhysRevD.103.L031901} {\bibfield  {journal} {\bibinfo  {journal}
  {Phys. Rev. D}\ }\textbf {\bibinfo {volume} {103}},\ \bibinfo {pages}
  {L031901} (\bibinfo {year} {2021})},\ \Eprint
  {http://arxiv.org/abs/1907.11808} {arXiv:1907.11808 [hep-ph]} \BibitemShut
  {NoStop}%
\bibitem [{\citenamefont {Zakharov}(1996)}]{Zakharov:1996fv}%
  \BibitemOpen
  \bibfield  {author} {\bibinfo {author} {\bibfnamefont {B.~G.}\ \bibnamefont
  {Zakharov}},\ }\href {\doibase 10.1134/1.567126} {\bibfield  {journal}
  {\bibinfo  {journal} {JETP Lett.}\ }\textbf {\bibinfo {volume} {63}},\
  \bibinfo {pages} {952} (\bibinfo {year} {1996})},\ \Eprint
  {http://arxiv.org/abs/hep-ph/9607440} {arXiv:hep-ph/9607440} \BibitemShut
  {NoStop}%
\bibitem [{\citenamefont {Arnold}\ \emph {et~al.}(2000)\citenamefont {Arnold},
  \citenamefont {Moore},\ and\ \citenamefont {Yaffe}}]{Arnold:2000dr}%
  \BibitemOpen
  \bibfield  {author} {\bibinfo {author} {\bibfnamefont {P.~B.}\ \bibnamefont
  {Arnold}}, \bibinfo {author} {\bibfnamefont {G.~D.}\ \bibnamefont {Moore}}, \
  and\ \bibinfo {author} {\bibfnamefont {L.~G.}\ \bibnamefont {Yaffe}},\ }\href
  {\doibase 10.1088/1126-6708/2000/11/001} {\bibfield  {journal} {\bibinfo
  {journal} {JHEP}\ }\textbf {\bibinfo {volume} {11}},\ \bibinfo {pages} {001}
  (\bibinfo {year} {2000})},\ \Eprint {http://arxiv.org/abs/hep-ph/0010177}
  {arXiv:hep-ph/0010177} \BibitemShut {NoStop}%
\bibitem [{\citenamefont {Arnold}\ \emph {et~al.}(2003)\citenamefont {Arnold},
  \citenamefont {Moore},\ and\ \citenamefont {Yaffe}}]{Arnold:2003zc}%
  \BibitemOpen
  \bibfield  {author} {\bibinfo {author} {\bibfnamefont {P.~B.}\ \bibnamefont
  {Arnold}}, \bibinfo {author} {\bibfnamefont {G.~D.}\ \bibnamefont {Moore}}, \
  and\ \bibinfo {author} {\bibfnamefont {L.~G.}\ \bibnamefont {Yaffe}},\ }\href
  {\doibase 10.1088/1126-6708/2003/05/051} {\bibfield  {journal} {\bibinfo
  {journal} {JHEP}\ }\textbf {\bibinfo {volume} {05}},\ \bibinfo {pages} {051}
  (\bibinfo {year} {2003})},\ \Eprint {http://arxiv.org/abs/hep-ph/0302165}
  {arXiv:hep-ph/0302165} \BibitemShut {NoStop}%
\bibitem [{\citenamefont {Gyulassy}\ \emph
  {et~al.}(2000{\natexlab{a}})\citenamefont {Gyulassy}, \citenamefont {Levai},\
  and\ \citenamefont {Vitev}}]{Gyulassy:1999zd}%
  \BibitemOpen
  \bibfield  {author} {\bibinfo {author} {\bibfnamefont {M.}~\bibnamefont
  {Gyulassy}}, \bibinfo {author} {\bibfnamefont {P.}~\bibnamefont {Levai}}, \
  and\ \bibinfo {author} {\bibfnamefont {I.}~\bibnamefont {Vitev}},\ }\href
  {\doibase 10.1016/S0550-3213(99)00713-0} {\bibfield  {journal} {\bibinfo
  {journal} {Nucl. Phys.}\ }\textbf {\bibinfo {volume} {B571}},\ \bibinfo
  {pages} {197} (\bibinfo {year} {2000}{\natexlab{a}})},\ \Eprint
  {http://arxiv.org/abs/hep-ph/9907461} {arXiv:hep-ph/9907461 [hep-ph]}
  \BibitemShut {NoStop}%
\bibitem [{\citenamefont {Gyulassy}\ \emph {et~al.}(2001)\citenamefont
  {Gyulassy}, \citenamefont {Levai},\ and\ \citenamefont
  {Vitev}}]{Gyulassy:2000er}%
  \BibitemOpen
  \bibfield  {author} {\bibinfo {author} {\bibfnamefont {M.}~\bibnamefont
  {Gyulassy}}, \bibinfo {author} {\bibfnamefont {P.}~\bibnamefont {Levai}}, \
  and\ \bibinfo {author} {\bibfnamefont {I.}~\bibnamefont {Vitev}},\ }\href
  {\doibase 10.1016/S0550-3213(00)00652-0} {\bibfield  {journal} {\bibinfo
  {journal} {Nucl. Phys. B}\ }\textbf {\bibinfo {volume} {594}},\ \bibinfo
  {pages} {371} (\bibinfo {year} {2001})},\ \Eprint
  {http://arxiv.org/abs/nucl-th/0006010} {arXiv:nucl-th/0006010} \BibitemShut
  {NoStop}%
\bibitem [{\citenamefont {Gyulassy}\ \emph
  {et~al.}(2000{\natexlab{b}})\citenamefont {Gyulassy}, \citenamefont {Levai},\
  and\ \citenamefont {Vitev}}]{Gyulassy:2000fs}%
  \BibitemOpen
  \bibfield  {author} {\bibinfo {author} {\bibfnamefont {M.}~\bibnamefont
  {Gyulassy}}, \bibinfo {author} {\bibfnamefont {P.}~\bibnamefont {Levai}}, \
  and\ \bibinfo {author} {\bibfnamefont {I.}~\bibnamefont {Vitev}},\ }\href
  {\doibase 10.1103/PhysRevLett.85.5535} {\bibfield  {journal} {\bibinfo
  {journal} {Phys. Rev. Lett.}\ }\textbf {\bibinfo {volume} {85}},\ \bibinfo
  {pages} {5535} (\bibinfo {year} {2000}{\natexlab{b}})},\ \Eprint
  {http://arxiv.org/abs/nucl-th/0005032} {arXiv:nucl-th/0005032} \BibitemShut
  {NoStop}%
\bibitem [{\citenamefont {Wiedemann}(2000)}]{Wiedemann:2000za}%
  \BibitemOpen
  \bibfield  {author} {\bibinfo {author} {\bibfnamefont {U.~A.}\ \bibnamefont
  {Wiedemann}},\ }\href {\doibase 10.1016/S0550-3213(00)00457-0} {\bibfield
  {journal} {\bibinfo  {journal} {Nucl. Phys. B}\ }\textbf {\bibinfo {volume}
  {588}},\ \bibinfo {pages} {303} (\bibinfo {year} {2000})},\ \Eprint
  {http://arxiv.org/abs/hep-ph/0005129} {arXiv:hep-ph/0005129} \BibitemShut
  {NoStop}%
\bibitem [{\citenamefont {Qin}\ and\ \citenamefont
  {Majumder}(2013)}]{Qin:2012fua}%
  \BibitemOpen
  \bibfield  {author} {\bibinfo {author} {\bibfnamefont {G.-Y.}\ \bibnamefont
  {Qin}}\ and\ \bibinfo {author} {\bibfnamefont {A.}~\bibnamefont {Majumder}},\
  }\href {\doibase 10.1103/PhysRevC.87.024909} {\bibfield  {journal} {\bibinfo
  {journal} {Phys. Rev. C}\ }\textbf {\bibinfo {volume} {87}},\ \bibinfo
  {pages} {024909} (\bibinfo {year} {2013})},\ \Eprint
  {http://arxiv.org/abs/1205.5741} {arXiv:1205.5741 [hep-ph]} \BibitemShut
  {NoStop}%
\bibitem [{\citenamefont {Ovanesyan}\ and\ \citenamefont
  {Vitev}(2011)}]{Ovanesyan:2011xy}%
  \BibitemOpen
  \bibfield  {author} {\bibinfo {author} {\bibfnamefont {G.}~\bibnamefont
  {Ovanesyan}}\ and\ \bibinfo {author} {\bibfnamefont {I.}~\bibnamefont
  {Vitev}},\ }\href {\doibase 10.1007/JHEP06(2011)080} {\bibfield  {journal}
  {\bibinfo  {journal} {JHEP}\ }\textbf {\bibinfo {volume} {06}},\ \bibinfo
  {pages} {080} (\bibinfo {year} {2011})},\ \Eprint
  {http://arxiv.org/abs/1103.1074} {arXiv:1103.1074 [hep-ph]} \BibitemShut
  {NoStop}%
\bibitem [{\citenamefont {Ovanesyan}\ and\ \citenamefont
  {Vitev}(2012)}]{Ovanesyan:2011kn}%
  \BibitemOpen
  \bibfield  {author} {\bibinfo {author} {\bibfnamefont {G.}~\bibnamefont
  {Ovanesyan}}\ and\ \bibinfo {author} {\bibfnamefont {I.}~\bibnamefont
  {Vitev}},\ }\href {\doibase 10.1016/j.physletb.2011.11.040} {\bibfield
  {journal} {\bibinfo  {journal} {Phys. Lett. B}\ }\textbf {\bibinfo {volume}
  {706}},\ \bibinfo {pages} {371} (\bibinfo {year} {2012})},\ \Eprint
  {http://arxiv.org/abs/1109.5619} {arXiv:1109.5619 [hep-ph]} \BibitemShut
  {NoStop}%
\bibitem [{\citenamefont {Kang}\ \emph {et~al.}(2015)\citenamefont {Kang},
  \citenamefont {Lashof-Regas}, \citenamefont {Ovanesyan}, \citenamefont
  {Saad},\ and\ \citenamefont {Vitev}}]{Kang:2014xsa}%
  \BibitemOpen
  \bibfield  {author} {\bibinfo {author} {\bibfnamefont {Z.-B.}\ \bibnamefont
  {Kang}}, \bibinfo {author} {\bibfnamefont {R.}~\bibnamefont {Lashof-Regas}},
  \bibinfo {author} {\bibfnamefont {G.}~\bibnamefont {Ovanesyan}}, \bibinfo
  {author} {\bibfnamefont {P.}~\bibnamefont {Saad}}, \ and\ \bibinfo {author}
  {\bibfnamefont {I.}~\bibnamefont {Vitev}},\ }\href {\doibase
  10.1103/PhysRevLett.114.092002} {\bibfield  {journal} {\bibinfo  {journal}
  {Phys. Rev. Lett.}\ }\textbf {\bibinfo {volume} {114}},\ \bibinfo {pages}
  {092002} (\bibinfo {year} {2015})},\ \Eprint {http://arxiv.org/abs/1405.2612}
  {arXiv:1405.2612 [hep-ph]} \BibitemShut {NoStop}%
\bibitem [{\citenamefont {Kang}\ \emph {et~al.}(2014)\citenamefont {Kang},
  \citenamefont {Wang}, \citenamefont {Wang},\ and\ \citenamefont
  {Xing}}]{Kang:2013raa}%
  \BibitemOpen
  \bibfield  {author} {\bibinfo {author} {\bibfnamefont {Z.-B.}\ \bibnamefont
  {Kang}}, \bibinfo {author} {\bibfnamefont {E.}~\bibnamefont {Wang}}, \bibinfo
  {author} {\bibfnamefont {X.-N.}\ \bibnamefont {Wang}}, \ and\ \bibinfo
  {author} {\bibfnamefont {H.}~\bibnamefont {Xing}},\ }\href {\doibase
  10.1103/PhysRevLett.112.102001} {\bibfield  {journal} {\bibinfo  {journal}
  {Phys. Rev. Lett.}\ }\textbf {\bibinfo {volume} {112}},\ \bibinfo {pages}
  {102001} (\bibinfo {year} {2014})},\ \Eprint {http://arxiv.org/abs/1310.6759}
  {arXiv:1310.6759 [hep-ph]} \BibitemShut {NoStop}%
\bibitem [{\citenamefont {Kang}\ \emph
  {et~al.}(2016{\natexlab{a}})\citenamefont {Kang}, \citenamefont {Wang},
  \citenamefont {Wang},\ and\ \citenamefont {Xing}}]{Kang:2014ela}%
  \BibitemOpen
  \bibfield  {author} {\bibinfo {author} {\bibfnamefont {Z.-B.}\ \bibnamefont
  {Kang}}, \bibinfo {author} {\bibfnamefont {E.}~\bibnamefont {Wang}}, \bibinfo
  {author} {\bibfnamefont {X.-N.}\ \bibnamefont {Wang}}, \ and\ \bibinfo
  {author} {\bibfnamefont {H.}~\bibnamefont {Xing}},\ }\href {\doibase
  10.1103/PhysRevD.94.114024} {\bibfield  {journal} {\bibinfo  {journal} {Phys.
  Rev. D}\ }\textbf {\bibinfo {volume} {94}},\ \bibinfo {pages} {114024}
  (\bibinfo {year} {2016}{\natexlab{a}})},\ \Eprint
  {http://arxiv.org/abs/1409.1315} {arXiv:1409.1315 [hep-ph]} \BibitemShut
  {NoStop}%
\bibitem [{\citenamefont {Kang}\ \emph
  {et~al.}(2016{\natexlab{b}})\citenamefont {Kang}, \citenamefont {Qiu},
  \citenamefont {Wang},\ and\ \citenamefont {Xing}}]{Kang:2016ron}%
  \BibitemOpen
  \bibfield  {author} {\bibinfo {author} {\bibfnamefont {Z.-B.}\ \bibnamefont
  {Kang}}, \bibinfo {author} {\bibfnamefont {J.-W.}\ \bibnamefont {Qiu}},
  \bibinfo {author} {\bibfnamefont {X.-N.}\ \bibnamefont {Wang}}, \ and\
  \bibinfo {author} {\bibfnamefont {H.}~\bibnamefont {Xing}},\ }\href {\doibase
  10.1103/PhysRevD.94.074038} {\bibfield  {journal} {\bibinfo  {journal} {Phys.
  Rev. D}\ }\textbf {\bibinfo {volume} {94}},\ \bibinfo {pages} {074038}
  (\bibinfo {year} {2016}{\natexlab{b}})},\ \Eprint
  {http://arxiv.org/abs/1605.07175} {arXiv:1605.07175 [hep-ph]} \BibitemShut
  {NoStop}%
\bibitem [{\citenamefont {Zhang}\ \emph {et~al.}(2018)\citenamefont {Zhang},
  \citenamefont {Hou},\ and\ \citenamefont {Qin}}]{Zhang:2018kkn}%
  \BibitemOpen
  \bibfield  {author} {\bibinfo {author} {\bibfnamefont {L.}~\bibnamefont
  {Zhang}}, \bibinfo {author} {\bibfnamefont {D.-F.}\ \bibnamefont {Hou}}, \
  and\ \bibinfo {author} {\bibfnamefont {G.-Y.}\ \bibnamefont {Qin}},\ }\href
  {\doibase 10.1103/PhysRevC.98.034913} {\bibfield  {journal} {\bibinfo
  {journal} {Phys. Rev. C}\ }\textbf {\bibinfo {volume} {98}},\ \bibinfo
  {pages} {034913} (\bibinfo {year} {2018})},\ \Eprint
  {http://arxiv.org/abs/1804.00470} {arXiv:1804.00470 [nucl-th]} \BibitemShut
  {NoStop}%
\bibitem [{\citenamefont {Zhang}\ \emph
  {et~al.}(2019{\natexlab{a}})\citenamefont {Zhang}, \citenamefont {Hou},\ and\
  \citenamefont {Qin}}]{Zhang:2018nie}%
  \BibitemOpen
  \bibfield  {author} {\bibinfo {author} {\bibfnamefont {L.}~\bibnamefont
  {Zhang}}, \bibinfo {author} {\bibfnamefont {D.-F.}\ \bibnamefont {Hou}}, \
  and\ \bibinfo {author} {\bibfnamefont {G.-Y.}\ \bibnamefont {Qin}},\ }\href
  {\doibase 10.1103/PhysRevC.100.034907} {\bibfield  {journal} {\bibinfo
  {journal} {Phys. Rev. C}\ }\textbf {\bibinfo {volume} {100}},\ \bibinfo
  {pages} {034907} (\bibinfo {year} {2019}{\natexlab{a}})},\ \Eprint
  {http://arxiv.org/abs/1812.11048} {arXiv:1812.11048 [hep-ph]} \BibitemShut
  {NoStop}%
\bibitem [{\citenamefont {Zhang}\ \emph
  {et~al.}(2019{\natexlab{b}})\citenamefont {Zhang}, \citenamefont {Qin},\ and\
  \citenamefont {Wang}}]{Zhang:2019toi}%
  \BibitemOpen
  \bibfield  {author} {\bibinfo {author} {\bibfnamefont {Y.-Y.}\ \bibnamefont
  {Zhang}}, \bibinfo {author} {\bibfnamefont {G.-Y.}\ \bibnamefont {Qin}}, \
  and\ \bibinfo {author} {\bibfnamefont {X.-N.}\ \bibnamefont {Wang}},\ }\href
  {\doibase 10.1103/PhysRevD.100.074031} {\bibfield  {journal} {\bibinfo
  {journal} {Phys. Rev. D}\ }\textbf {\bibinfo {volume} {100}},\ \bibinfo
  {pages} {074031} (\bibinfo {year} {2019}{\natexlab{b}})},\ \Eprint
  {http://arxiv.org/abs/1905.12699} {arXiv:1905.12699 [hep-ph]} \BibitemShut
  {NoStop}%
\bibitem [{\citenamefont {Klasen}\ and\ \citenamefont
  {Kova\v{r}\'\i{}k}(2018)}]{Klasen:2018gtb}%
  \BibitemOpen
  \bibfield  {author} {\bibinfo {author} {\bibfnamefont {M.}~\bibnamefont
  {Klasen}}\ and\ \bibinfo {author} {\bibfnamefont {K.}~\bibnamefont
  {Kova\v{r}\'\i{}k}},\ }\href {\doibase 10.1103/PhysRevD.97.114013} {\bibfield
   {journal} {\bibinfo  {journal} {Phys. Rev. D}\ }\textbf {\bibinfo {volume}
  {97}},\ \bibinfo {pages} {114013} (\bibinfo {year} {2018})},\ \Eprint
  {http://arxiv.org/abs/1803.10985} {arXiv:1803.10985 [hep-ph]} \BibitemShut
  {NoStop}%
\bibitem [{\citenamefont {Guzey}\ and\ \citenamefont
  {Klasen}(2020)}]{Guzey:2020zza}%
  \BibitemOpen
  \bibfield  {author} {\bibinfo {author} {\bibfnamefont {V.}~\bibnamefont
  {Guzey}}\ and\ \bibinfo {author} {\bibfnamefont {M.}~\bibnamefont {Klasen}},\
  }\href {\doibase 10.1103/PhysRevC.102.065201} {\bibfield  {journal} {\bibinfo
   {journal} {Phys. Rev. C}\ }\textbf {\bibinfo {volume} {102}},\ \bibinfo
  {pages} {065201} (\bibinfo {year} {2020})},\ \Eprint
  {http://arxiv.org/abs/2003.09129} {arXiv:2003.09129 [hep-ph]} \BibitemShut
  {NoStop}%
\bibitem [{\citenamefont {Marquet}(2007)}]{Marquet:2007vb}%
  \BibitemOpen
  \bibfield  {author} {\bibinfo {author} {\bibfnamefont {C.}~\bibnamefont
  {Marquet}},\ }\href {\doibase 10.1016/j.nuclphysa.2007.09.001} {\bibfield
  {journal} {\bibinfo  {journal} {Nucl. Phys. A}\ }\textbf {\bibinfo {volume}
  {796}},\ \bibinfo {pages} {41} (\bibinfo {year} {2007})},\ \Eprint
  {http://arxiv.org/abs/0708.0231} {arXiv:0708.0231 [hep-ph]} \BibitemShut
  {NoStop}%
\bibitem [{\citenamefont {Dominguez}\ \emph {et~al.}(2011)\citenamefont
  {Dominguez}, \citenamefont {Marquet}, \citenamefont {Xiao},\ and\
  \citenamefont {Yuan}}]{Dominguez:2011wm}%
  \BibitemOpen
  \bibfield  {author} {\bibinfo {author} {\bibfnamefont {F.}~\bibnamefont
  {Dominguez}}, \bibinfo {author} {\bibfnamefont {C.}~\bibnamefont {Marquet}},
  \bibinfo {author} {\bibfnamefont {B.-W.}\ \bibnamefont {Xiao}}, \ and\
  \bibinfo {author} {\bibfnamefont {F.}~\bibnamefont {Yuan}},\ }\href {\doibase
  10.1103/PhysRevD.83.105005} {\bibfield  {journal} {\bibinfo  {journal} {Phys.
  Rev. D}\ }\textbf {\bibinfo {volume} {83}},\ \bibinfo {pages} {105005}
  (\bibinfo {year} {2011})},\ \Eprint {http://arxiv.org/abs/1101.0715}
  {arXiv:1101.0715 [hep-ph]} \BibitemShut {NoStop}%
\bibitem [{\citenamefont {Altinoluk}\ \emph {et~al.}(2016)\citenamefont
  {Altinoluk}, \citenamefont {Armesto}, \citenamefont {Beuf},\ and\
  \citenamefont {Rezaeian}}]{Altinoluk:2015dpi}%
  \BibitemOpen
  \bibfield  {author} {\bibinfo {author} {\bibfnamefont {T.}~\bibnamefont
  {Altinoluk}}, \bibinfo {author} {\bibfnamefont {N.}~\bibnamefont {Armesto}},
  \bibinfo {author} {\bibfnamefont {G.}~\bibnamefont {Beuf}}, \ and\ \bibinfo
  {author} {\bibfnamefont {A.~H.}\ \bibnamefont {Rezaeian}},\ }\href {\doibase
  10.1016/j.physletb.2016.05.032} {\bibfield  {journal} {\bibinfo  {journal}
  {Phys. Lett. B}\ }\textbf {\bibinfo {volume} {758}},\ \bibinfo {pages} {373}
  (\bibinfo {year} {2016})},\ \Eprint {http://arxiv.org/abs/1511.07452}
  {arXiv:1511.07452 [hep-ph]} \BibitemShut {NoStop}%
\bibitem [{\citenamefont {Hatta}\ \emph {et~al.}(2016)\citenamefont {Hatta},
  \citenamefont {Xiao},\ and\ \citenamefont {Yuan}}]{Hatta:2016dxp}%
  \BibitemOpen
  \bibfield  {author} {\bibinfo {author} {\bibfnamefont {Y.}~\bibnamefont
  {Hatta}}, \bibinfo {author} {\bibfnamefont {B.-W.}\ \bibnamefont {Xiao}}, \
  and\ \bibinfo {author} {\bibfnamefont {F.}~\bibnamefont {Yuan}},\ }\href
  {\doibase 10.1103/PhysRevLett.116.202301} {\bibfield  {journal} {\bibinfo
  {journal} {Phys. Rev. Lett.}\ }\textbf {\bibinfo {volume} {116}},\ \bibinfo
  {pages} {202301} (\bibinfo {year} {2016})},\ \Eprint
  {http://arxiv.org/abs/1601.01585} {arXiv:1601.01585 [hep-ph]} \BibitemShut
  {NoStop}%
\bibitem [{\citenamefont {M\"antysaari}\ \emph {et~al.}(2020)\citenamefont
  {M\"antysaari}, \citenamefont {Mueller}, \citenamefont {Salazar},\ and\
  \citenamefont {Schenke}}]{Mantysaari:2019hkq}%
  \BibitemOpen
  \bibfield  {author} {\bibinfo {author} {\bibfnamefont {H.}~\bibnamefont
  {M\"antysaari}}, \bibinfo {author} {\bibfnamefont {N.}~\bibnamefont
  {Mueller}}, \bibinfo {author} {\bibfnamefont {F.}~\bibnamefont {Salazar}}, \
  and\ \bibinfo {author} {\bibfnamefont {B.}~\bibnamefont {Schenke}},\ }\href
  {\doibase 10.1103/PhysRevLett.124.112301} {\bibfield  {journal} {\bibinfo
  {journal} {Phys. Rev. Lett.}\ }\textbf {\bibinfo {volume} {124}},\ \bibinfo
  {pages} {112301} (\bibinfo {year} {2020})},\ \Eprint
  {http://arxiv.org/abs/1912.05586} {arXiv:1912.05586 [nucl-th]} \BibitemShut
  {NoStop}%
\bibitem [{\citenamefont {Salazar}\ and\ \citenamefont
  {Schenke}(2019)}]{Salazar:2019ncp}%
  \BibitemOpen
  \bibfield  {author} {\bibinfo {author} {\bibfnamefont {F.}~\bibnamefont
  {Salazar}}\ and\ \bibinfo {author} {\bibfnamefont {B.}~\bibnamefont
  {Schenke}},\ }\href {\doibase 10.1103/PhysRevD.100.034007} {\bibfield
  {journal} {\bibinfo  {journal} {Phys. Rev. D}\ }\textbf {\bibinfo {volume}
  {100}},\ \bibinfo {pages} {034007} (\bibinfo {year} {2019})},\ \Eprint
  {http://arxiv.org/abs/1905.03763} {arXiv:1905.03763 [hep-ph]} \BibitemShut
  {NoStop}%
\bibitem [{\citenamefont {Hatta}\ \emph {et~al.}(2021)\citenamefont {Hatta},
  \citenamefont {Xiao}, \citenamefont {Yuan},\ and\ \citenamefont
  {Zhou}}]{Hatta:2020bgy}%
  \BibitemOpen
  \bibfield  {author} {\bibinfo {author} {\bibfnamefont {Y.}~\bibnamefont
  {Hatta}}, \bibinfo {author} {\bibfnamefont {B.-W.}\ \bibnamefont {Xiao}},
  \bibinfo {author} {\bibfnamefont {F.}~\bibnamefont {Yuan}}, \ and\ \bibinfo
  {author} {\bibfnamefont {J.}~\bibnamefont {Zhou}},\ }\href {\doibase
  10.1103/PhysRevLett.126.142001} {\bibfield  {journal} {\bibinfo  {journal}
  {Phys. Rev. Lett.}\ }\textbf {\bibinfo {volume} {126}},\ \bibinfo {pages}
  {142001} (\bibinfo {year} {2021})},\ \Eprint
  {http://arxiv.org/abs/2010.10774} {arXiv:2010.10774 [hep-ph]} \BibitemShut
  {NoStop}%
\bibitem [{\citenamefont {Boer}\ \emph {et~al.}(2011)\citenamefont {Boer} \emph
  {et~al.}}]{Boer:2011fh}%
  \BibitemOpen
  \bibfield  {author} {\bibinfo {author} {\bibfnamefont {D.}~\bibnamefont
  {Boer}} \emph {et~al.},\ }\href@noop {} {\  (\bibinfo {year} {2011})},\
  \Eprint {http://arxiv.org/abs/1108.1713} {arXiv:1108.1713 [nucl-th]}
  \BibitemShut {NoStop}%
\bibitem [{\citenamefont {Kang}\ \emph {et~al.}(2021)\citenamefont {Kang},
  \citenamefont {Reiten}, \citenamefont {Shao},\ and\ \citenamefont
  {Terry}}]{Kang:2020xgk}%
  \BibitemOpen
  \bibfield  {author} {\bibinfo {author} {\bibfnamefont {Z.-B.}\ \bibnamefont
  {Kang}}, \bibinfo {author} {\bibfnamefont {J.}~\bibnamefont {Reiten}},
  \bibinfo {author} {\bibfnamefont {D.~Y.}\ \bibnamefont {Shao}}, \ and\
  \bibinfo {author} {\bibfnamefont {J.}~\bibnamefont {Terry}},\ }\href
  {\doibase 10.1007/JHEP05(2021)286} {\bibfield  {journal} {\bibinfo  {journal}
  {JHEP}\ }\textbf {\bibinfo {volume} {05}},\ \bibinfo {pages} {286} (\bibinfo
  {year} {2021})},\ \Eprint {http://arxiv.org/abs/2012.01756} {arXiv:2012.01756
  [hep-ph]} \BibitemShut {NoStop}%
\bibitem [{\citenamefont {del Castillo}\ \emph {et~al.}(2021)\citenamefont {del
  Castillo}, \citenamefont {Echevarria}, \citenamefont {Makris},\ and\
  \citenamefont {Scimemi}}]{delCastillo:2020omr}%
  \BibitemOpen
  \bibfield  {author} {\bibinfo {author} {\bibfnamefont {R.~F.}\ \bibnamefont
  {del Castillo}}, \bibinfo {author} {\bibfnamefont {M.~G.}\ \bibnamefont
  {Echevarria}}, \bibinfo {author} {\bibfnamefont {Y.}~\bibnamefont {Makris}},
  \ and\ \bibinfo {author} {\bibfnamefont {I.}~\bibnamefont {Scimemi}},\ }\href
  {\doibase 10.1007/JHEP01(2021)088} {\bibfield  {journal} {\bibinfo  {journal}
  {JHEP}\ }\textbf {\bibinfo {volume} {01}},\ \bibinfo {pages} {088} (\bibinfo
  {year} {2021})},\ \Eprint {http://arxiv.org/abs/2008.07531} {arXiv:2008.07531
  [hep-ph]} \BibitemShut {NoStop}%
\bibitem [{\citenamefont {Ji}\ \emph {et~al.}(2005)\citenamefont {Ji},
  \citenamefont {Ma},\ and\ \citenamefont {Yuan}}]{Ji:2004wu}%
  \BibitemOpen
  \bibfield  {author} {\bibinfo {author} {\bibfnamefont {X.-D.}\ \bibnamefont
  {Ji}}, \bibinfo {author} {\bibfnamefont {J.-P.}\ \bibnamefont {Ma}}, \ and\
  \bibinfo {author} {\bibfnamefont {F.}~\bibnamefont {Yuan}},\ }\href {\doibase
  10.1103/PhysRevD.71.034005} {\bibfield  {journal} {\bibinfo  {journal} {Phys.
  Rev. D}\ }\textbf {\bibinfo {volume} {71}},\ \bibinfo {pages} {034005}
  (\bibinfo {year} {2005})},\ \Eprint {http://arxiv.org/abs/hep-ph/0404183}
  {arXiv:hep-ph/0404183} \BibitemShut {NoStop}%
\bibitem [{\citenamefont {Hirai}\ \emph {et~al.}(2007)\citenamefont {Hirai},
  \citenamefont {Kumano}, \citenamefont {Nagai},\ and\ \citenamefont
  {Sudoh}}]{Hirai:2007cx}%
  \BibitemOpen
  \bibfield  {author} {\bibinfo {author} {\bibfnamefont {M.}~\bibnamefont
  {Hirai}}, \bibinfo {author} {\bibfnamefont {S.}~\bibnamefont {Kumano}},
  \bibinfo {author} {\bibfnamefont {T.~H.}\ \bibnamefont {Nagai}}, \ and\
  \bibinfo {author} {\bibfnamefont {K.}~\bibnamefont {Sudoh}},\ }\href
  {\doibase 10.1103/PhysRevD.75.094009} {\bibfield  {journal} {\bibinfo
  {journal} {Phys. Rev. D}\ }\textbf {\bibinfo {volume} {75}},\ \bibinfo
  {pages} {094009} (\bibinfo {year} {2007})},\ \Eprint
  {http://arxiv.org/abs/hep-ph/0702250} {arXiv:hep-ph/0702250} \BibitemShut
  {NoStop}%
\bibitem [{\citenamefont {Eskola}\ \emph {et~al.}(2009)\citenamefont {Eskola},
  \citenamefont {Paukkunen},\ and\ \citenamefont {Salgado}}]{Eskola:2009uj}%
  \BibitemOpen
  \bibfield  {author} {\bibinfo {author} {\bibfnamefont {K.~J.}\ \bibnamefont
  {Eskola}}, \bibinfo {author} {\bibfnamefont {H.}~\bibnamefont {Paukkunen}}, \
  and\ \bibinfo {author} {\bibfnamefont {C.~A.}\ \bibnamefont {Salgado}},\
  }\href {\doibase 10.1088/1126-6708/2009/04/065} {\bibfield  {journal}
  {\bibinfo  {journal} {JHEP}\ }\textbf {\bibinfo {volume} {04}},\ \bibinfo
  {pages} {065} (\bibinfo {year} {2009})},\ \Eprint
  {http://arxiv.org/abs/0902.4154} {arXiv:0902.4154 [hep-ph]} \BibitemShut
  {NoStop}%
\bibitem [{\citenamefont {Osborne}\ and\ \citenamefont
  {Wang}(2002)}]{Osborne:2002st}%
  \BibitemOpen
  \bibfield  {author} {\bibinfo {author} {\bibfnamefont {J.}~\bibnamefont
  {Osborne}}\ and\ \bibinfo {author} {\bibfnamefont {X.-N.}\ \bibnamefont
  {Wang}},\ }\href {\doibase 10.1016/S0375-9474(02)01085-0} {\bibfield
  {journal} {\bibinfo  {journal} {Nucl. Phys. A}\ }\textbf {\bibinfo {volume}
  {710}},\ \bibinfo {pages} {281} (\bibinfo {year} {2002})},\ \Eprint
  {http://arxiv.org/abs/hep-ph/0204046} {arXiv:hep-ph/0204046} \BibitemShut
  {NoStop}%
\bibitem [{\citenamefont {Hautmann}\ \emph {et~al.}(2014)\citenamefont
  {Hautmann}, \citenamefont {Jung}, \citenamefont {Kr\"amer}, \citenamefont
  {Mulders}, \citenamefont {Nocera}, \citenamefont {Rogers},\ and\
  \citenamefont {Signori}}]{Hautmann:2014kza}%
  \BibitemOpen
  \bibfield  {author} {\bibinfo {author} {\bibfnamefont {F.}~\bibnamefont
  {Hautmann}}, \bibinfo {author} {\bibfnamefont {H.}~\bibnamefont {Jung}},
  \bibinfo {author} {\bibfnamefont {M.}~\bibnamefont {Kr\"amer}}, \bibinfo
  {author} {\bibfnamefont {P.~J.}\ \bibnamefont {Mulders}}, \bibinfo {author}
  {\bibfnamefont {E.~R.}\ \bibnamefont {Nocera}}, \bibinfo {author}
  {\bibfnamefont {T.~C.}\ \bibnamefont {Rogers}}, \ and\ \bibinfo {author}
  {\bibfnamefont {A.}~\bibnamefont {Signori}},\ }\href {\doibase
  10.1140/epjc/s10052-014-3220-9} {\bibfield  {journal} {\bibinfo  {journal}
  {Eur. Phys. J. C}\ }\textbf {\bibinfo {volume} {74}},\ \bibinfo {pages}
  {3220} (\bibinfo {year} {2014})},\ \Eprint {http://arxiv.org/abs/1408.3015}
  {arXiv:1408.3015 [hep-ph]} \BibitemShut {NoStop}%
\bibitem [{\citenamefont {Bermudez~Martinez}\ \emph {et~al.}(2019)\citenamefont
  {Bermudez~Martinez}, \citenamefont {Connor}, \citenamefont {Jung},
  \citenamefont {Lelek}, \citenamefont {{\v{Z}}leb{\v{c}}{\'\i}k},
  \citenamefont {Hautmann},\ and\ \citenamefont {Radescu}}]{Martinez:2018jxt}%
  \BibitemOpen
  \bibfield  {author} {\bibinfo {author} {\bibfnamefont {A.}~\bibnamefont
  {Bermudez~Martinez}}, \bibinfo {author} {\bibfnamefont {P.}~\bibnamefont
  {Connor}}, \bibinfo {author} {\bibfnamefont {H.}~\bibnamefont {Jung}},
  \bibinfo {author} {\bibfnamefont {A.}~\bibnamefont {Lelek}}, \bibinfo
  {author} {\bibfnamefont {R.}~\bibnamefont {{\v{Z}}leb{\v{c}}{\'\i}k}},
  \bibinfo {author} {\bibfnamefont {F.}~\bibnamefont {Hautmann}}, \ and\
  \bibinfo {author} {\bibfnamefont {V.}~\bibnamefont {Radescu}},\ }\href
  {\doibase 10.1103/PhysRevD.99.074008} {\bibfield  {journal} {\bibinfo
  {journal} {Phys. Rev. D}\ }\textbf {\bibinfo {volume} {99}},\ \bibinfo
  {pages} {074008} (\bibinfo {year} {2019})},\ \Eprint
  {http://arxiv.org/abs/1804.11152} {arXiv:1804.11152 [hep-ph]} \BibitemShut
  {NoStop}%
\bibitem [{\citenamefont {Collins}(2003)}]{Collins:2003fm}%
  \BibitemOpen
  \bibfield  {author} {\bibinfo {author} {\bibfnamefont {J.~C.}\ \bibnamefont
  {Collins}},\ }\href@noop {} {\bibfield  {journal} {\bibinfo  {journal} {Acta
  Phys. Polon. B}\ }\textbf {\bibinfo {volume} {34}},\ \bibinfo {pages} {3103}
  (\bibinfo {year} {2003})},\ \Eprint {http://arxiv.org/abs/hep-ph/0304122}
  {arXiv:hep-ph/0304122} \BibitemShut {NoStop}%
\bibitem [{\citenamefont {Collins}\ \emph {et~al.}(2016)\citenamefont
  {Collins}, \citenamefont {Gamberg}, \citenamefont {Prokudin}, \citenamefont
  {Rogers}, \citenamefont {Sato},\ and\ \citenamefont
  {Wang}}]{Collins:2016hqq}%
  \BibitemOpen
  \bibfield  {author} {\bibinfo {author} {\bibfnamefont {J.}~\bibnamefont
  {Collins}}, \bibinfo {author} {\bibfnamefont {L.}~\bibnamefont {Gamberg}},
  \bibinfo {author} {\bibfnamefont {A.}~\bibnamefont {Prokudin}}, \bibinfo
  {author} {\bibfnamefont {T.~C.}\ \bibnamefont {Rogers}}, \bibinfo {author}
  {\bibfnamefont {N.}~\bibnamefont {Sato}}, \ and\ \bibinfo {author}
  {\bibfnamefont {B.}~\bibnamefont {Wang}},\ }\href {\doibase
  10.1103/PhysRevD.94.034014} {\bibfield  {journal} {\bibinfo  {journal} {Phys.
  Rev. D}\ }\textbf {\bibinfo {volume} {94}},\ \bibinfo {pages} {034014}
  (\bibinfo {year} {2016})},\ \Eprint {http://arxiv.org/abs/1605.00671}
  {arXiv:1605.00671 [hep-ph]} \BibitemShut {NoStop}%
\bibitem [{\citenamefont {Gamberg}\ \emph {et~al.}(2018)\citenamefont
  {Gamberg}, \citenamefont {Metz}, \citenamefont {Pitonyak},\ and\
  \citenamefont {Prokudin}}]{Gamberg:2017jha}%
  \BibitemOpen
  \bibfield  {author} {\bibinfo {author} {\bibfnamefont {L.}~\bibnamefont
  {Gamberg}}, \bibinfo {author} {\bibfnamefont {A.}~\bibnamefont {Metz}},
  \bibinfo {author} {\bibfnamefont {D.}~\bibnamefont {Pitonyak}}, \ and\
  \bibinfo {author} {\bibfnamefont {A.}~\bibnamefont {Prokudin}},\ }\href
  {\doibase 10.1016/j.physletb.2018.03.024} {\bibfield  {journal} {\bibinfo
  {journal} {Phys. Lett. B}\ }\textbf {\bibinfo {volume} {781}},\ \bibinfo
  {pages} {443} (\bibinfo {year} {2018})},\ \Eprint
  {http://arxiv.org/abs/1712.08116} {arXiv:1712.08116 [hep-ph]} \BibitemShut
  {NoStop}%
\bibitem [{\citenamefont {Buckley}\ \emph {et~al.}(2015)\citenamefont
  {Buckley}, \citenamefont {Ferrando}, \citenamefont {Lloyd}, \citenamefont
  {Nordstr\"om}, \citenamefont {Page}, \citenamefont {R\"ufenacht},
  \citenamefont {Sch\"onherr},\ and\ \citenamefont {Watt}}]{Buckley:2014ana}%
  \BibitemOpen
  \bibfield  {author} {\bibinfo {author} {\bibfnamefont {A.}~\bibnamefont
  {Buckley}}, \bibinfo {author} {\bibfnamefont {J.}~\bibnamefont {Ferrando}},
  \bibinfo {author} {\bibfnamefont {S.}~\bibnamefont {Lloyd}}, \bibinfo
  {author} {\bibfnamefont {K.}~\bibnamefont {Nordstr\"om}}, \bibinfo {author}
  {\bibfnamefont {B.}~\bibnamefont {Page}}, \bibinfo {author} {\bibfnamefont
  {M.}~\bibnamefont {R\"ufenacht}}, \bibinfo {author} {\bibfnamefont
  {M.}~\bibnamefont {Sch\"onherr}}, \ and\ \bibinfo {author} {\bibfnamefont
  {G.}~\bibnamefont {Watt}},\ }\href {\doibase 10.1140/epjc/s10052-015-3318-8}
  {\bibfield  {journal} {\bibinfo  {journal} {Eur. Phys. J. C}\ }\textbf
  {\bibinfo {volume} {75}},\ \bibinfo {pages} {132} (\bibinfo {year} {2015})},\
  \Eprint {http://arxiv.org/abs/1412.7420} {arXiv:1412.7420 [hep-ph]}
  \BibitemShut {NoStop}%
\bibitem [{\citenamefont {Gribov}\ \emph {et~al.}(1983)\citenamefont {Gribov},
  \citenamefont {Levin},\ and\ \citenamefont {Ryskin}}]{Gribov:1984tu}%
  \BibitemOpen
  \bibfield  {author} {\bibinfo {author} {\bibfnamefont {L.~V.}\ \bibnamefont
  {Gribov}}, \bibinfo {author} {\bibfnamefont {E.~M.}\ \bibnamefont {Levin}}, \
  and\ \bibinfo {author} {\bibfnamefont {M.~G.}\ \bibnamefont {Ryskin}},\
  }\href {\doibase 10.1016/0370-1573(83)90022-4} {\bibfield  {journal}
  {\bibinfo  {journal} {Phys. Rept.}\ }\textbf {\bibinfo {volume} {100}},\
  \bibinfo {pages} {1} (\bibinfo {year} {1983})}\BibitemShut {NoStop}%
\bibitem [{\citenamefont {Mueller}\ and\ \citenamefont
  {Qiu}(1986)}]{Mueller:1985wy}%
  \BibitemOpen
  \bibfield  {author} {\bibinfo {author} {\bibfnamefont {A.~H.}\ \bibnamefont
  {Mueller}}\ and\ \bibinfo {author} {\bibfnamefont {J.-W.}\ \bibnamefont
  {Qiu}},\ }\href {\doibase 10.1016/0550-3213(86)90164-1} {\bibfield  {journal}
  {\bibinfo  {journal} {Nucl. Phys. B}\ }\textbf {\bibinfo {volume} {268}},\
  \bibinfo {pages} {427} (\bibinfo {year} {1986})}\BibitemShut {NoStop}%
\bibitem [{\citenamefont {Mueller}(1990)}]{Mueller:1989st}%
  \BibitemOpen
  \bibfield  {author} {\bibinfo {author} {\bibfnamefont {A.~H.}\ \bibnamefont
  {Mueller}},\ }\href {\doibase 10.1016/0550-3213(90)90173-B} {\bibfield
  {journal} {\bibinfo  {journal} {Nucl. Phys. B}\ }\textbf {\bibinfo {volume}
  {335}},\ \bibinfo {pages} {115} (\bibinfo {year} {1990})}\BibitemShut
  {NoStop}%
\bibitem [{\citenamefont {McLerran}\ and\ \citenamefont
  {Venugopalan}(1994{\natexlab{a}})}]{McLerran:1993ni}%
  \BibitemOpen
  \bibfield  {author} {\bibinfo {author} {\bibfnamefont {L.~D.}\ \bibnamefont
  {McLerran}}\ and\ \bibinfo {author} {\bibfnamefont {R.}~\bibnamefont
  {Venugopalan}},\ }\href {\doibase 10.1103/PhysRevD.49.2233} {\bibfield
  {journal} {\bibinfo  {journal} {Phys. Rev. D}\ }\textbf {\bibinfo {volume}
  {49}},\ \bibinfo {pages} {2233} (\bibinfo {year} {1994}{\natexlab{a}})},\
  \Eprint {http://arxiv.org/abs/hep-ph/9309289} {arXiv:hep-ph/9309289}
  \BibitemShut {NoStop}%
\bibitem [{\citenamefont {McLerran}\ and\ \citenamefont
  {Venugopalan}(1994{\natexlab{b}})}]{McLerran:1993ka}%
  \BibitemOpen
  \bibfield  {author} {\bibinfo {author} {\bibfnamefont {L.~D.}\ \bibnamefont
  {McLerran}}\ and\ \bibinfo {author} {\bibfnamefont {R.}~\bibnamefont
  {Venugopalan}},\ }\href {\doibase 10.1103/PhysRevD.49.3352} {\bibfield
  {journal} {\bibinfo  {journal} {Phys. Rev. D}\ }\textbf {\bibinfo {volume}
  {49}},\ \bibinfo {pages} {3352} (\bibinfo {year} {1994}{\natexlab{b}})},\
  \Eprint {http://arxiv.org/abs/hep-ph/9311205} {arXiv:hep-ph/9311205}
  \BibitemShut {NoStop}%
\bibitem [{\citenamefont {Mueller}(1999)}]{Mueller:1999wm}%
  \BibitemOpen
  \bibfield  {author} {\bibinfo {author} {\bibfnamefont {A.~H.}\ \bibnamefont
  {Mueller}},\ }\href {\doibase 10.1016/S0550-3213(99)00394-6} {\bibfield
  {journal} {\bibinfo  {journal} {Nucl. Phys. B}\ }\textbf {\bibinfo {volume}
  {558}},\ \bibinfo {pages} {285} (\bibinfo {year} {1999})},\ \Eprint
  {http://arxiv.org/abs/hep-ph/9904404} {arXiv:hep-ph/9904404} \BibitemShut
  {NoStop}%
\bibitem [{\citenamefont {Qiu}(1987)}]{Qiu:1986wh}%
  \BibitemOpen
  \bibfield  {author} {\bibinfo {author} {\bibfnamefont {J.-W.}\ \bibnamefont
  {Qiu}},\ }\href {\doibase 10.1016/0550-3213(87)90494-9} {\bibfield  {journal}
  {\bibinfo  {journal} {Nucl. Phys. B}\ }\textbf {\bibinfo {volume} {291}},\
  \bibinfo {pages} {746} (\bibinfo {year} {1987})}\BibitemShut {NoStop}%
\bibitem [{\citenamefont {Brodsky}\ and\ \citenamefont
  {Lu}(1990)}]{Brodsky:1989qz}%
  \BibitemOpen
  \bibfield  {author} {\bibinfo {author} {\bibfnamefont {S.~J.}\ \bibnamefont
  {Brodsky}}\ and\ \bibinfo {author} {\bibfnamefont {H.~J.}\ \bibnamefont
  {Lu}},\ }\href {\doibase 10.1103/PhysRevLett.64.1342} {\bibfield  {journal}
  {\bibinfo  {journal} {Phys. Rev. Lett.}\ }\textbf {\bibinfo {volume} {64}},\
  \bibinfo {pages} {1342} (\bibinfo {year} {1990})}\BibitemShut {NoStop}%
\bibitem [{\citenamefont {Eskola}\ \emph {et~al.}(1994)\citenamefont {Eskola},
  \citenamefont {Qiu},\ and\ \citenamefont {Wang}}]{Eskola:1993mb}%
  \BibitemOpen
  \bibfield  {author} {\bibinfo {author} {\bibfnamefont {K.~J.}\ \bibnamefont
  {Eskola}}, \bibinfo {author} {\bibfnamefont {J.-W.}\ \bibnamefont {Qiu}}, \
  and\ \bibinfo {author} {\bibfnamefont {X.-N.}\ \bibnamefont {Wang}},\ }\href
  {\doibase 10.1103/PhysRevLett.72.36} {\bibfield  {journal} {\bibinfo
  {journal} {Phys. Rev. Lett.}\ }\textbf {\bibinfo {volume} {72}},\ \bibinfo
  {pages} {36} (\bibinfo {year} {1994})},\ \Eprint
  {http://arxiv.org/abs/nucl-th/9307025} {arXiv:nucl-th/9307025} \BibitemShut
  {NoStop}%
\bibitem [{\citenamefont {Ashman}\ \emph {et~al.}(1988)\citenamefont {Ashman}
  \emph {et~al.}}]{Ashman:1988bf}%
  \BibitemOpen
  \bibfield  {author} {\bibinfo {author} {\bibfnamefont {J.}~\bibnamefont
  {Ashman}} \emph {et~al.} (\bibinfo {collaboration} {European Muon}),\ }\href
  {\doibase 10.1016/0370-2693(88)91872-2} {\bibfield  {journal} {\bibinfo
  {journal} {Phys. Lett. B}\ }\textbf {\bibinfo {volume} {202}},\ \bibinfo
  {pages} {603} (\bibinfo {year} {1988})}\BibitemShut {NoStop}%
\bibitem [{\citenamefont {Arneodo}\ \emph {et~al.}(1988)\citenamefont {Arneodo}
  \emph {et~al.}}]{Arneodo:1988aa}%
  \BibitemOpen
  \bibfield  {author} {\bibinfo {author} {\bibfnamefont {M.}~\bibnamefont
  {Arneodo}} \emph {et~al.} (\bibinfo {collaboration} {European Muon}),\ }\href
  {\doibase 10.1016/0370-2693(88)91900-4} {\bibfield  {journal} {\bibinfo
  {journal} {Phys. Lett. B}\ }\textbf {\bibinfo {volume} {211}},\ \bibinfo
  {pages} {493} (\bibinfo {year} {1988})}\BibitemShut {NoStop}%
\bibitem [{\citenamefont {Kharzeev}\ and\ \citenamefont
  {Levin}(2001)}]{Kharzeev:2001gp}%
  \BibitemOpen
  \bibfield  {author} {\bibinfo {author} {\bibfnamefont {D.}~\bibnamefont
  {Kharzeev}}\ and\ \bibinfo {author} {\bibfnamefont {E.}~\bibnamefont
  {Levin}},\ }\href {\doibase 10.1016/S0370-2693(01)01309-0} {\bibfield
  {journal} {\bibinfo  {journal} {Phys. Lett. B}\ }\textbf {\bibinfo {volume}
  {523}},\ \bibinfo {pages} {79} (\bibinfo {year} {2001})},\ \Eprint
  {http://arxiv.org/abs/nucl-th/0108006} {arXiv:nucl-th/0108006} \BibitemShut
  {NoStop}%
\bibitem [{\citenamefont {Dumitru}\ \emph {et~al.}(2012)\citenamefont
  {Dumitru}, \citenamefont {Kharzeev}, \citenamefont {Levin},\ and\
  \citenamefont {Nara}}]{Dumitru:2011wq}%
  \BibitemOpen
  \bibfield  {author} {\bibinfo {author} {\bibfnamefont {A.}~\bibnamefont
  {Dumitru}}, \bibinfo {author} {\bibfnamefont {D.~E.}\ \bibnamefont
  {Kharzeev}}, \bibinfo {author} {\bibfnamefont {E.~M.}\ \bibnamefont {Levin}},
  \ and\ \bibinfo {author} {\bibfnamefont {Y.}~\bibnamefont {Nara}},\ }\href
  {\doibase 10.1103/PhysRevC.85.044920} {\bibfield  {journal} {\bibinfo
  {journal} {Phys. Rev. C}\ }\textbf {\bibinfo {volume} {85}},\ \bibinfo
  {pages} {044920} (\bibinfo {year} {2012})},\ \Eprint
  {http://arxiv.org/abs/1111.3031} {arXiv:1111.3031 [hep-ph]} \BibitemShut
  {NoStop}%
\bibitem [{\citenamefont {Golec-Biernat}\ and\ \citenamefont
  {Wusthoff}(1999)}]{GolecBiernat:1999qd}%
  \BibitemOpen
  \bibfield  {author} {\bibinfo {author} {\bibfnamefont {K.~J.}\ \bibnamefont
  {Golec-Biernat}}\ and\ \bibinfo {author} {\bibfnamefont {M.}~\bibnamefont
  {Wusthoff}},\ }\href {\doibase 10.1103/PhysRevD.60.114023} {\bibfield
  {journal} {\bibinfo  {journal} {Phys. Rev. D}\ }\textbf {\bibinfo {volume}
  {60}},\ \bibinfo {pages} {114023} (\bibinfo {year} {1999})},\ \Eprint
  {http://arxiv.org/abs/hep-ph/9903358} {arXiv:hep-ph/9903358} \BibitemShut
  {NoStop}%
\bibitem [{\citenamefont {Liou}\ \emph {et~al.}(2013)\citenamefont {Liou},
  \citenamefont {Mueller},\ and\ \citenamefont {Wu}}]{Liou:2013qya}%
  \BibitemOpen
  \bibfield  {author} {\bibinfo {author} {\bibfnamefont {T.}~\bibnamefont
  {Liou}}, \bibinfo {author} {\bibfnamefont {A.~H.}\ \bibnamefont {Mueller}}, \
  and\ \bibinfo {author} {\bibfnamefont {B.}~\bibnamefont {Wu}},\ }\href
  {\doibase 10.1016/j.nuclphysa.2013.08.005} {\bibfield  {journal} {\bibinfo
  {journal} {Nucl. Phys. A}\ }\textbf {\bibinfo {volume} {916}},\ \bibinfo
  {pages} {102} (\bibinfo {year} {2013})},\ \Eprint
  {http://arxiv.org/abs/1304.7677} {arXiv:1304.7677 [hep-ph]} \BibitemShut
  {NoStop}%
\bibitem [{\citenamefont {Wu}(2014)}]{Wu:2014nca}%
  \BibitemOpen
  \bibfield  {author} {\bibinfo {author} {\bibfnamefont {B.}~\bibnamefont
  {Wu}},\ }\href {\doibase 10.1007/JHEP12(2014)081} {\bibfield  {journal}
  {\bibinfo  {journal} {JHEP}\ }\textbf {\bibinfo {volume} {12}},\ \bibinfo
  {pages} {081} (\bibinfo {year} {2014})},\ \Eprint
  {http://arxiv.org/abs/1408.5459} {arXiv:1408.5459 [hep-ph]} \BibitemShut
  {NoStop}%
\bibitem [{\citenamefont {Blaizot}\ and\ \citenamefont
  {Mehtar-Tani}(2014)}]{Blaizot:2014bha}%
  \BibitemOpen
  \bibfield  {author} {\bibinfo {author} {\bibfnamefont {J.-P.}\ \bibnamefont
  {Blaizot}}\ and\ \bibinfo {author} {\bibfnamefont {Y.}~\bibnamefont
  {Mehtar-Tani}},\ }\href {\doibase 10.1016/j.nuclphysa.2014.05.018} {\bibfield
   {journal} {\bibinfo  {journal} {Nucl. Phys. A}\ }\textbf {\bibinfo {volume}
  {929}},\ \bibinfo {pages} {202} (\bibinfo {year} {2014})},\ \Eprint
  {http://arxiv.org/abs/1403.2323} {arXiv:1403.2323 [hep-ph]} \BibitemShut
  {NoStop}%
\bibitem [{\citenamefont {Ghiglieri}\ and\ \citenamefont
  {Teaney}(2015)}]{Ghiglieri:2015zma}%
  \BibitemOpen
  \bibfield  {author} {\bibinfo {author} {\bibfnamefont {J.}~\bibnamefont
  {Ghiglieri}}\ and\ \bibinfo {author} {\bibfnamefont {D.}~\bibnamefont
  {Teaney}},\ }\href {\doibase 10.1142/S0218301315300131} {\bibfield  {journal}
  {\bibinfo  {journal} {Int. J. Mod. Phys. E}\ }\textbf {\bibinfo {volume}
  {24}},\ \bibinfo {pages} {1530013} (\bibinfo {year} {2015})},\ \Eprint
  {http://arxiv.org/abs/1502.03730} {arXiv:1502.03730 [hep-ph]} \BibitemShut
  {NoStop}%
\bibitem [{\citenamefont {Blaizot}\ and\ \citenamefont
  {Dominguez}(2019)}]{Blaizot:2019muz}%
  \BibitemOpen
  \bibfield  {author} {\bibinfo {author} {\bibfnamefont {J.-P.}\ \bibnamefont
  {Blaizot}}\ and\ \bibinfo {author} {\bibfnamefont {F.}~\bibnamefont
  {Dominguez}},\ }\href {\doibase 10.1103/PhysRevD.99.054005} {\bibfield
  {journal} {\bibinfo  {journal} {Phys. Rev. D}\ }\textbf {\bibinfo {volume}
  {99}},\ \bibinfo {pages} {054005} (\bibinfo {year} {2019})},\ \Eprint
  {http://arxiv.org/abs/1901.01448} {arXiv:1901.01448 [hep-ph]} \BibitemShut
  {NoStop}%
\bibitem [{\citenamefont {Liu}\ \emph {et~al.}(2019)\citenamefont {Liu},
  \citenamefont {Ringer}, \citenamefont {Vogelsang},\ and\ \citenamefont
  {Yuan}}]{Liu:2018trl}%
  \BibitemOpen
  \bibfield  {author} {\bibinfo {author} {\bibfnamefont {X.}~\bibnamefont
  {Liu}}, \bibinfo {author} {\bibfnamefont {F.}~\bibnamefont {Ringer}},
  \bibinfo {author} {\bibfnamefont {W.}~\bibnamefont {Vogelsang}}, \ and\
  \bibinfo {author} {\bibfnamefont {F.}~\bibnamefont {Yuan}},\ }\href {\doibase
  10.1103/PhysRevLett.122.192003} {\bibfield  {journal} {\bibinfo  {journal}
  {Phys. Rev. Lett.}\ }\textbf {\bibinfo {volume} {122}},\ \bibinfo {pages}
  {192003} (\bibinfo {year} {2019})},\ \Eprint
  {http://arxiv.org/abs/1812.08077} {arXiv:1812.08077 [hep-ph]} \BibitemShut
  {NoStop}%
\bibitem [{\citenamefont {Abdul~Khalek}\ \emph {et~al.}(2021)\citenamefont
  {Abdul~Khalek} \emph {et~al.}}]{AbdulKhalek:2021gbh}%
  \BibitemOpen
  \bibfield  {author} {\bibinfo {author} {\bibfnamefont {R.}~\bibnamefont
  {Abdul~Khalek}} \emph {et~al.},\ }\href@noop {} {\  (\bibinfo {year}
  {2021})},\ \Eprint {http://arxiv.org/abs/2103.05419} {arXiv:2103.05419
  [physics.ins-det]} \BibitemShut {NoStop}%
\bibitem [{DIS(2010)}]{DIS}%
  \BibitemOpen
  \href {https://wiki.bnl.gov/eic/index.php/DIS_Kinematics} {\enquote {\bibinfo
  {title} {{DIS Kinematics - EIC}},}\ } (\bibinfo {year} {2010})\BibitemShut
  {NoStop}%
\bibitem [{\citenamefont {Hautmann}\ \emph {et~al.}(2018)\citenamefont
  {Hautmann}, \citenamefont {Jung}, \citenamefont {Lelek}, \citenamefont
  {Radescu},\ and\ \citenamefont {Zlebcik}}]{Hautmann:2017fcj}%
  \BibitemOpen
  \bibfield  {author} {\bibinfo {author} {\bibfnamefont {F.}~\bibnamefont
  {Hautmann}}, \bibinfo {author} {\bibfnamefont {H.}~\bibnamefont {Jung}},
  \bibinfo {author} {\bibfnamefont {A.}~\bibnamefont {Lelek}}, \bibinfo
  {author} {\bibfnamefont {V.}~\bibnamefont {Radescu}}, \ and\ \bibinfo
  {author} {\bibfnamefont {R.}~\bibnamefont {Zlebcik}},\ }\href {\doibase
  10.1007/JHEP01(2018)070} {\bibfield  {journal} {\bibinfo  {journal} {JHEP}\
  }\textbf {\bibinfo {volume} {01}},\ \bibinfo {pages} {070} (\bibinfo {year}
  {2018})},\ \Eprint {http://arxiv.org/abs/1708.03279} {arXiv:1708.03279
  [hep-ph]} \BibitemShut {NoStop}%
\bibitem [{\citenamefont {Cronin}\ \emph {et~al.}(1975)\citenamefont {Cronin},
  \citenamefont {Frisch}, \citenamefont {Shochet}, \citenamefont {Boymond},
  \citenamefont {Mermod}, \citenamefont {Piroue},\ and\ \citenamefont
  {Sumner}}]{Cronin:1974zm}%
  \BibitemOpen
  \bibfield  {author} {\bibinfo {author} {\bibfnamefont {J.~W.}\ \bibnamefont
  {Cronin}}, \bibinfo {author} {\bibfnamefont {H.~J.}\ \bibnamefont {Frisch}},
  \bibinfo {author} {\bibfnamefont {M.~J.}\ \bibnamefont {Shochet}}, \bibinfo
  {author} {\bibfnamefont {J.~P.}\ \bibnamefont {Boymond}}, \bibinfo {author}
  {\bibfnamefont {R.}~\bibnamefont {Mermod}}, \bibinfo {author} {\bibfnamefont
  {P.~A.}\ \bibnamefont {Piroue}}, \ and\ \bibinfo {author} {\bibfnamefont
  {R.~L.}\ \bibnamefont {Sumner}},\ }\href {\doibase 10.1103/PhysRevD.11.3105}
  {\bibfield  {journal} {\bibinfo  {journal} {Phys. Rev. D}\ }\textbf {\bibinfo
  {volume} {11}},\ \bibinfo {pages} {3105} (\bibinfo {year}
  {1975})}\BibitemShut {NoStop}%
\end{thebibliography}%
\end{document}